\documentclass[12pt,onecolumn]{IEEEtran}

\makeatletter
\def\ps@headings{%
\def\@oddhead{\mbox{}\scriptsize\rightmark \hfil \thepage}%
\def\@evenhead{\scriptsize\thepage \hfil \leftmark\mbox{}}%
\def\@oddfoot{}%
\def\@evenfoot{}}
\makeatother 
\usepackage{mathrsfs}
\usepackage{amsfonts}
\usepackage{array}
\usepackage{amssymb}
\usepackage{amsmath}
\usepackage{graphicx}
\usepackage{multirow}
\usepackage{amsthm}

\usepackage[lined,ruled,commentsnumbered]{algorithm2e}
\usepackage{epstopdf}
\usepackage{amssymb,amsmath}
\usepackage{multirow,url}
\usepackage{color,cite}

\theoremstyle{plain}
\newtheorem{theorem}{Theorem}
\newtheorem{lemma}{Lemma}

\newtheorem{definition}{Definition}
\newtheorem{proposition}{Proposition}

\newtheorem{assumption}{Assumption}

\ifodd 121
\newcommand{\rev}[1]{{\color{blue}#1}} %revise of the text
\newcommand{\com}[1]{\textbf{\color{red} (COMMENT: #1) }} %comment of the text
\newcommand{\comg}[1]{\textbf{\color{green} (COMMENT: #1)}}
\newcommand{\response}[1]{\textbf{\color{green} (RESPONSE: #1)}} %response to comment
\else
\newcommand{\rev}[1]{#1}

\newcommand{\com}[1]{}
\newcommand{\comg}[1]{}
\newcommand{\response}[1]{}
\fi
\def\Ex{\mathrm{E}}

\def\db{{database}}

\def\dbs{{databases}}

\def\eu{{WSD}}
\def\eus{{WSDs}}
\def\Eu{{WSD}}
\def\Eus{{WSDs}}

\def\M{M}            %%% the number of DB
\def\Mset{\mathcal{\M}}   %%% the set of DBs
\def\m{m}           %% index of datbase
\def\n{n}           %% index of datbase

\def\dbbm{{\m}}
\def\dbbmm{{\m-1}}
\def\dbbo{{1}}

\def\dbbM{{\M}}

\def\ch{{channel}}
\def\chs{{channels}}

\def\p{\pi}           %% advance service subcription fee
\def\bp{\boldsymbol{\p}}
\def\bpm{\boldsymbol{\p}_{-\m}}
\def\pm{\p_{\m}}
\def\pM{\p_{\M}}
\def\pn{\p_{\n}}
\def\po{\p_{1}}
\def\ps{c}
\def\pmm{\p_{\m - 1}}
\def\co{\c_{1}}
\def\cm{\c_{\m}}

\def\gy{g}  %% network effect
\def\s{l}             %% service type

\def\B{\textsf{b}}  %% choose basic service
\def\A{\textsf{a}} %% choose dbm's advanced service
\def\Am{\textsf{\m}} %% choose dbm's advanced service
\def\Ao{{1}}
\def\S{\textsf{s}}  %% choose sensing service

\def\RB{B}  %% expected data rate of random
\def\RA{A}  %% expected data rate of random
\def\RAm{A_{\m}}  %% expected data rate of DB1's advanced serivce
\def\RAn{A_{\n}}  %% expected data rate of DB1's advanced serivce
\def\RAmo{A_{\m-1}}  %% expected data rate of DB1's advanced serivce
\def\RAo{A_{1}}  %% expected data rate of DB1's advanced serivce
\def\RS{S}  %% expected data rate of DB2's advanced serivce
\def\RAM{A_{\M}}  %% expected data rate of DB1's advanced serivce

\def\th{\theta}      %% end-users preference
\def\thab{\th_{\textsc{AB}}}  %% threshold between random and advanced choice
\def\thsa{\th_{\textsc{SA}}}  %% threshold between advanced choice and sensing
\def\thsb{\th_{\textsc{SB}}}  %% threshold between random and sensing choice

\def\thb{\th_{\textsc{B}}}  %% threshold between random and advanced choice
\def\ths{\th_{\textsc{S}}}  %% threshold between advanced choice and sensing
\def\tham{\th_{{\m}}}  %% threshold between different db
\def\thamm{\th_{{\m+1}}}  %% threshold between different db
\def\thaM{\th_{{\M}}}  %% threshold between different db

\def\U{\Pi}        %% end-user's utility
\def\Ueu{\U^{\textsc{eu}}}
\def\Udb{\U^{\textsc{db}}}
\def\Udbo{\U^{\textsc{db}}_{1}}
\def\Udbm{\U^{\textsc{db}}_{\m}}
\def\Udbm{\U^{\textsc{db}}_{\m}}
\def\Ur{\widetilde{\U}}
\def\Urdb{\Ur^{\textsc{db}}}
\def\Urdbm{\Ur^{\textsc{db}}_{\m}}

\def\Prob{\eta}      %% percentage of end-users subscribe advanced servie
\def\BProb{\boldsymbol{\eta}}      %% the set of percentage of end-users subscribe advanced servie
\def\Probs{\Prob_{s}}      %% fixed percentage of end-users subscribe advanced servie of db2
\def\Probb{\Prob_{b} }     %% fixed percentage of end-users subscribe basic service

\def\BProba{\boldsymbol{\eta}}      %% the set of percentage of end-users subscribe advanced servie
\def\BProbam{\boldsymbol{\eta}_{-\m}}      %% the set of percentage of end-users subscribe advanced servie
\def\Probam{\Prob_{{\m}}}      %% fixed percentage of end-users subscribe advanced servie of db1
\def\Proban{\Prob_{{\n}}}      %% fixed percentage of end-users subscribe advanced servie of db1
\def\Probamo{\Prob_{{\m-1}}}      %% fixed percentage of end-users subscribe advanced servie of db1
\def\Probamm{\Prob_{{\m+1}}}      %% fixed percentage of end-users subscribe advanced servie of db1
\def\Proba{\Prob}      %% fixed percentage of end-users subscribe advanced servie of db1
\def\Probao{\Prob_{{1}}}      %% fixed percentage of end-users subscribe advanced servie of db1
\def\Probaoo{\Prob_{{2}}}      %% fixed percentage of end-users subscribe advanced servie of db1
\def\ProbaM{\Prob_{{\M}}}      %% fixed percentage of end-users subscribe advanced servie of db1
\def\ProbaMM{\Prob_{{\M - 1}}}      %% fixed percentage of end-users subscribe advanced servie of db1

\def\tv{{channel}}
\def\tvs{{channels}}
\def\tvn{{TV white space network}}

\def\eq{\triangleq}
\def\t{t}

\def\InfTV{U}         %% interference coming form tv station
\def\InfOut{V}        %% interference coming form other DB' users
\def\InfEU{W}         %% interference coming form same network users
\def\InfKnown{X}      %% interference known by DB
\def\InfKnownMin{ \InfKnown_{(1)} }      %% interference known by DB
\def\InfUnknown{Y}    %% interference unknown by DB
\def\InfTot{Z}        %% total interference
\def\InfTotMin{ \InfTot_{(1)} }        %% minimum total interference
\def\InfTotA{ \InfTot_{[\A]} }        %% total interference for advanced users

\def\K{K}                %%% the number of TV channels
\def\Kset{\mathcal{\K}}   %%% the set of TV channels
\def\k{k}

\def\ch{{channel}}
\def\chs{{channels}}

\def\tvch{{TV \ch}}

\def\Nkset{\Nset_{\k}}

\def\N{N}            %%% the number of users
\def\Nset{\mathcal{\N}}   %%% the set of users
\def\n{n}             %%% SU's index

\def\Probb{\Prob_{\textsc{b}}}      %% fixed percentage of end-users subscribe basic service
\def\BProba{\BProb_{\textsc{a}}}      %% fixed percentage of end-users subscribe advanced servie
\def\l{s}             %% service type

\def\R{R}
\def\Ra{\R_{\A}}

\def\t{t}
\def\c{c}   %% cost of database

\def\ut{U}
\def\rt{\mathcal{R}}

\makeatletter
\let\@copyrightspace\relax
\makeatother

\begin{document}

%\title{Information Market for TV White Space Network}
\title{MINE GOLD to Deliver Green Cognitive Communications}

\author{
	Yuan~Luo,
	Lin~Gao, and
	Jianwei~Huang
	\IEEEcompsocitemizethanks{
%		\IEEEcompsocthanksitem
%		Manuscript received date: March 30, 2015.
%		Manuscript revised dates: July 16, 2015 and Sept 10, 2015
		\IEEEcompsocthanksitem
		This work is supported by the General Research Funds (Project Number CUHK 412713 and 14202814) established under the University Grant Committee of the Hong Kong Special Administrative Region, China.
		\IEEEcompsocthanksitem
		Yuan~Luo and Jianwei~Huang {(corresponding author)} are with Network Communications and Economics Lab (NCEL),
		Department of Information Engineering, The Chinese University of Hong Kong, HK,
		E-mail: \{ly011, jwhuang\}@ie.cuhk.edu.hk.
		\IEEEcompsocthanksitem
		Lin~Gao is with the Harbin Institute of Technology (HIT) Shenzhen Graduate School, E-mail: gaolin@hitsz.edu.cn.}
	\vspace{-5mm}
}

\maketitle
%\thispagestyle{empty}
%\vspace{-18mm}
\begin{abstract}
Geo-location database-assisted TV white space network reduces the need of energy-intensive processes (such as spectrum sensing), hence can achieve green cognitive communication effectively.
%can effectively reduce the energy consumption in cognitive communications by reducing the need of energy-intensive processes (such as spectrum sensing), hence can help achieving the goal of green cognitive communication.
The success of such a  network relies on a proper business model that provides incentives for all parties involved.
In this paper, we propose MINE GOLD (a Model of INformation markEt for GeO-Location Database), which enables databases to sell the spectrum information to unlicensed white space devices ({\eus}) for profit.
Specifically, we focus on an oligopoly information market with multiple databases,
and study the interactions among databases and WSDs using a two-stage hierarchical model.
In Stage I, databases compete to sell information to {\eus} by optimizing their information prices.
In Stage II, each {\eu} decides whether and from which database to purchase the information, to maximize his benefit of using the TV white space.
We first characterize how the WSDs' purchasing behaviors dynamically evolve, and what is the equilibrium point under fixed information prices from the databases.
%s, and how the initial market share determines the eventual market equilibrium.
We then analyze how the system parameters and the databases' pricing decisions affect the market equilibrium, and what is the equilibrium of the database price competition.
Our numerical results show that, perhaps counter-intuitively, the databases' aggregate revenue is not monotonic with the number of databases.
%Hence it is important to properly balance a trade-off between the decreasing equilibrium prices and the increasing market shares for the databases.
Moreover, numerical results show that a large degree of positive network externality would improve the databases' revenues and the system performance.

%\begin{IEEEkeywords}
%	TV White Space, Information Market, Oligopoly Competition,  Game Theory
%\end{IEEEkeywords}

\end{abstract}
%\IEEEpeerreviewmaketitle

%\footnotetext[1]{This work is supported by the General Research Funds
%(Project Number CUHK412710 and CUHK412511) established under the
%University Grant Committee of the Hong Kong Special Administrative
%Region, China. Author emails: \{ly011,\ lgao,\ jwhuang\}@ie.cuhk.edu.hk.}

\IEEEpeerreviewmaketitle

%\addtolength{\abovedisplayskip}{-2mm}
%\addtolength{\belowdisplayskip}{-2mm}

%\input{Section_introduction}
%!TEX root = main_pure_information_journal.tex
%SourceDoc main_pure_information_journal.tex

\section{Introduction}\label{sec:intro}
%%%%%%%%%%%%%%%%%%%%%%%%%%%%%%%%%%%%%%%%%%%%%%%%%%

\vspace{-2mm}
\subsection{Motivations}
%Energy-efficient systems become increasingly important for protecting our environment, coping with global warming, and facilitating sustainable development.
%However, with the explosive growth of telecommunication industry, the associated energy consumption in telecommunication has been increasing at an approximate speed of $16\%-20\%$ per annum \cite{Gur2011}.
%Moreover, $3\%$ of worldwide energy consumption and $2\%$ of the worldwide $CO_2$ emissions have been caused by the information and communication technology (ICT) infrastructures \cite{climatae2008report}.
%Hence, energy optimization of wireless communications, ranging from equipment manufacturing to core functionalities, is an important research issue today.

With the explosive growth of telecommunication industry, $3\%$ of worldwide energy consumption and $2\%$ of the worldwide $CO_2$ emissions have been caused by the information and communication technology (ICT) infrastructures \cite{climatae2008report}.
%Hence, energy optimization of wireless communications, ranging from equipment manufacturing to core functionalities, is an important research issue today.
Cognitive communication is a promising paradigm for achieving energy-efficient communications,
as a cognitive radio device is able to adapt its configuration and transmission decision to the radio environment.
Such an adaptability enables it to select the best reconfiguration operation that balances the energy consumption and communication quality.
One of the promising commercial realizations of such cognitive communication technology is the \emph{TV white space network}, where unlicensed wireless devices (called white space devices, WSDs) opportunistically exploit the under-utilized broadcast television spectrum (called TV white space, TVWS\footnote{For convenience, we will refer to TV white space as ``TV channel''.}) via a third-party \emph{geo-location} white space database \cite{federal2012third,Ofcom2010geo}.

Cognitive communication and TV white space network rely on the accurate detection of radio environment (e.g., locating the idle channels). However, relying on the mobile device to sense radio environment usually consumes significant energy.
%The higher accuracy, the higher computational burden on mobile devices, and thus the higher energy consumption.
In order to save energy consumption and guarantee the performance of cognitive communication,
some spectrum regulators (\emph{e.g.,} FCC in the USA and Ofcom in the UK)
%, together with industrial organizations\footnote{
%	The real world white space database systems deployed by Google (\url{http://www.google.org/spectrum/whitespace/}), Microsoft (\url{http://whitespaces.msresearch.us/}), and Spectrum Bridge (\url{http://www.spectrumbridge.com/}).},
have advocated a \emph{database-assisted} TV white space network architecture.
%In order to save energy consumption and guarantee the performance of cognitive communication,
%some spectrum regulators (such as FCC in the USA and Ofcom in the UK), together with standards bodies and industrial organizations\footnote{
%Example include IEEE 802.22 WRAN standard (\url{http://www.ieee802.org/22/}), and the real world white space database systems deployed by Google (\url{http://www.google.org/spectrum/whitespace/}), Microsoft (\url{http://whitespaces.msresearch.us/}), and Spectrum Bridge (\url{http://www.spectrumbridge.com/}).},
%have advocated a \emph{database-assisted} TV white space network architecture.

Specifically, the white space database (also called geo-location database)  houses a global repository of TV licensees, and updates the licensees' channel occupations periodically.
Each WSD obtains the available TV channel information via querying a geo-location database, rather than sensing the wireless environment that can consume quite some energy.
WSDs and databases communicate with each other through the Internet.
In such a database-assisted TV white space network, WSDs perform the necessary local computations (e.g., identifying the locations)
and databases implement the complex data processing (e.g., computing the available TV channels for each WSD).
Such a network architecture can effectively reduce the energy consumptions of WSDs, and create a green communication ecosystem.

\begin{figure}
  \centering
    \includegraphics[width=4in]{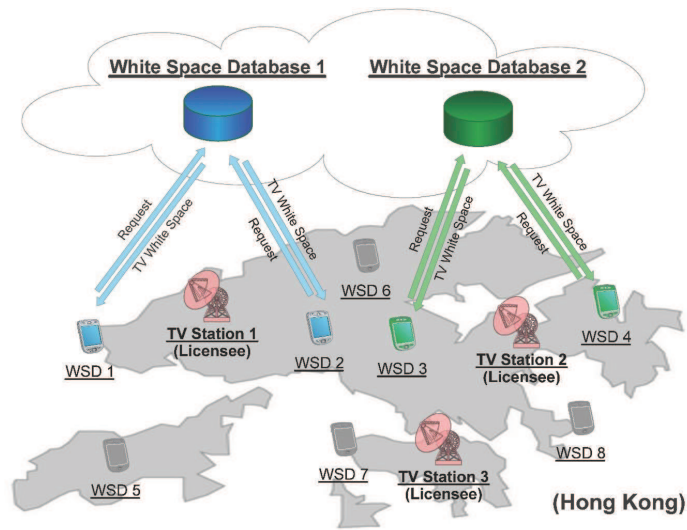}
   \vspace{-2mm}
  \caption{Illustration of a database-assisted TV white space network. To access a TV channel, each WSD first reports its location to a white space database (request), and then the database returns the available channel list to the WSD. }
  \vspace{-4mm}
\label{fig:model}
\end{figure}

Figure \ref{fig:model} illustrates such a database-assisted TV white space network, with 3 licensed TV stations and 8 unlicensed WSDs. Here WSDs $1$ and $2$ query the available channel information from database 1, WSDs $3$ and $4$ query the available channel information from database 2, and {WSDs $5$ to $8$ remain inactive (hence are not connected with any database in the figure).}

The geo-location databases are usually operated by third-party companies, such as Google and SpectrumBridge.
Hence, the commercial deployment of such a database-assisted network requires a proper business model, which offers sufficient incentives to the database operators to cover their capital expense (CapEx) and operating expense (OpEx).
%However, the commercial deployment of such a network requires a proper business model, which offers sufficient economic incentives to the database operators to cover their capital expense (CapEx) and operating expense (OpEx). Such an issue has not been extensively discussed in the existing literature.
%Hence, in line with green-related efforts in cognitive communication, the main interest of this paper  here
%is to combine two eco-friendly systems to produce
%a hybrid technology. We hope that this
%innovation can bring more effective use of radio
%resources while allowing a broader market to be
%explored
The existing business modeling of {\tvn} mainly focused on the \emph{spectrum market} \cite{niyato2009dynamic,luo2012,Bogucka2012, feng2013database,luo2013}, where the database operators, acting as spectrum brokers or agents, sell the TV white spaces to unlicensed WSDs for profit.
However, the TV spectrum market model may \emph{not} be suitable in practice due to some regulatory considerations.
For example, TV white spaces (especially those not licensed to any TV stations) are usually treated as \emph{public} resources, and \emph{shared} by unlicensed devices.
%designated by regulators for \emph{public and shared} usage by unlicensed devices.
Therefore, it may not be suitable for TV white spaces to be traded in a spectrum market like other licensed spectrum bands.
To this end, a new business model \emph{without} involving spectrum trading is highly desirable.

The world first white space database operator certified by FCC --
Spectrum Bridge --
proposed an alternative business model called ``White Space Plus".
The basic idea is to sell some advanced information (regarding the quality of TV channels) to WSDs, such that the latter can choose and operate on the high quality {\tvs}.
An example of such information is the degree of interference on every available \tv.
This essentially leads to an \emph{information market}, where WSDs purchase the information regarding the {\tv} quality from the database, instead of purchasing the channel.
Clearly, the successful deployment of such an information market requires (i) an accurate model to evaluate the value of   information for WSDs (buyers), and (ii) a carefully designed pricing strategy for each database.
However, none of these two issues has been considered in the current White Space Plus.\footnote{Currently Spectrum Bridge just offers a one year free trial to use this White Space Plus service.}
%In our previous work \cite{luo2014wiopt}, we proposed and studied a monopoly  information market with a single database, without considering the competition among multiple databases in a more practical scenario.
This motivates us to study the oligopoly information market model for white space databases in this paper.

\subsection{Contributions}
In this paper, we present and study a Model of INformation markEt for GeO-Location Database (MINE GOLD),
%model and study an \emph{oligopoly} competitive {information market},
where multiple databases (sellers) compete to sell the advanced information regarding the quality of TV channels to {\eus}.
The WSDs (buyers) decide whether and from which database to purchase the information. This leads to the following two-stage hierarchical model. In Stage I, each database determines the information price to {\eus}. In Stage II, {\eus} decide the best purchasing decisions, given the information prices of all databases.
Note that the {\eus}' behaviors dynamically evolve due to the \emph{positive} network externality in the information market, as more {\eus} purchasing the information increases the quality/value of the database's information and improves the {\eus} performance.
Such a performance change further stimulates {\eus} to adjust their behaviors in the future, hence the {\eus}' behaviors dynamically evolve.
%We will show how the hybrid market dynamically evolves along user behaving, and what is the \emph{market equilibrium} point.

Through such a two-stage hierarchical model, we will provide insights regarding the databases' and {\eus}' strategic decisions. Specifically, we will study the following problems systematically:
\begin{itemize}
\item
\emph{How should each database determine the information price (in Stage I) to maximize his expected revenue, considering the competition from other databases?}
\item
\emph{How will the {\eus}' optimal purchasing behavior (in Stage II) dynamically evolve over time, and what is the stable market shares\footnote{The market shares is the percentage of {\eus} purchasing information from the database.} of databases (also called market equilibrium)?}
\end{itemize}

Both problems are challenging due to the following reasons.
First, there is lack of a unified framework to evaluate the value of information to {\eus}.
In particular, one database's known information may not be the same as the others, and no database has the global information.
To this end, we propose a general framework to evaluate the value of information for {\eus}.
The framework considers not only the potential error of the information provided by databases, but also the heterogeneity of {\eus}.

Second, the information market has the property of {\emph{positive network externality}},
i.e., the more {\eus} purchasing information from the same database, the higher value of that database's information for each buyer.
This is quite different from traditional spectrum markets which are usually congestion-oriented, i.e., the more users purchasing and using the same spectrum, the less value of spectrum for each buyer due to interferences.
Here the positive correlation between the information value and  market share complicates the market behavior analysis, as the change of a single {\eu}'s purchasing behavior may affect the information evaluation and purchasing decisions of other {\eus}.
{We  show how the market share of each database dynamically evolves over time, and what is the  market equilibrium it eventually converges to.}

%\com{Should we add a third point, explaining the challenging in analyzing the competition among databases? This is a key differentiator from our JSAC paper. }

Third, the competition among multiple databases makes the analysis even more challenging, especially when considering the positive network externality. This is different from most prior price competition analysis in the wireless literature, where the users' decisions are either decoupled \cite{niyato2008competitive,Kasbekar2012game} or negatively correlated \cite{shetty2010congestion}.
% \cite{zhu2014game,shetty2010congestion}.
% This price competition model here is quite different from the traditional price competition models where the decisions of buyers are usually decoupled or negative-correlated. In our model, however, the decisions of WSDs (buyers) are positive-correlated due to the positive network externality of the information market. This greatly complicates the databases' pricing decisions.
Nevertheless, we are able to characterize the conditions for the existence and uniqueness of the price equilibrium.

As far as we know, this is the first work that systematically studies an oligopoly information market for TV white space networks.
In summary, the key contributions of this paper are summarized as follows.

\begin{itemize}

\item
\emph{Novelty and Practical Significance.}
We consider an oligopoly information market and propose a two-stage hierarchical business model, which captures the positive network externality of the TV white space network.
Comparing with the traditional spectrum market model, this information market model better fits the regulatory requirements and industry practice.

%\item
%\emph{Market Equilibrium Analysis.}
%We first derive the WSD¡\UTF{00D8}s best purchasing behavior, and show how the market share of each database dynamically evolves over time.
%We first characterize the equilibrium market shares of databases ({market equilibrium}) systematically given the information prices of databases.
%Based on this, we further derive the
%equilibrium prices of two databases (game equilibrium) in the oligopoly price competition game.
%the optimal information pricing plan that maximizes the database operator's revenue.

\item
\emph{Market Equilibrium Analysis.}
We characterize the equilibrium of the proposed information market systematically.
Our analysis indicates that given the prices of databases, there may be multiple market equilibria, and which one will actually emerge depends on the initial market shares of databases.
We further show that some equilibria are stable, in the sense that a small fluctuation on the equilibrium will drive the market back to the equilibrium, while others are not.

\item
\emph{Competition among databases.}
We formulate the competition among databases as a price competition game, and study the existence and uniqueness of the price equilibrium. To do this, we first transform the price competition game into an equivalent market share competition game. Then we analyze the existence and uniqueness of the equilibrium of the transformed game systematically using supermodular game theory.

\item
\emph{Observations and Insights.}
Our numerical results show that, perhaps counter-intuitively, the databases' aggregate revenue first increases and then decreases with the number of databases.
Intuitively, there is a trade-off between the decreasing equilibrium prices and the increasing market shares for the databases.
Extensive simulations show that having two databases will maximize the databases' aggregate revenue under a wide range of system parameters.
Moreover, our numerical results show that a large degree of positive network externality would improve the databases' revenues and the system performance.
%, and results in the increasing of databases' profit.

\end{itemize}

The rest of the paper is organized as follows.
In Section \ref{sec:related}. we review the related literature.
In Section~\ref{sec:model},
we present the system model.
In Sections \ref{sec:monopoly} and \ref{sec:oligopoly}, we study the monopoly and competitive network scenarios, respectively.
In Section, we provide numerical results.
Finally, we conclude in Section \ref{sec:conclusion}.
We provide the detail proofs in the appendix.
%\rev{We need to tell the readers where to find all the missing proofs. }

%\input{Section_literature}
%!TEX root = main_pure_information_journal.tex
%SourceDoc main_pure_information_journal.tex

\vspace{-2mm}
\section{Related Work}
\label{sec:related}

%While most prior studies focused on the technical issues such
%Most of the existing studies in TV white space networks focused on the technical issue such as the white space exploration, database deployments and network optimization
%\cite{gurney2008geo, murty2011senseless, goncalves2011value, feng2011database, chen2012}.
%In \cite{gurney2008geo}, Gurney \emph{et al.} discussed the computation of the protection areas for TV stations.
%In \cite{murty2011senseless}, Murty \emph{et al.} presented and evaluated a database driven white space network using a more accurate propagation model with terrain data.
%In \cite{goncalves2011value}, Goncalves \emph{et al.} compared the geo-location database approach with sensing approach in terms of the technical and business values.
%In \cite{feng2011database}, Feng \emph{et al.} presented the design and implementation of a multi-cell infrastructure-based TV white space network.
%In \cite{chen2012},
%Chen \emph{et al.} considered the joint channel selection and access point association problem in white space networks.
Most of the existing studies on cognitive green communications
focused on the technical issues such as spectrum sharing, resource optimization, and platform implementation \cite{palicot2009,Gur2011,Ji2013}.
In \cite{palicot2009}, Palicot demonstrated how to apply cognitive radio technology to achieve green radio communications.
In \cite{Gur2011}, G$\ddot{u}t$ \emph{et al.} studied   the trade-off among energy efficiency, performance,  and practicality in cognitive radio network.
In \cite{Ji2013}, Ji \emph{et al.} proposed a platform to explore TV white space in order to achieve green communication in cognitive radio network.
%In \cite{Xie2012}, Xie \emph{et al.} studied the problem of spectrum sharing and power allocation in heterogeneous cognitive radio networks considering the issue of energy efficiency.
However, none of the above studies considered the incentives issues in implementing such a cognitive radio network. Without a proper business model to provide sufficient incentives to the involved parties such as spectrum licensees and the network operators, it is difficult to envision strong commercialisation of this new technology.

Prior studies related to the business modeling of TV white space networks mainly focused on the spectrum trading market \cite{niyato2009dynamic,luo2012,Bogucka2012, feng2013database,luo2013,luo20132}.
In \cite{niyato2009dynamic}, Niyato \emph{et al.} proposed a hierarchical spectrum trading model to analyze the interaction among service providers, TV licensees, and users.
In \cite{luo2012}, Luo \emph{et al.} studied the (dedicated) TVWS reservation problem for a single database.
In \cite{Bogucka2012}, Bogucka \emph{et al.} discussed a spectrum trading mechanism implemented by the spectrum broker in TV white spaces.
In \cite{feng2013database}, Feng \emph{et al.} studied the hybrid pricing scheme for the database manager.
In \cite{luo20132}, Luo \emph{et al.} discussed the price-Inventory competition game among multiple databases.
%In \cite{liu2013}, Liu \emph{et al.} presented a new auction mechanism for the database to preserve the location privacy.
The key idea of this spectrum trading market is to let the databases, acting as spectrum brokers or agents, sell the TV white spaces to {\eus} for profit. However, TV spectrum trading may not always be possible as TV white spaces are sometimes considered as \emph{public} spectrum resources and need to be used in a shared fashion.
%which sometimes are treated as the \emph{public} spectrum resource and are utilized for public and {shared} usage.
Some recent studies \cite{luo2014wiopt,luo2014SDP,luo2015magazine,luo2015INFOCOM} proposed the pure and hybrid information models for TV white spaces.
%information market model and the hybrid information and spectrum market model for TV white spaces.
However, these studies focused either on a single database or two competitive databases. In this work, we consider a more general oligopoly market with many competitive databases.
%Our preliminary work \cite{luo2014wiopt} and \cite{luo2014SDP} proposed the information market for TV white space networks, without involving the trading of spectrum.
%However, these two works did not consider the competitions among multiple databases in selling information to \eus.

Price competition in a market can be modeled as a non-cooperative game.
%[14] considered competition among multiple primary users to sell spectrum to the secondary users.
In \cite{niyato2008competitive}, Niyato \emph{et al.} studied the problem of spectrum trading with multiple licensed users selling spectrum opportunities to multiple unlicensed users, and proposed an iterative algorithm to achieve the Nash equilibrium in this competitive network.
In \cite{Kasbekar2012game}, Kasbekar \emph{et al.} analyzed price competition in spectrum trading market, jointly considering both bandwidth uncertainty and spatial reuse.
%In \cite{Min2012game}, Min \emph{et al.} studied two wireless service providers' pricing competition by considering spectrum heterogeneity.
%In \cite{zhu2014game}, Zhu \emph{et al.} studied pricing competition among macrocell service providers via a two-stage multi-leader-follow game.
%\com{Yuan: I can only find one with Linyang Song and Zhu Han with price competition in cognitve radio, i.e. \cite{Niyato2009game}...}
However, in all of the above works, the market is usually assumed to be associated with the negative network externality or non-externality. In our work, as will be discussed later, the information market is associated with the \emph{positive} network externality. This makes our  market analysis quite different with the above works.

%\input{Section_model}
%!TEX root = main_pure_information_journal.tex
%SourceDoc main_pure_information_journal.tex

\vspace{-2mm}
\section{System Model}\label{sec:model}
We consider a database-assisted TV white space network with a set $\Mset = \{ 1, \ldots, \M \}$ of \emph{geo-location {\dbs}}
%(denoted by $\dbbm, \m \in \Mset$ )
and a set of $N$ \emph{unlicensed users} (devices) operating on TV channels.
The {\dbs} hold the list of TV licensees, update the licensees' channel occupations information periodically, and calculate a set of available TV channels (i.e., unlicensed TV channels or those are not occupied by the licensees).
%available TV channels set (i.e., unlicensed TV channels or those are not occupied by the licensees).
The available TV channels are also called TV white spaces, which can be used by unlicensed users freely in a shared manner (e.g., using CDMA or CSMA).
%There exist some unlicensed TV channels (TV white spaces), which can be  used by unlicensed users freely in a shared manner (e.g., using CDMA or CSMA).
%%Let $\Kset = \{ 1,\ldots, \K \}$ denote the set of available TV channels in the area of the network.
Each \eu~queries a database for the available TV channel set, and  can only operate on \emph{one} of the available \chs~at any time.

%\begin{figure}
%%\vspace{-6mm}
%  \centering
%    \includegraphics[width=3in]{model-new}
%\vspace{-5mm}
%  \caption{Illustration of a database-assistant TV white space network. To access a TV channel, each WSD first reports its location to a white space database (request), and then the database returns the available channel list to the WSD. }
%\label{fig:model}
%\vspace{-5mm}
%\end{figure}

\vspace{-2mm}
\subsection{Geo-location Database}
Motivated by t he current commercial examples, the database provides the following two services to the WSDs.

\subsubsection{Basic Service}
According to the regulation policy (e.g., \cite{federal2012third}), it is mandatory for a geo-location white space {\db} to provide the following information for any unlicensed {\eu}: (i) the list of TV white spaces (i.e., unlicensed TV channels), and (ii) the transmission constraint (e.g., maximum transmission power) on each channel in the list.
% and (iii) other optional requirements.
The database needs to provide this \emph{basic (information) service} free of charge for any unlicensed user.

\subsubsection{Advanced Service}
Beyond the basic information, each {\db} can also provide certain advanced information regarding the quality of TV channels (as SpectrumBridge does in White Space Plus), which we call the \emph{advanced (information) service}, as long as it does not conflict with the free basic service.
Such an advanced information can be rather general, and a typical example is the ``interference level on each channel''.
% used in \cite{luo2014wiopt, luo2014SDP}.
%Motivated by the practice of SpectrumBridge \cite{SpectrumBridgeCommericial} and the model proposed by Luo \emph{et al.} in \cite{luo2014wiopt, luo2014SDP}, we define the following advanced information:
%%consider such an advanced service, where the {\db} provides the following advanced information to every {\eu} $n$ subscribing to its advanced service (i.e., purchasing the advanced information):
%$$
%\boldsymbol{\InfTot} = \{\InfTot_{\k}\}_{k\in\Kset},
%$$
%where $\InfTot_{\k} $ is the interference level on TV {\ch} $k$ for a particular \eu.\footnote{Note that $\InfTot_{\k} $ may be different for different users. Here, we omit the user index for the writing convenience.}
%When subscribing to the advanced service of a \db,
With the advanced information, the {\eu} is able to choose a channel with the  highest quality (e.g., with the lowest interference level).
%\footnote{{We assume that {\eus} are   separated geographically, and hence experience different interferences from TV stations and from other WSDs operating on the same channel.
%Thus, the information provided by the database to one WSD may not be applicable to other WSDs.}}
Hence, the {\db} can \emph{sell} this advanced information to users for profit.
This leads to an \emph{information market}.
For convenience, let $\p_{\m} \geq 0$ denote the (advanced) {information price} of database $\m$, $\m \in \Mset$.
%\footnote{\rev{Notice that databases announce their price at the beginning of a frame which lasts for certain time (e.g., 2 hours) and the databases can change their prices at the beginning of the next frame.}}.

WSDs need to interact with databases periodically for the basic service or advanced service.
The length of each interaction period (called \emph{frame}) will be subject to the regulatory constraint, e.g., 15 minutes according to the latest Ofcom rule.
In this work, we focus on the interactions of WSDs and databases in a particular frame, where databases announce their prices at the beginning of the frame, and then WSDs choose actions that last for the entire frame.

\vspace{-3mm}
\subsection{White Space Devices}

%\com{All the databases in our model provide the same available channel set (hence choice b or s leads to the same result no matter which database to inquire) ? Need to clarify. }
After obtaining the available channel list through the free basic service, each {\eu} has $\M + 2$ choices (denoted by $\s$) in terms of channel selection:
\begin{itemize}
\item[(i)] $\s = \B$:
{Inquires one database and chooses the basic service (i.e., randomly chooses an available channel) provided by the chosen database;}
\item[(ii)] $\s = \S$:
{Inquires one database to obtain the list of available channels and senses all the available channels to determine the best one at the cost $\ps$;\footnote{{The sensing cost $c$ can be used for characterizing the energy consumption cost of a WSD for performing spectrum sensing.}}}
\item[(iii)] $\s = \Am$:
Subscribes to database $\m$'s advanced service, and picks the channel with the best quality indicated by database $\dbbm$.
\end{itemize}

Here we assume that the sensing is perfect without errors, hence a {\eu} can always choose the best channel when choosing the sensing service.
We further denote $\RB$, $\RS$, and $\RAm$ as the expected \emph{utility} that a {\eu} can achieve from choosing the basic service ($\s = \B$), sensing ($\s = \S$), and the advanced service of database $\m$~($\s = \Am$), respectively. As all the databases provide the same basic information (i.e., the available channel set), WSDs would achieve the same expected utility from choosing the basic service (i.e., $\s = \B$) or from choosing sensing (i.e., $\s = \S$), no matter which database they inquire.
However, the expected utility from choosing different databases' advanced services (i.e., $\s = \Am$) can be different, as databases may hold  different  qualities of information.

The \emph{payoff} of a {\eu} is defined as the difference between the achieved utility and the service cost (i.e., the  {information price} when choosing the advanced service, or the  {sensing cost} if choosing sensing by itself).
Let $\th$ denote the \eu's evaluation for the achieved utility.
Then, the payoff of a {\eu} with an evaluation factor $\th$ is
\begin{equation}\label{eq:utility-basic}
\textstyle
\Ueu_{\th} = \left\{
  \begin{aligned}
  &\textstyle  \th \cdot \RB ,      &&  \ \text{if} ~ \s = \B, \\
  &\textstyle  \th \cdot \RS - \ps ,      &&  \ \text{if} ~ \s = \S, \\
  &\textstyle  \th \cdot \RAm   -  \pm , &&  \  \text{if} ~ \s  = \Am.
   \end{aligned}
\right.
\end{equation}

Each {\eu} is rational and will choose a strategy $\s \in \{\B, \S, \Am \}$ that maximizes its payoff.
Note that different {\eus} may have different values of $\th$ (e.g., depending on application types), hence have different choices. That is, {\eus} are heterogeneous in term of $\th$.
For convenience, we assume that $\th$ is uniformly distributed in $[0,1]$ for all \eus\footnote{This assumption is commonly used in the existing literature, e.g., \cite{shetty2010congestion}. Relaxing to more general distributions often does not change the main insights.}.

Let $\Probb$, $\Probs$, and $\Probam$ denote the the fraction of {\eus} choosing the basic service, sensing, and the advanced service of database $\m$, respectively.
For convenience, we refer the fraction of {\eus} choosing particular service as the \emph{market share} of such service.
%For convenience, we refer to $\Probb$, $\Probs$, and $\Probam$ as the \emph{market shares} of the basic service, sensing, and the advanced service, respectively.
Obviously, $\Probb, \Probs, \Probam \geq 0$ and $\Probb + \Probs + \sum_{\m \in \Mset}\Probam  = 1$.
Hence, the \emph{payoff} of the database $\m \in \Mset$, which is defined as the difference between the revenue obtained by providing the advanced service and the cost, is
\begin{equation}\label{eq:db-profit}
\begin{aligned}
\Udb = ( \p_m - \cm) \cdot \Probam \cdot N,
\end{aligned}
\end{equation}
where $\cm$ denotes  database $\m$'s energy consumption cost when providing the advance service to one {\eu}.
%under the revenue sharing scheme (Scheme I), and
%\begin{equation}\label{eq:u2}
%\left\{
%\begin{aligned}
%\Usl \eq \Uslwp &= (\pl -\w)  \Probl ,
%\\
% \Udb \eq \Udbwp &= \pa \Proba + \w   \Probl ,
%\end{aligned}
%\right.
%\end{equation}
%under the wholesale pricing scheme (Scheme II).

\subsection{Positive Network Externality}
Note that the information market has the property of \emph{positive network externality}. This is because the more {\eus} subscribing to the advanced service,  the more accurate the database's information is, and further the more benefit for the {\eus} subscribing to the advanced service.
%\rev{We further assume that $\RA(\Prob)$ is concave in $\Prob$, which is verified by simulations.}
Next we analytically quantify this positive network externality.
We first list important assumptions made in this paper to clarify the scenario on which we focus. All these assumptions have been verified to be reasonable through extensive simulations,
%are made from the simulations observation,
where the advanced information is the interference level on each channel.
We provide more detailed modeling and formulation for such interference information in Appendix.
%Appendix \ref{sec:information_market_appendix}.}
%We provide more detailed discussions about modeling the information of interference level in our online technical report \cite{report}.

\vspace{-1mm}
\begin{assumption}\label{assume:random}
$\RB$ and $\RS$ are independent of $\Probb $, $\Probam$, and $ \Probs $.
\end{assumption}
\vspace{-2mm}

The reason for this assumption is as follows.  From the system perspective, each WSD will access a channel randomly and independently.
First, {\eus} choosing the basic service will access one TV channel randomly.
Second, WSDs choosing sensing will always access their best TV channels, and the best channels for different WSDs are independent.
Hence, from the system perspective, all the WSDs will be  randomly and uniformly distributed in all the channels.
Hence, the utility provided by the basic service or sensing depends on the average number of {\eus} in each channel, while not on the detailed numbers of WSDs using different services.

%\begin{assumption}\label{assume:sense}
%$\RS$ is independent of $\Probb $, $\Probam$, and $ \Probs $.
%\end{assumption}

%This assumption is because the {\eus} will sense all the available channels and choose the best one. Hence, the utility provided by the sensing service \rev{only depends on the total number of {\eus} using the TV channels.}

\vspace{-1mm}
\begin{assumption}\label{assume:advance}
$\RAm$ is non-decreasing in $\Probam$.
\end{assumption}
\vspace{-2mm}

This assumption actually reflects the \textbf{positive network externality} in the information market. Namely, the more users subscribing to database $\m$'s advanced service, the higher quality of the database $\m$'s information.
For more details, please refer to Appendix.

\vspace{-1mm}
\begin{assumption}\label{assume:utility_relationship1}
$\RS \geq \RAm \geq \RB $.
\end{assumption}
\vspace{-2mm}

The reason behind $\RS \geq \RAm$ is that sensing is perfect and can enable a {\eu} to locate the optimal TV channel.\footnote{{We will study the impact of imperfect sensing in our future work.}}
%This means that $\RS$ is always larger than $\RAm $ and $\RB$\footnote{{As our focus on studying the {\eus}' choice for different service, we assume that perfect sensing in this paper. Taking false alarm and missed detection in {\eus}'s sensing is left for future work.}}.
The reason behind $\RAm \geq \RB$ is that WSDs can achieve additional performance gains from the advanced information provided by any database. Note that if $\RAm < \RB$, then we have the trivial case that WSDs will never choose the advanced service even when the information price $\pm = 0$.
%We can further assume that $\RAm$ is larger than $\RB$ due to additional advanced information.

%\com{This paragraph also contains two assumptions. We should put them in the Assumption environment.}

For convenience,
we write $\RAm$ as a non-decreasing function of  $ \Probam$, $\m \in \Mset$, i.e.,
%i.e.,\footnote{Later we will also write $\Proba + \Probb $ as $1- \Probl$, since $\Proba + \Probb  + \Probl  = 1$.}
$
\RAm(\Probam) \triangleq \gy(\Probam).
$
%and write $\RA$ as the combination of a
%non-increasing function of $ \Proba + \Probb $ and a non-decreasing function of $\Proba $, i.e.,
%$$
%\RA \triangleq \fx(\Proba + \Probb) + \gy(\Proba) ,
%$$
%%where $\fx(\cdot)$ is a non-increasing function, and $\gy(\cdot)$ is a non-decreasing function.
%Note that function $\fx (\cdot)$ reflects the congestion effect, and is identical for $\RB $ and  $\RA$ (as users experience the same congestion effect in both basic and advanced services).

Note that function $\gy(\cdot)$ reflects the performance gain induced by the advanced information, i.e., the (advanced) information value.
%\footnote{
%Note that if we set $\RL < \RB$, then users will never choose the leasing service even with a zero channel price $\pl$.
%In this case, our model degenerates to the pure information market, similar as that in \cite{luo2014wiopt}.
%Moreover, if we set $\RA = \RB$, then users will never choose the advanced service even with a zero information price $\pa$.
%In this case, our model degenerates to a monopoly spectrum market (where the licensee is the monopolist).
%In this sense, our hybrid market model generalizes both the pure spectrum market and pure information market.}.
%To facilitate the later analysis,
We further introduce the following assumptions on functions $\gy (\cdot)$.
%\vspace{-1mm}
%\begin{assumption}\label{assum:congestion}
%Function $\fx(\cdot)$ is non-negative, non-increasing, convex, and continuously differentiable.
%\end{assumption}

\vspace{-1mm}
\begin{assumption}\label{assum:positive}
Function $\gy(\cdot)$ is non-negative, non-decreasing, concave, and continuously differentiable.
\end{assumption}
\vspace{-2mm}
%The non-increasing and convexity assumption of $\fx(\cdot)$ is due to the increasing of marginal performance degradation under congestion, and is widely used to model the network congestion effect in wireless networks (see, e.g., \cite{shetty2010congestion,johari2010congestion} and references therein).
Assumption \ref{assum:positive} results form the diminishing marginal performance improvement induced by the advanced information.
Such a generic function $\gy(\cdot)$ can cover a wide range of application specific definitions of the advanced information, e.g., the interference level on each channel.
The detailed discussion is provided in Appendix.
%Due to space limit, we leave the detailed discussion in \cite{report}.

\begin{figure}[t]
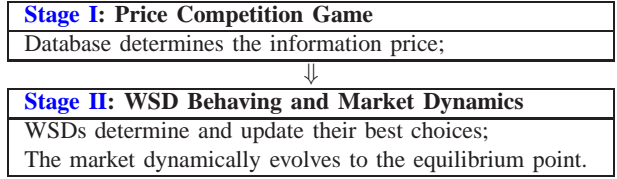

\centering
\footnotesize
\begin{tabular}{|m{3in}|}
\hline
%\textbf{\revh{Layer I}: Commission Negotiation}
%\\
%\hline
%The database and the spectrum licensee negotiate the commission charge details (i.e.,
%$\delta $ under RSS or $w$ under WPS).
%\\
%\hline
%\multicolumn{1}{c}{$\Downarrow$} \\
%\hline
\textbf{\rev{Stage I}: Price Competition Game}
\\
\hline
Database determines the information price;
\\
%The spectrum licensee determines the channel price $\pl$.
%\\
\hline
\multicolumn{1}{c}{$\Downarrow$} \\
\hline
\textbf{\rev{Stage II}: {\Eu} Behaving and Market Dynamics}
\\
\hline
{\Eus} determine and update their best choices;
\\
The market dynamically evolves to the equilibrium point.
\\
\hline
\end{tabular}
\caption{Two-stage Interaction Model}
\label{fig:layer}
\vspace{-5mm}
\end{figure}

\vspace{-3mm}
\subsection{Two-Stage Interaction Model}

Based on the above discussion, an information market captures the interactions among the geo-location databases and the {\eus}.
Hence, we formulate the interactions as a two-stage hierarchical model illustrated in Figure \ref{fig:layer}.
Specifically, in Stage I, each database determines the advanced information price $\pm$.
In Stage II, {\eus} determine their best choices, and dynamically update their choices based on the current market shares.
Accordingly, the market dynamically evolves and finally reaches the equilibrium point.

In Section \ref{sec:monopoly}, we will first analyze a simple monopoly information market with a single database to facilitate the understanding of this market.
Then in Section \ref{sec:oligopoly}, we will analyze a more general oligopoly information market with multiple databases.

\section{Monopoly Information Market}\label{sec:monopoly}
We first consider a simple monopoly scenario, where a single database provides TV white space information service to {\eus}.
We will denote the monopoly database as database $1$. This case study will serve as a benchmark for the later discussions of the oligopoly market in Section \ref{sec:oligopoly}.
%In this monopoly information market, the monopoly database acts as the leader and determines the information price first. Then, \eus~act  as the followers and make their subscription decisions based on their sensing costs and the information price chosen by the database.
%We will analyze the equilibrium of such a Stackelberg game by backward induction.
In what follows, we study the two-stage model by backward induction.
Namely,
we first study the {\eus}'s subscription behaviour and market equilibrium in Stage II.
Then, based on the market analysis, we study the monopoly database's best pricing decision that maximizes its revenue in Stage I.

\subsection{{\Eus}' Best Strategy}\label{sec:user-choice-monopoly}
As Assumption \ref{assume:advance} shows that the utility provided by the advanced service of database $1$ is varying with the database's market share, each WSD will form a belief on the utility of database $1$ and make a subscription decision.
For convenience, we introduce a virtual time-discrete system with slots $\t=1,2,\ldots$, where {\eus} change their decisions at the beginning of every slot, based on the derived market shares in the previous slot.\footnote{{The main purpose of
	introducing the virtual time-discrete system is to characterize the relation between the price and the market equilibrium, and to facilitate the calculation of
	database's optimal price strategy later. Such an analysis technique
%	(i.e., using a dynamic system to understand the outcome of a one-shot system)
	has been extensively adopted in the existing literature, e.g.,
	\cite{shetty2010congestion,khan2012}.}}
%	 \cite{manshaei2008evolution,zemlianov,khan2012,shaolei2011}.}}
%\footnote{We would like to clarify that the WSD subscription game
%is a one-shot static game, where all WSDs choose their strategies once and simultaneously.
%The main purpose of
%introducing the virtual time-discrete system, similar as the best response iterative algorithm in classic game theoretic analysis,
%is to characterize the relation between the price and the market equilibrium, and to facilitate the calculation of
%database's optimal price strategy later.
%In other words, such a virtual time-discrete system is only for theoretical
%analysis, and is not actually used in the real system. Hence, there is no correspondence to the physical time for each
%time slot in the virtual discrete-time system.
%Such an analysis technique (i.e., using a dynamic system to understand the outcome of a one-shot system)
%has been extensively adopted in the existing literature, e.g., \cite{manshaei2008evolution,zemlianov,khan2012,shaolei2011}.}
Let $\Probao^{\t}$ denote the market share derived at the end of slot $t$.
%(which serves as the initial market share in the next slot $t+1$.
Then we consider a WSD's best strategy at the end of slot $\t$, given the market share $\{\Probb^{\t}, \Probs^{\t}, \Probao^{\t} \}$ where $\Probb^{\t} + \Probs^{\t} + \Probao^{\t}  = 1$.

When there is only one database operating in the TV white space market, each  {\eu}  has three choices:
(i) chooses the basic service by randomly choosing a channel from the available channel set, i.e., $\l = \B$, with zero cost;
(ii) senses all the available channels to determine the best one, i.e., $\l = \S$ with the sensing cost $\ps$; and
(iii) subscribes to the advanced service of the (only) database $\dbbo$, i.e., $\l = \Ao$, and pays the database $\dbbo$ an information price $\po$.
Notice that each {\eu} will choose a strategy that maximizes its payoff defined in (\ref{eq:utility-basic}).
Hence, given the market share $\{\Probb^{\t}, \Probs^{\t}, \Probao^{\t} \}$ with $\Probb^{\t} + \Probs^{\t} + \Probao^{\t}  = 1$,
%Hence, given the initial market share $\{\Probb^0, \Probs^0, \Probao^0 \}$ where $\Probb^0 + \Probs^0 + \Probao^0  = 1$,
%Hence, given the market share $\{\Probb, \Probs, \Probao \}$ where $\Probb + \Probs + \Probao  = 1$,
a type-$\th$ {\eu}'s best strategy is
\footnote{{Here, ``iff'' stands for ``if and only if''.
We omit the cases of $\th \cdot \RB  = \max\{ \th \cdot \RS - \ps, \th \cdot \RAo(\Probao^{\t})  -  \po \}$, $\th \cdot \RS -  \ps = \max\{ \th \cdot \RB ,\ \th \cdot \RAo(\Probao^{\t})  -  \po \}$, and $\th \cdot \RAo(\Probao^{\t}) -  \po = \max\{ \th \cdot \RS  -  \ps ,\ \th \cdot \RB \}$, which are negligible
%(i.e., occurring with zero probability)
due to the continuous distribution assumption of $\th$.}}
\begin{equation}\label{eq:utility_function}
\textstyle
\left\{
\begin{aligned}
 & \l_{\th}^* = \B, \mbox{~~~~iff~~} \th \cdot \RB > \max\{ \th \cdot \RAo(\Probao^{\t})  -  \po ,\ \th \cdot \RS - \ps \}
\\
& \l_{\th}^* = \S, \mbox{~~~~iff~~} \th \cdot \RS  -  \ps > \max\{ \th \cdot \RB , \ \th \cdot \RAo(\Probao^{\t})  -  \po \}
\\
& \l_{\th}^* = \Ao, \mbox{~~~iff~~} \th \cdot \RAo(\Probao^{\t})  -  \po >  \max\{ \th \cdot \RB, \ \th \cdot \RS  -  \ps \}
\end{aligned}
\right.
\end{equation}
%where $\RAo =  \gy(\Probao^0)$, and $\RB <\RAo(\Probao^0)< \RS$.
where $\RAo(\Probao^{\t})=  \gy(\Probao^{\t})$, and $\RB <\RAo(\Probao^{\t})< \RS$ based on Assumptions \ref{assume:random} and \ref{assume:advance}.

\begin{figure}
\centering
  \includegraphics[width=4in]{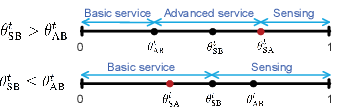}
  \vspace{-3mm}
  \caption{Illustration of $\thsb^{\t}$, $\thab^{\t}$, and $\thsa^{\t}$ in slot $\t$.}\label{fig:threshold}
  \vspace{-5mm}
\end{figure}

To better illustrate the above best strategy, we introduce the following notations:
\begin{equation*}\label{eq:p-thres}
\textstyle
\thsb^{\t} \eq \frac{ \ps}{ \RS- \RB },
~~~~
\thab^{\t} \eq \frac{ \po}{ \RAo(\Probao^{\t})- \RB },
~~~~
\thsa^{\t} \eq \frac{\ps-\po}{\RS - \RAo(\Probao^{\t})}.
\end{equation*}
Intuitively, $\thsb^{\t}$ denotes the smallest $\th$ such that a type-$\th$ \eu~prefers sensing than the basic service; $\thab^{\t}$ denotes the smallest $\th$ such that a type-$\th$ \eu~prefers the advanced service than the basic service; and $\thsa^{\t}$ denotes the smallest $\th$ such that a type-$\th$ \eu~prefers sensing than the advanced service.
Notice that $\RAo(\Probao^{\t})$ is a function of the market share $\Probao^{\t}$. Hence, $\thab^{\t} $ and $\thsa^{\t}$ are also functions of $\Probao^{\t}$.

Figure \ref{fig:threshold} illustrates two possible relationships of $\thsb^{\t}$, $\thab^{\t}$, and $\thsa^{\t}$\footnote{{Note that we only need to compare the value of $\thsb^{\t}$ and $\thab^{\t}$ to get the relationship of $\thsb^{\t}$, $\thab^{\t}$, and $\thsa^{\t}$.}}.
Intuitively, Figure \ref{fig:threshold} implies that WSDs with a high utility evaluation factor $\th$ are more willing to choose sensing in order to achieve the maximum utility.
WSDs with a low utility evaluation factor $\th$ are more willing to choose the basic service, so that they will pay zero cost.
WSDs with a middle utility evaluation factor $\th$ are willing to choose the advanced service, in order to achieve a relatively large utility with a relatively low service cost.
Notice that when the information price $\po$ is high or the information value (i.e., $\RAo(\Probao^{\t})-\RB$) is low, we could have $\thsb^{\t} < \thab^{\t}$, in which no users will choose the advanced service (as illustrated in the lower subfigure of Figure \ref{fig:threshold}).

Next we characterize the market shares in slot $\t + 1$,
%(called the \emph{derived market shares})
resulting from the {\eus}' best choices in slot $\t$.
Such derived market shares are important for analyzing
%the {\eu} behavior dynamics and
the market evolution in the next subsection.
Assume that all \eus~update the best strategies once and simultaneously.
Recall that $\th$ is uniformly distributed in $[0,1]$.
Then, given any market share $\Probao^{\t}$ in slot $\t$, the market share $\Probao^{\t+1}$ in slot $\t + 1$ is
\begin{itemize}
\item
If $\thsb^{\t} > \thab^{\t}$, then $\Probao^{\t+1} = \thsa^{\t} - \thab^{\t}$;
\item
If $\thsb^{\t} \leq \thab^{\t}$, then $\Probao^{\t+1}  = 0$.
\end{itemize}

%To facilitate the characterization of the derived market share $\Probao$, we introduce the following two critical prices:\footnote{Intuitively, the critical price $\pbs$ corresponds to the case that $\thba = \thas$ (which must be same as $\thbs$), i.e., three curves in Figure \ref{fig:threshold_division} intersect at the same point.
%The critical price $\pas$ corresponds to the case that $\thas = 1$, i.e., the green and red curves in Figure \ref{fig:threshold_division} intersect at $\th = 1$.}
%\begin{equation}\label{eq:p-thres-BS}
%\textstyle
%\pbs = \frac{ c \cdot ( \RA - \RB )  }{ \RS - \RB  },\quad \textstyle
%\pas = c - ( \RS - \RA ).
%\end{equation}
%Then, under a particular information price $\p$ and an initial market share, the derived market share $\Prob(\p)$ is given by
%\begin{itemize}
%\item
%If $\p > \pbs$, then $\Prob(\p) = 0$ (as $\thas < \thba$);
%\item
%If $ \pas\leq \p \leq \pbs$, then $\Prob(\p) = \thas-\thba = \frac{ \c - \p }{ \RS - \RA }- \frac{ \p }{ \RA - \RB } $;
%\item
%If $ 0\leq \p \leq \pas$, then $\Prob(\p) = 1 - \thba = 1 -\frac{ \p }{ \RA - \RB } .$
%\end{itemize}

Formally, we have the following market share in slot $\t +1$.
%\vspace{-2mm}
\begin{lemma}\label{lemma:market-share}
Given database $\dbbo$'s market share $\Probao^{\t}$ at the end of slot $\t$, the market share $\Probao^{\t + 1}$ in slot $\t+1$ is given by
\begin{equation}\label{eq:user-prob-1}
\textstyle
\begin{aligned}
\Probao^{\t+1}  &
= \max \big\{ \min\{ \thsa^{\t} ,1 \} - \thab^{\t} ,\ 0  \big\}.
\end{aligned}
\end{equation}
\end{lemma}

%The results in Lemma \ref{lemma:market-share} assume that all \eus~update the best strategies once and simultaneously.
As $\thab^{\t}$ and $\thsa^{\t}$ are functions of the market share $\Probao^{\t}$, the market share $\Probao^{\t+1} $ in slot $\t+1$  is also a function of  $\Probao^{\t}$, and hence can be written as $\Probao^{\t+1}(\Probao^{\t})$.

%\com{This is a comment that should be given on page 11, when we first talk about the initial market share with the superscript 0. In fact, such a superscript represents "initial", which is confusing as it is not consistent with the later discussions that we use time t to index the market evolution.
%In other words, the change of
%$\eta_1^0 -> \eta_1$
%has the same meaning of the change of
%$\eta_1^t ->\eta_1^{t+1}$
%This is quite confusing. I am not sure what is the best way to fix this issue. }

\subsection{Market Dynamics and Equilibrium}\label{sec:market-dynamic-monopoly}
When the market share of database $1$ changes,
the {\eus}' payoffs (when choosing the advanced service) change accordingly, as $\RAo(\Probao^{\t})$ changes.
As a result, {\eus} will update their best strategies continuously, hence the market shares will evolve dynamically, until reaching a {stable} point (called \emph{market equilibrium}).
In this subsection, we will study such a market dynamics and equilibrium, under a fixed price  $\po$.

%For convenience, we introduce a virtual time-discrete system with slots $\t=1,2,\ldots$, where {\eus} change their decisions at the beginning of every slot, based on the derived market shares in the previous slot.
Base on analysis in Section \ref{sec:user-choice-monopoly},
let $\Probao^{0}$ denote the \emph{initial market share} in slot $\t = 0$ and $\Probao^{\t}$ denote the market share derived at the end of slot $t$.
%Base on analysis in Section \ref{sec:user-choice-monopoly}, $\thab $ and $\thsa$ in next slot $t + 1$ are functions of $\Probao^{\t}$.
We further denote $\triangle \Probao $ as the change of market share between two successive time slots, e.g., $\t$ and $\t+1$, that is,
\begin{equation}\label{eq:user-prob-diff}
\begin{aligned}
\triangle \Probao(\Probao^{\t}) &= \Probao^{\t+1} - \Probao^{\t },
\end{aligned}
\end{equation}
where $\Probao^{\t+1}$ is the derived market share in slot $\t+1$, which can be computed by Lemma \ref{lemma:market-share}.
Obviously, if $\triangle \Probao $ is zero in  slot $\t+1$, i.e., $\Probao^{\t+1} = \Probao^{\t} $, then {\eus} will no longer change their strategies in the future. This implies that the market achieves a stable state, which we call the \emph{market equilibrium}.
Formally,

\begin{definition}[Monopoly Market Equilibrium]\label{def:stable-pt}
A market share  $\Probao^{\t}$ in slot $\t$ is a market equilibrium iff
\begin{equation}\label{eq:equilibriu_pt_mono}
\triangle \Probao (\Probao^{\t}) = 0.
\end{equation}
\end{definition}

Definition \ref{def:stable-pt} implies that once the market share satisfies (\ref{eq:equilibriu_pt_mono}) in slot $\t$, the market share remains the same from that time slot on.
For notational convenience, we will also denote the market equilibrium by $  \Probao^{*}  $.
%We denote the market equilibrium point as $\Probao^{*} = \Probao^{\t}$, where $\Probao^{\t}$ satisfies (\ref{eq:equilibriu_pt_mono}).
%\com{This definition does not seem to be complete, without specifying what is the choice of t.
%Same comment for Definition 2 (which is even more vague as t does not appear). }

%\begin{figure}[t]
%  \centering
%  \includegraphics[width=2.8in]{fig_dynami_eta_mono}
%  \caption{Illustration of Market Dynamics and Market Equilibrium.}
%%  \com{change the label}
%\vspace{-2mm}
%  \label{fig:percent-dyna-monopoly}
%  \vspace{-2mm}
%\end{figure}

Next, we study the existence of the market equilibrium, and further characterize the market equilibrium analytically.
Specifically,
%we will show that under a given price, there may be one or multiple \emph{tipping points} of the initial market share, around which a slight change will leads to a significant change on the market equilibrium.
we will show that  under a fixed price $\po$,
there may be \emph{multiple} equilibria, and which one will eventually emerge depends on database's initial market share (i.e., market share in slot $\t = 0$).
Besides, some equilibria are \emph{stable} in the sense that a small fluctuation around these equilibria will not drive the market share away from the equilibria,
while some equilibria  are \emph{unstable} in the sense that a tiny fluctuation on these equilibria will drive the market share to a different equilibrium.

%A key system parameter that affects the characterization of the market equilibria is the sensing cost $\ps$. Next we will consider both low and high sensing cost.
%For convenience, we denote the the magnitude of the {\eus}'s sensing cost as $\afa \triangleq \frac{\ps}{ \RS - \RB}$, where $\afa \in [0,1]$.
\vspace{-2mm}
\begin{proposition}[Existence]\label{lemma:existence-eq_pt_mon}
Given any fixed price $\po$ and sensing cost $\ps$,
there exists at least one market equilibrium.
\end{proposition}
\vspace{-3mm}
\begin{proposition}[Uniqueness]\label{lemma:uniqueness-eq_pt_mon}
Given any  fixed price $\po$ and sensing cost $\ps$,
there exists a unique market equilibrium $\Probao^*$ if
\begin{equation}\label{eq:stable_condition_mon2}
\textstyle
\max_{\Probao \in [0,1]} \frac{ \RAo^{\prime}(\Probao) }{ \RAo(\Probao) - \RB } \cdot \frac{  \RS - \RB }{ \RS - \RAo(\Probao) }  \leq \kappa_2,
\end{equation}
where $\kappa_2 = 1/{\max_{\Probao \in [0,1]} \thsa(\Probao)}$.
\end{proposition}
\vspace{-1mm}

Recall that $\gy(\Probao)$ is concave in $\Probao$ and $\RAo(\Probao) = \gy(\Probao)$ by Assumption \ref{assum:positive}. Hence,
a practical implication of (\ref{eq:stable_condition_mon2}) is that if the information value $\RAo(\Probao)$ (i.e., the positive network externality) increases   slowly with $\Probao$, then there exists a unique equilibrium.
Note that the condition (\ref{eq:stable_condition_mon2}) is sufficient but not necessary for the uniqueness.
Simulations show that the market converges to a unique equilibrium for a wide range of prices, under which the condition (\ref{eq:stable_condition_mon2}) can be violated.
Nevertheless, the  condition in (\ref{eq:stable_condition_mon2}) leads to the insight that if the change of positive network externality is slow, there   exists a unique equilibrium point.

%We use Figure \ref{fig:percent-dyna-monopoly} as an illustration, where there are three equilibrium points $\Prob^{NE}_{1}$, $\Prob^{NE}_{2}$, and $\Prob^{NE}_{3}$.
%%$\Prob_{A1}$, $\Prob_{A2}$, and $\Prob_{A3}$.
%If the initial market share $\Probao^0 < \Prob^{NE}_{1}$, then the market share will gradually increase to $\Prob^{NE}_{1}$ as $\triangle \Probao>0$.
%Similarly, if
% $\Prob^{NE}_{1}< \Probao^0 < \Prob^{NE}_{2}$, then the market share will gradually decrease to $\Prob^{NE}_{1}$ as $\triangle \Probao < 0$.
%Only if  $\Probao^0 > \Prob^{NE}_{2}$, the highest stable equilibrium $\Prob^{NE}_{3}$ will emerge eventually.
%Notice that given the price $\po$, the database always prefers the highest stable equilibrium if multiple equilibria exist.
%Thus, some incentive mechanism is necessary to motivate more {\eus} subscribing to the advanced service earlier, so as to construct a higher initial market share and achieve a higher stable equilibrium.

Suppose the uniqueness condition (\ref{eq:stable_condition_mon2})  is satisfied.
%Let $\RAo^{\text{max}}$ be the maximum expected utility that a {\eu} can achieve from choosing the advanced service of monopoly database $1$, i.e., when all {\eus} choose database $1$'s advanced service.
%Correspondingly, let $\thsa^{\text{max}} =\frac{\ps-\po}{\RS - \RAo^{\text{max}}}$ be the corresponding value when $\RAo = \RAo^{\text{max}}$.
Let $\RAo^{\text{min}}$ be the minimum expected utility that a {\eu} can achieve from choosing the advanced service of monopoly database $1$. Correspondingly, let $\thab^{\text{max}} = \frac{ \po}{ \RAo^{\text{min}}- \RB }$ be the corresponding value when $\RAo = \RAo^{\text{min}}$.
We characterize the unique equilibrium by the following theorem.
\vspace{-2mm}
\begin{theorem}[Market Equilibrium]\label{thrm:stable-eq_pt-mon}
Suppose the uniqueness condition (\ref{eq:stable_condition_mon2}) holds.
Then, for any price $\po$ and sensing cost $\ps$, the unique market equilibrium is given by
\begin{itemize}
\item[(a)]
%If $ \left. \thab \right|_{\Probao = 0} \leq \thsb $, then there is a unique market equilibrium  $ \Probao^{*} $ given by
%If $ \thab (\Probao = 0) \leq \thsb(\Probao) $, then there is a unique market equilibrium  $ \Probao^{*} $ given by
If $\thab^{\text{max}} < \thsb$,
%If $ \thsa^{\text{max}} < 1 $,
then there is a unique market equilibrium  $ \Probao^{*} $ given by
    \begin{equation}\label{eq:NE-pt-mon1}
%    \textstyle  \Probao^{\dag} = 1 - \frac{ \po }{ \RAo(\Probao^{\dag}) - \RB } .
      \textstyle \Probao^{*} = \min\{ \thsa(\Probao^{*}), 1 \} - \thab(\Probao^{*}).
%
%       \min\{ \thsa(\Probao^{*}), 1 \} - \thab(\Probao^{*}).
    \end{equation}

\item[(b)]
%If $ \left. \thab \right|_{\Probao = 0} > \thsb $ , then there is a unique market equilibrium  $\Probao^{\dag} $ given by
%If $ \thab (\Probao = 0) > \thsb(\Probao) $ , then there is a unique market equilibrium  $\Probao^{\dag} $ given by
%If $ \thsa^{\text{max}} \geq 1 $,
If $\thab^{\text{max}} \geq \thsb$,
then there is a unique market equilibrium  $\Probao^{\dag} $ given by
    \begin{equation}\label{eq:NE-pt-mon2}
%%    \Probao^{\dag} =  1 -  \thab(\Probao^{*}).
%    \textstyle  \Probao^{*} = \thsa(\Probao^{*}) - \thab(\Probao^{*}).
      \textstyle \Probao^{\dag} = 0.
    %\textstyle  \Probao^{*} = \frac{ \ps - \po }{ \RS - \RAo(\Probao^{*}) } - \frac{ \po }{ \RAo(\Probao^{*}) - \RB }.
    \end{equation}
\end{itemize}
\end{theorem}

Theorem \ref{thrm:stable-eq_pt-mon} shows that if the information value (i.e., $\RAo^{\text{min}}-\RB$) is low, then $\thab^{\text{max}} \geq \thsb$ and no WSDs will choose the advanced service in the market equilibrium.
Only when the information value is high enough (i.e., $\thab^{\text{max}} < \thsb$ ), the database $1$ can obtain the positive market equilibrium.

%the utility provided by the advanced service is high enough (i.e., $\thsa^{\text{max}} \geq 1$), the database $1$'s advanced service will attract WSDs with middle and high value of $\th$ (i.e., all WSDs with $\th \geq \thab(\Probao^{*})$). Otherwise, the database $1$'s advanced service can only attract WSDs with WSDs with middle value of $\th$. WSDs with high value of $\th$ will choose sensing service and WSDs with low value of $\th$ will choose basic service.
%%\com{Explain the physical meanings of these two cases. }

\subsection{Revenue Maximization}
{
Based on the market equilibrium analysis in the previous subsection, we will study the optimal information
pricing strategy of the monopoly database $\dbbo$ that maximizes its payoff, i.e.,
\begin{equation}\label{eq:db-profit-mono}
\Udbo(\po) = ( \po - \co) \cdot \Probao^{*}(\po)
\end{equation}
where $\co$ is the operational cost of the database that characterizes the energy consumption of the database to provide the advanced service, and $\Probao^{*}$ is the equilibrium point of the {\eu} subscription dynamics at price $\po$ given by Theorem \ref{thrm:stable-eq_pt-mon}.
}

Directly solving the optimal price that maximizes (\ref{eq:db-profit-mono})  is very challenging, due to the difficulty in analytically characterizing the market equilibrium $\Probao(\po)$ under a particular price pair $\po$.
To this end, we transform the original price maximization problem into an
equivalent \emph{market~share~maximization problem}.
The key idea is to view the market share as the strategy of the database, and the price as a function of the market share.

Furthermore, under the uniqueness condition (\ref{eq:stable_condition_mon2}), there is a \emph{one-to-one} correspondence between the market equilibrium $\Probao^{*} $ and the prices  $ \po$.
In this sense, once the monopoly database $\dbbo$ chooses the prices $\po$, it has equivalently chosen the market share $\Probao^{*} $ (given the fixed sensing cost).
Hence, we obtain the equivalent {market share maximization problem}, where the strategy of the database is its market share (i.e., $\Probao$), and the prices $\po$ is the function of the market share $\Probao$.
Let $\RAo^{\text{max}}$ be the maximum expected utility that a {\eu} can achieve from choosing the advanced service of monopoly database $1$, i.e., when all {\eus} choose database $1$'s advanced service.
Then substitute $\thsa=\frac{\ps-\po}{\RS - \RAo(\Probao)}$ and $\thab=\frac{ \po}{ \RAo(\Probao)- \RB }$ into (\ref{eq:NE-pt-mon1}) and (\ref{eq:NE-pt-mon2}), we can derive the inverse function of (\ref{eq:NE-pt-mon1}) and (\ref{eq:NE-pt-mon2}), where price is a function of market share, i.e.,
\begin{itemize}
\item[(a)]
%\emph{ Low sensing cost:} $ \ps < \RS - \left.\RAo(\Probao)\right|_{\Probao = 1}$,
\emph{ Low sensing cost:} $ \ps < \RS - \RAo^{\text{max}}$,
    \begin{equation}\label{eq:p-vs-share-mon1}
    \textstyle
       \po(\Probao ) = \frac{\RS - \RAo(\Probao)}{ \RS - \RB }\cdot \big( \frac{ \ps }{\RS - \RAo(\Probao)} - \Probao \big)\cdot \big[ \RAo(\Probao) - \RB \big].
    \end{equation}

\item[(b)]
%\emph{ High sensing cost:} $ \ps \geq \RS - \left.\RAo(\Probao)\right|_{\Probao = 1}$,
\emph{ High sensing cost:} $ \ps \geq \RS - \RAo^{\text{max}}$,
    \begin{equation}\label{eq:p-vs-share-mon2}
    \textstyle
       \po(\Probao ) = (1 - \Probao) \cdot \big[ \RAo(\Probao) - \RB \big] .
    \end{equation}
\end{itemize}

Accordingly, the revenue of the monopoly database can be written as:
\begin{equation}\label{eq:db-profit-mono-new}
\textstyle
\Udbo(\Probao) = ( \po(\Probao) - \co) \cdot \Probao
\end{equation}

We first show the equivalence between the price maximization problem and the market share maximization problem.
\vspace{-1mm}
\begin{proposition}[Equivalence]\label{lemma:game_tranform_mono}
If $\Proba^*$ is an optimal solution of (\ref{eq:db-profit-mono-new}), then $ \po^{*}$ calculated by substituting $\Proba^*$ into (\ref{eq:p-vs-share-mon1}) or (\ref{eq:p-vs-share-mon2}) is an optimal solution of (\ref{eq:db-profit-mono}).
\end{proposition}
\vspace{-1mm}

We can easily check that the database's revenue in (\ref{eq:db-profit-mono-new}) is monotonic in $\Probao \in [0,1]$, hence we have:
\vspace{-1mm}
\begin{proposition}[Optimal Information Pricing]\label{lemma:db-profit-mono}
There exists a unique optimal solution $\po^*$ for the database, where for
\begin{itemize}
\item[(a)]
%\emph{ low sensing cost:} $ \ps < \RS - \left.\RAo(\Probao)\right|_{\Probao = 1}$,
\emph{ low sensing cost:} $ \ps < \RS - \RAo^{\text{max}}$,
    \begin{equation}\label{eq:p-vs-share-mon11}
    \textstyle
%    \po^* \eq \frac{ \RA(\Probao^{\dag}) - \RB }{ \RS - \RB } \cdot \big[ \ps - ( \RS  - \RA(\Probao^{\dag}) )\cdot \Probao^{\dag} \big].
    \po^* \eq   \frac{\RS - \RAo(\Probao^{\dag})}{ \RS - \RB }\cdot \big( \frac{ \ps }{\RS - \RAo(\Probao^{\dag})} - \Probao^{\dag} \big)\cdot \big[ \RAo(\Probao^{\dag}) - \RB \big],
    %\frac{ \mathrm{d} }{ \mathrm{d} \Probao } \big(  \frac{ \RAo(\Probao) - \RB }{ \RS - \RB } \cdot[ \ps \cdot \Probao -( \RS - \RAo (\Probao)) \cdot \Probao^2] \big) = 0 .
    \end{equation}

\vspace{-2mm}
\item[(b)]
\emph{ high sensing cost:} $ \ps \geq \RS - \RAo^{\text{max}}$,
\begin{equation}\label{eq:p-vs-share-mon22}
\textstyle
%    \RAo(\Probao) - \RB  + \frac{( 1 - \Probao ) \cdot \Probao}{ 1 - 2 \Probao} \cdot \frac{ \mathrm{d} \RAo(\Probao) }{ \mathrm{d} \Probao } = 0 .
    \po^* \eq ( 1 - \Probao^{\ddag}) \cdot \big[ \RA(\Probao^{\ddag}) - B \big],
%    \po^* \eq (1 - \Probao^{*}) \cdot \big[ \RAo(\Probao^{*}) - \RB \big] .
    \end{equation}
%    where $\Probao^{\ddag}$ is the solution of $\RA(\Probao) - \RB  + \frac{( 1 - \Probao ) \cdot \Probao}{ 1 - 2 \Probao} \cdot \frac{ \mathrm{d} \RA(\Probao) }{ \mathrm{d} \Probao } = 0$.
\end{itemize}
where $\Probao^{\dag}$ is the solution of
$\textstyle \frac{  \RAo(\Probao)  - \RB}{ \RS - \RB}\c + \big( \frac{ \ps }{\RS - \RB} + \frac{  2 \RAo(\Probao) - \RS - \RB   }{  \RS - \RB  } \Probao \big) \frac{ \mathrm{d} \RA(\Probao) }{ \mathrm{d} \Probao } \Probao - 2  \Probao \frac{\RS - \RAo(\Probao)}{ \RS - \RB }   \big[ \RAo(\Probao) - \RB \big]  - \c_1 = 0,$
%$ \left( \frac{ \ps }{\RS - \RB} - \Probao \right)\cdot \frac{ \mathrm{d} \RA(\Probao) }{ \mathrm{d} \Probao }  - 2 \cdot \frac{\RS - \RAo(\Probao)}{ \RS - \RB } \cdot  \big[ \RAo(\Probao) - \RB \big]  = 0$ ,
%    $\RA(\Probao) - \RB  + \frac{( 1 - \Probao ) \cdot \Probao}{ 1 - 2 \Probao} \cdot \frac{ \mathrm{d} \RA(\Probao) }{ \mathrm{d} \Probao } = 0$,
    and $\textstyle \Probao^{\ddag}$ is the solution of $\textstyle \RA(\Probao) - \RB  + \frac{( 1 - \Probao ) \cdot \Probao}{ 1 - 2 \Probao} \cdot \frac{ \mathrm{d} \RA(\Probao) }{ \mathrm{d} \Probao }  - \frac{\c_1}{ 1 - 2 \Probao}= 0$.
\end{proposition}
\vspace{-1mm}

\section{Oligopoly Information Market}\label{sec:oligopoly}
In this section, we study the general competition scenario, where $\M$ databases compete for selling information to the same pool of {\eus}.
In such an oligopoly information market, $M$ databases (the leaders) first choose their own information prices independently.
 Then, {\eus} (the followers) subscribe to different services accordingly.
%We similarly analyze this Multiple-Leader Stackelberg game by backward induction.
%We will analyze this two-layer interaction model under the competitive network using backward induction.
Similar as in the monopoly scenario, we will
%analyze this Multiple-Leader Stackelberg game by backward induction.
study the oligopoly information market by backward induction.

\vspace{-4mm}
\subsection{Stage II - Users Behavior and Market Equilibrium}\label{sec:user-dyanm-oligopoly}
Similar as in Sections \ref{sec:user-choice-monopoly} and \ref{sec:market-dynamic-monopoly}, we study the {\eu} behavior and market dynamics in this section, given the databases' information prices $\pm$, $\m \in \Mset$, and the sensing cost $\ps$.

We first consider a WSD's best strategy at the end of slot $\t$, where the market shares are $\{\Probb^{\t}, \Probs^{\t}, \Probam^{\t} \}$ with $\Probb^{\t} + \Probs^{\t} + \sum_{\m=1}^{\M} \Probam^{\t}  = 1$.
A type-$\th$ {\eu} at the time slot $\t + 1$ will
\\
(i) subscribes to the basic service and randomly chooses a channel, i.e., $\l_{\th}^* = \B$, iff
%\vspace{-1mm}
\begin{equation}\label{eq:eu-choice-random-oligopoly}
\begin{aligned}
 \th \cdot \RB > \max\{ \th \cdot \RS - \ps,\ \ \max_{m\in\M} (\th \cdot  \RAm(\Probam^{\t})  -  \pm) \},
%\mbox{~for~} \m = 1,2,\ldots, \M,
%\vspace{-2mm}
\end{aligned}
\end{equation}
%\vspace{-1mm}
(ii) senses all the available channels to determine the best one, i.e., $\l_{\th}^* = \S$, iff
\begin{equation}\label{eq:eu-choice-sensing-oligopoly}
\begin{aligned}
 \th \cdot \RS  -  \ps > \max\{ \th \cdot \RB , \ \ \max_{m\in\M} (\th \cdot \RAm(\Probam^{\t})  -  \pm) \},
%\mbox{~for~} \m = 1,2,\ldots, \M,
\end{aligned}
\end{equation}
%\vspace{-1mm}
(iii) subscribes to the database $\dbbm$'s advanced service, i.e., $\l_{\th}^* = \Am$, iff
\begin{equation}\label{eq:eu-choice-db-oligopoly}
\begin{aligned}
\th \cdot \RAm(\Probam^{\t})  -  \pm >  & \max\{  \th \cdot \RB,\ \ \th \cdot \RS  -  \ps, \max_{n\in\M,n\neq m} (\th \cdot \RAn(\Proba_{n}^{\t})  -  \pn)  \},
%\mbox{~for~} \n = 1,2,\ldots, \M, \n \neq \m,
\end{aligned}
\end{equation}
%\begin{equation}
%\left\{
%\begin{aligned}
%& \l_{\th}^* = \B, \mbox{~~~~iff~~} \th \cdot \RB > \max\{ \th \cdot \RAm  -  \pm ,\ \th \cdot \RS - \ps \}
%\\
%& \l_{\th}^* = \S, \mbox{~~~~iff~~} \th \cdot \RS  -  \ps > \max\{ \th \cdot \RB , \ \th \cdot \RAm  -  \pm \}
%\\
%& \l_{\th}^* = \Am, \mbox{~~~iff~~} \th \cdot \RAm  -  \pm >  \max\{ \th \cdot \RB, \ \th \cdot \RS, \ \th \cdot \RAn  -  \pn \}
%\end{aligned}
%\right.
%\end{equation}
%where $\RAm =  \gy(\Probam^{\t})$,
where $\m  = 1,2,\ldots, \M$.
%\com{not clear.
%As the definition of $A_m$ only involves m, why do we need to mention n here? We should mention the notation of n in each of the above inequalities (15), (16), and (17)? }

%\com{We need to start associate time index t with the notations. }
Without loss of generality,
we suppose that the market shares of $\M$ databases in time slot $\t$ are ordered as: $\ProbaM^{\t} > \ProbaMM^{\t} > \ldots > \Probao^{\t}$,
%we suppose that the market shares at time $\t$ are  ordered as: $\ProbaM^{\t} > \ProbaMM^{\t} > \ldots > \Probao^{\t}$,
and accordingly, we have: $ \RS > \RA_{\M}(\ProbaM^{\t}) > \RA_{\M-1}(\ProbaMM^{\t}) > \ldots > \RA_{1}(\Probao^{\t})$.
%(as $\RAm$ is a non-decreasing function of $ \Probam$, $\m \in \Mset$).
%We further notice that $\RS > \RA_m$, $\forall m\in \mathcal{M}$, and hence we can easily find that no {\eu} will choose the advanced service of a database $m$ if $\p_m > \ps$.
Notice that no {\eu} would like to choose a service with a lower QoS and a higher price.
Therefore, we will consider the non-trivial scenario with $\ps > \p_{\M} > \p_{\M-1} > \ldots > \p_{1} $.
%Note that no {\eu} will choose an advance service of database $\dbbn$ if its service's QoS is lower but more expensive than the advance service of database $\dbb_{\n+1}$, we therefore only consider prices with $\ps > \p_{\M} > \p_{\M-1} > \ldots > \p_{1} $.
%Moreover, we

%For notational convenience, we denote $\BProba^{\t} = (\Probao^{\t},\ldots, \ProbaM^{\t} )$ as the vector of all databases' market shares at time slot $\t$. Besides, we denote $\BProbam^{\t} = (\Probao^{\t},\ldots, \Probamo^{\t},\Probamm^{\t},\ldots,\ProbaM^{\t})$ as the market shares vectors of all databases except database $\dbbm$ at time slot $\t$.
To better illustrate the best strategy in (\ref{eq:eu-choice-random-oligopoly})-(\ref{eq:eu-choice-db-oligopoly}), we introduce the following notations:
%\begin{equation*}\label{eq:p-thres-oligopoly}
%\ths(\BProba^{\t}) \eq \frac{ \ps - \pM}{ \RS- \RAM(\ProbaM^{\t}) },
%~~~~
%\thb(\Probao^{\t}) \eq \frac{ \po}{ \RAo(\Probao^{\t})- \RB },
%~~~~
%\tham(\BProba^{\t}) \eq \frac{\pm -\pmm}{\RAm(\Probam^{\t}) - \RAmo(\Proba_{\m - 1}^{\t})},\ m=2,3,...,M,
%\end{equation*}
\begin{equation*}\label{eq:p-thres-oligopoly}
\begin{aligned}
\textstyle
&\ths^{\t} \eq \frac{ \ps - \pM}{ \RS- \RAM(\ProbaM^{\t}) },
~~~~
\thb^{\t} \eq \frac{ \po}{ \RAo(\Probao^{\t})- \RB },
~~~~
\tham^{\t} \eq \frac{\pm -\pmm}{\RAm(\Probam^{\t}) - \RAmo(\Proba_{\m - 1}^{\t})},\ m=2,3,...,M,
\end{aligned}
\end{equation*}
where $\ths^{\t}$ denotes the marginal {\eu} who is indifferent between sensing all the available channels or subscribing to the advanced service of database $\dbbM$;
$\thb^{\t}$ denotes the marginal {\eu} who is indifferent between randomly choosing a channel or subscribing to the advanced service of database $\dbbo$;
and $\tham^{\t}$ denotes the marginal {\eu} who is indifferent between subscribing to the advanced service of database $\dbbm$ or database $\dbbmm$.

%Intuitively, the {\eus} with a high value $\th$ are more willing to choose sensing in order to achieve a large utility, while the {\eus} with a low value $\th$ are more willing to choose basic service so that they will pay zero service cost. For $\M$ databases, they compete for the {\eus} with a middle value $\th$.

Next we characterize the market shares in slot $\t+ 1$, resulting from the {\eus}' best choices in slot $\t$.
We assume that all {\eus} update the best strategies once and simultaneously.
Recall that $\th$ is uniformly distributed in $[0,1]$. Then, we have the following market shares in slot $\t + 1$.
%\begin{equation*}\label{eq:prob-sense}
%\begin{aligned}
%\Probs = 1 - \ths;
%\end{aligned}
%\end{equation*}
%(ii) basic service
%\begin{equation*}\label{eq:prob-basic}
%\begin{aligned}
%\Probb = \thb;
%\end{aligned}
%\end{equation*}
%and (iii) the advanced service of $\dbbm$
%\begin{equation*}\label{eq:prob-dbm}
%\begin{aligned}
%\Probam = \thamm - \tham;
%\end{aligned}
%\end{equation*}
%\begin{equation}\label{eq:market-share-oligopoly}
%\left\{
%\begin{aligned}
%\Probb = \thb,~~~~~~~~~~~~~~~~~ &\mbox{for basic service}
%\\
%\Probs = 1 - \ths, ~~~~~~~~~~~~&\mbox{for sensing service}
%\\
%\Probam = \thamm - \tham,~ &\mbox{for $\dbbm$'s advanced service}
%\end{aligned}
%\right.
%\end{equation}
%the newly derived market shares of basic service $\Probb$, the sensing service $\Probs$, and $\dbbm$'s advanced service $\Probam$ are:
%$\Probb = \thb$, $\Probs = 1 - \ths$, $\ProbaM = \ths - \thaM$, and $\Probam = \thamm - \tham$, $\m = 1,2,\ldots, \M-1$.
%\begin{equation}\label{eq:market-share-oligopoly}
%\left\{
%\begin{aligned}
%&\Probb = \thb,
%\\
%&\Probs = 1 - \ths,
%\\
%&\ProbaM = \ths - \thaM,
%\\
%&\Probam = \thamm - \tham, \m = 1,2,\ldots, \M-1
%\end{aligned}
%\right.
%\end{equation}

\vspace{-2mm}
\begin{lemma}\label{lemma:market-share-oligopoly}
%Given any initial market shares
Given market shares
%$\Probs^{0}$, $\Probb^{0}$, and
$\{ \Probao^{\t},\Probaoo^{\t},\ldots, \ProbaM^{\t} \}$ in slot $\t$ with
% $\Proba_1^{\t} > \Proba_2^{\t} > \ldots > \Proba_{\M}^{\t}$,
 $\ProbaM^{\t} > \ProbaMM^{\t} > \ldots > \Probao^{\t}$,
the newly market shares in slot $\t + 1$ are:
%\begin{equation}\label{eq:market-share-oligopoly}
%\left\{
%\begin{aligned}
%%\Probb &= \thb,
%%\\
%%\Probs &= 1 - \ths,
%%\\
%\Proba_1^{\t} &= \ths^{\t} - \thaM^{\t},
%\\
%\Probam^{\t} &= \thamm^{\t} - \tham^{\t}, \ \m = 2,3,\ldots, \M-1
%\\
%\Probao^{\t} &= \th^{\t}_{2} - \thb^{\t}.
%\end{aligned}
%\right.
%\end{equation}
\begin{equation}\label{eq:market-share-oligopoly}
\left\{
\begin{aligned}
%\Probb &= \thb,
%\\
%\Probs &= 1 - \ths,
%\\
\ProbaM^{\t} &= \ths^{\t} - \thaM^{\t},
\\
\Probam^{\t} &= \thamm^{\t} - \tham^{\t}, \ \m = 2,3,\ldots, \M-1
\\
\Probao^{\t} &= \th^{\t}_{2} - \thb^{\t}.
\end{aligned}
\right.
\end{equation}
\end{lemma}
%The results in Lemma \ref{lemma:market-share-oligopoly} assume that all {\eus} update the best strategies once and simultaneously.
For notational convenience, we denote $\BProba^{\t} = (\Probao^{\t},\ldots, \ProbaM^{\t} )$ as the vector of all databases' market shares in slot $\t$. Besides, we denote $\BProbam^{\t} = (\Probao^{\t},\ldots, \Probamo^{\t},\Probamm^{\t},\ldots,\ProbaM^{\t})$ as the market shares vector of all databases except database $\dbbm$.
We also denote $\BProba^{0}$ as the \emph{initial market share} in slot $\t = 0$.
As $\ths^{\t}$, $\thb^{\t}$, and $\tham^{\t}$ are functions of the market shares $\BProba^{\t}$, the market shares $\BProba^{\t+1}$ in the next slot $\t+ 1$ are also functions of  $\BProba^{\t}$.
and hence can be written as  $\Probam^{\t+1}(\BProba^{\t})$, $\forall \m \in \Mset$.
%Similar as in Section \ref{sec:market-dynamic-monopoly}, we consider a discrete time-slotted system with $\t = 1,2,\ldots$, where {\eus} change their decision at the beginning of every slot, based on the derived market shares in the previous slot. We denote $\BProb^{t}$ as the market share derived at the end of slot $t$.
We further let $\triangle \Probam $ as the change of database $\dbbm$'s market share between two successive time slots, e.g., $\t$ and $\t+1$, that is,
\begin{equation}\label{eq:user-prob-diff-oligopoly}
\begin{aligned}
\textstyle
\triangle \Probam(\BProba^{\t}) &= \Probam^{\t+1}(\BProba^{\t}) - \Probam^{\t },
\end{aligned}
\end{equation}
where $\Probam^{\t+1}$ is the derived market share of the database $\m \in \Mset$ in slot $\t+1$, which can be computed by Lemma \ref{lemma:market-share-oligopoly}.
Then we give the definition of an equilibrium point, which is similar to Definition \ref{def:stable-pt}.

\vspace{-2mm}
\begin{definition}[Oligopoly Market Equilibrium]\label{def:stable-pt-oligopoly}
A set of market shares   $\BProb^{\t}$ in slot $\t$ is a market equilibrium iff
\begin{equation}\label{eq:equilibrium_pt_set}
\triangle \Probam (\BProba^{\t}) = 0, \forall \m \in \Mset.
\end{equation}
\end{definition}
\vspace{-2mm}

Definition \ref{def:stable-pt-oligopoly}
implies that once the market shares set satisfy (\ref{eq:equilibrium_pt_set}) in slot $\t$, the market share set remains the same from that time slot on.
We will denote the market equilibrium by $\BProba^{*}$.

Based on the Definition \ref{def:stable-pt-oligopoly}, we can characterize the equilibrium by the following theorem.
\vspace{-2mm}
\begin{theorem}[Market Equilibrium]\label{thrm:stable-eq_pt-oligopoly}
For any prices set $\{ \pm \}_{\m \in \Mset}$ and {\eus}' sensing cost $ \ps$, the market equilibrium is given by:
%\begin{itemize}
%\item[(a)]
%If $ \left. \RS - \RAo(\Probao)\right|_{\Probao = 1} \leq 1 $, then there is a unique market equilibrium  $ \Probao^{\dag} $ given by
%    \begin{equation}\label{eq:NE-pt-mon1}
%%    \textstyle  \Probao^{\dag} = 1 - \frac{ \po }{ \RAo(\Probao^{\dag}) - \RB } .
%    \textstyle  \Probao^{\dag} = 1 - \thab(\Probao^{\dag}) .
%    \end{equation}
%
%\item[(b)]
%If $ \left. \RS - \RAo(\Probao)\right|_{\Probao = 1} > 1 $, then there is a unique market equilibrium  $\Probao^{*} $ given by
%    \begin{equation}\label{eq:NE-pt-mon2}
%    \textstyle  \Probao^{*} = \thsa(\Probao^{*}) - \thab(\Probao^{*}).
%    %\textstyle  \Probao^{*} = \frac{ \ps - \po }{ \RS - \RAo(\Probao^{*}) } - \frac{ \po }{ \RAo(\Probao^{*}) - \RB }.
%    \end{equation}
%\end{itemize}
\begin{equation}\label{eq:NE-pt-oligopoly}
\left\{
\begin{aligned}
\ProbaM^{*} &= \ths(\BProba^{*}) - \thaM(\BProba^{*}),
\\
\Probam^{*} &= \thamm(\BProba^{*}) - \tham(\BProba^{*}), \m = 2,3,\ldots, \M-1
\\
\Probao^{*} &= \th_{2}(\BProba^{*}) - \thb(\BProba^{*}).
\end{aligned}
\right.
\end{equation}
\end{theorem}
\vspace{-1mm}

Proving the uniqueness  of maker equilibrium is challenging due to the difficulty in anaylzing (\ref{eq:NE-pt-oligopoly}). However, extensive simulations show that,
even if there exist multiple market share equilibrium points, the market always converges to a unique one under fixed initial market shares.
%\footnote{We will provide the extensive simulations results in \cite{report}}.
Hence, we have the following proposition which is important in analyzing the price competition game in Stage I.
\vspace{-2mm}
\begin{proposition}\label{prop:unique_NE_oligopoly}
Given the initial market shares (i.e., the market shares achieved in slot $\t = 0$), the market always converges to a unique market share equilibrium.
\end{proposition}
%\com{Yuan: How to explain why this is the case? We can use figure to illustrate that....}
\vspace{-5mm}

\subsection{Stage I - Price Competition Game Equilibrium}\label{sec:price-compete-oligopoly}
In this section, we study the interaction among $\M$ databases in Stage I. Specifically, in this section we will formulate the interactions among databases as a {price competition game}, and study the Nash equilibrium systematically.

We first define the \emph{price competition game (PCG)}, denoted by $\Gamma = ( \Mset, \{ \pm \}_{ \m \in \Mset }, \{ \Udbm  \}_{ \m \in \Mset  })$, where
\begin{itemize}
\item
%\emph{Players:} The database and the spectrum licensee;
$\Mset$ is the set of game players (databases);
\item
%\emph{Strategies:} The database's strategy is the price  $\pa$ of its advanced information, and the licensee's strategy is the price  $\pl$ of its licensed channels;
$\pm$ is the strategy of database $\m$, where $\pm \geq 0$;
\item
%\emph{Payoffs:} The payoffs of players are defined in (\ref{eq:db-profit}).
$\Udbm$ is the revenue of database $\m$ defined in (\ref{eq:db-profit}).
\end{itemize}

For notational convenience, we denote $\bp= (\p_1,\ldots, \p_M )$ as the vector of all databases' information prices. Besides, we denote $\bpm = (\p_1,\ldots, \p_{\m-1},\p_{\m+1},\ldots,\p_{\M})$ as the price vectors of all databases except $m$.
We also write the (assuming unique)
% \com{The uniqueness issue has not been explained clearly. Is it always unique under certain assumption, or we are assuming that it is unique (under what conditions?). Need to explain clearly. }
 market equilibrium $\BProb^{*} = \{ \Probao^{*}, \ldots, \ProbaM^{*} \}$ in Stage II as functions of prices $\bp = \{ \p_1, \ldots, \p_{\M} \}$, i.e., $\BProb^{*}( \bp )$.
Intuitively, we can interpret $\Probam^{*} (\cdot)$ as the \emph{demand} functions of database $\m$. Moreover, the database $\m$'s market share $\Probam$ depends not only on its own price $\pm$, but also on other databases' price $\bpm$. By (\ref{eq:db-profit}), the revenue of the database $\m$ is:
\vspace{-2mm}
\begin{equation}\label{eq:sl-profit-dynamic-rv}
\begin{aligned}
\Udbm(\pm , \bpm) & = ( \pm - \cm) \cdot \Probam^{*}(\pm, \bpm).
\end{aligned}
\end{equation}

\vspace{-3mm}
\begin{definition}[Price Equilibrium]\label{def:nash-price-oligopoly}
A price profile $\{ \pm^{*} \}_{\m \in \Mset}$ is called a price equilibrium, if
\begin{equation}\label{eq:db-price-dynamic}
\begin{aligned}
  \pm^{*} & = \arg \max_{\pm \geq 0}\ \Udbm(\pm , \bpm^{*}),~~\forall \m \in \Mset
\end{aligned}
\end{equation}
\end{definition}
\vspace{-2mm}

Directly solving the price equilibrium in (\ref{eq:db-price-dynamic}) is very challenging, due to the difficulty in analytically characterizing the market equilibrium $\{\BProba^{*}(\bp) \}$ under a particular price pair $\bp$.
Hence, we transform the price competition game (PCG) into an
equivalent \emph{market~share~competition game} (MSCG).
The key idea is to view the market shares as the strategy of databases, and the prices as functions of the market shares.

Based on Proposition \ref{prop:unique_NE_oligopoly}, there is a \emph{one-to-one} correspondence between the market equilibrium $\{\BProbam^{*} \}$ and the prices  $\{\bp\}$ given fixed the initial market shares.
In this sense, once the databases choose the prices $\{ \pm \}_{\m \in \Mset}$, they have equivalently chosen the market shares $\{\BProbam^{*}\}_{\m \in \Mset}$.
Hence, we obtain the equivalent {market share competition game}---MSCG, where the strategy of each player is its market share (i.e., $\Probam$ for the database $\m$), and the prices $\{\pm \}_{\m \in \Mset}$ are functions of the market shares $\{\Probam \}_{\m \in \Mset}$.
Substitute $\ths \eq \frac{ \ps - \pM}{ \RS- \RAM(\BProbam) }$, $\thb \eq \frac{ \po}{ \RAo(\BProbam)- \RB }$, and $\tham \eq \frac{\pm -\pmm}{\RAm(\BProbam) - \RAmo(\BProbam)}$ into (\ref{eq:NE-pt-oligopoly}), we can derive the inverse function of (\ref{eq:NE-pt-oligopoly}) by recursion, where prices are functions of market shares, i.e.,\footnote{We omit the trivial case where some databases have a zero market share, as this will never the case at the pricing equilibrium of Stage I.}
\begin{equation}\label{eq:price-market-share-oligopoly}
\textstyle
%\left\{
\begin{aligned}
 \pm   =  \sum_{\m = 1}^{\M+1} \bigg[ \big( 1 - \sum_{\n = \m}^{\M+1}\Proba_{\n} \big) \cdot (  \gy(\Proba_{\m}) - \gy(\Proba_{\m-1}) ) \bigg]
\end{aligned}
%   \right.
\end{equation}
where $\Prob_{\M+1} = \Probs$, $\gy(\Prob_{\M+1}) = \RS$, and $\gy(\Prob_{0}) = B$.

Accordingly, the revenue of database $\m \in \Mset$ is:
\begin{equation}\label{eq:db-profit-oligpoly-market-share}
\begin{aligned}
\Urdbm(\Probam , \BProbam) & = ( \pm(\Probam, \BProbam) - \cm) \cdot \Probam .
\end{aligned}
\end{equation}
Similarly, a pair of market shares $\{\Probam^{*}\}_{\m \in \Mset}$ is called a Nash equilibrium of MSCG, if
$$\Probam^* = \arg \max_{\Probam} \Urdbm(\Probam, \BProbam^{*}),~ \forall \m \in \Mset.$$

We first show that the equivalence between the original PCG and the above MSCG.
\vspace{-3mm}
\begin{proposition}[Equivalence]\label{lemma:game_tranform}
If $\BProba^*$ is a Nash equilibrium of MSCG, then $ \bp^*$ given by (\ref{eq:price-market-share-oligopoly}) is a Nash equilibrium of the original price competition game PCG.
\end{proposition}
\vspace{-1mm}

We can check that $\Urdbm(\Probam , \BProbam)$ for $\m \in \Mset$ is a decreasing differential function\footnote{{A function $f(x_1,x_2)$ has decreasing differences in ($x_1, x_2$) if for all $x_1 \geq x'_1$, the difference $f(x_1, x_2) - f(x'_1, x_2)$ is nonincreasing in $x_2$. If the function $f$ is twice differentiable, the property is equivalent to $\partial^2{f}/{\partial{x_1}\partial{x_2}} \leq 0 $.}}. Hence, under duopoly databases scenario  (with two databases), the MSCG is a supermodular game (with a straightforward strategy transformation), and hence the market share equilibrium of MSCG can be easily obtained by using the supermodular game theory
  \cite{topkis1998supermodular}.
%Our key results about the existence of NE is below.
%\begin{lemma}
%The market share competition game is a supermodular game with respect to $\Proba$ and $-\Probl$.
%\end{lemma}

\vspace{-2mm}
\begin{lemma}[Existence of Market Equilibrium under Duopoly Scenario]\label{thrm:NE-existence-two}
A duopoly MSCG is a supermodular game with respect to $\Proba_{1}$ and $-\Proba_{2}$. Hence, there exists at least one market share equilibrium.
\end{lemma}
\vspace{-1mm}

Note that the MSCG under oligopoly scenario (i.e., the number of databases $\M \geq 3$) cannot be transformed into supermodular game. In order to study oligopoly scenario, we consider a special case where the positive network externality of database $\m \in \Mset$ is characterized as
\begin{equation}\label{eq:network_externality_specially}
\gy(\Prob_{\m}) = \alpha_{\m} + (\beta_{\m} - \alpha_{\m}) \cdot {\Prob_{\m}}^{\gamma_{\m}},
\end{equation}
where $\gamma_{\m}\in (0,1] $.
Then we can show that $\Urdbm(\Probam , \BProbam)$ under function $\gy_{\m}$ is quasiconcave in $\Probam$. This is sufficient for guaranteeing a pure-strategy Nash equilibrium \cite{Fudenberg1991game}.

The reasons that we use (\ref{eq:network_externality_specially}) to characterize the positive network externality are as follows.
$\alpha_{\m}$ denotes the minimum benefit brought by the database $\m$'s knowledge of licensees' channel occupation information, and $\beta_\m$ denotes the maximum benefit brought by the database's advanced information.
The parameter $\gamma_{\m} \in (0,1]$ characterizes the elasticity of the network externality.
Note that this function generalizes the linear network externality models in many existing literatures such as \cite{easley2010effect}.

%First, when $\Proba = 0$, the performance gain induced by the database $\m$'s advanced information is $\RAm = \gy(\Prob_{\m}) = \alpha_{\m}$, which depends on the database $\m$'s knowledge of licensees' channel occupation information. Second, ${\Prob_{\m}}^{\gamma_{\m}}$ increases in $\Prob_{\m}$, hence corresponds to a positive network externality.
%Third, this function allows us to use a single parameter $\gamma_{\m} \in (0,1]$ to model the elasticity of the network externality.
%A small $\gamma_{\m}$ means that the value of $\gy(\Prob_{\m})$ can be reasonably large even with a small $\Prob_{\m}$. This means that a small amount of database $\m$'s market share can be enough to bring a large performance gain to WSDs by utilizing database $\m$'s advanced information. Note that this function generalizes the linear network externality models in many existing literatures such as \cite{easley2010effect}.
%%We next show that the MSCG is a supermodular game (with minor strategy transformation), and then derive the Nash equilibrium using the supermodular game theory \cite{topkis1998supermodular}.

\vspace{-2mm}
\begin{proposition}[Existence of Market Equilibrium under Oligopoly Scenario]\label{thrm:NE-existence}
Given the positive network externally function (\ref{eq:network_externality_specially}), the revenue function $\Urdbm(\Probam , \BProbam)$, $\forall \m \in \Mset$ in MSCG is quasi-concave in $\Probam$. Hence, there exists a pure-strategy Nash equilibrium $\BProba^*$.
\end{proposition}
\vspace{-1mm}

Proof of Proposition \ref{thrm:NE-existence} is similar to the analysis of the price competition game model in \cite{gallego2006quasi}.

%\com{This is vague.
%We should say that
%Proposition 7 can be proved similarly as ???? in [??], using ???? in [?]. }
%%\com{The proof of Proposition 6 is very easy. Just mention the corresponding result in the game theory textbook. }

We then apply the contraction mapping method to establish the uniqueness of the Nash equilibrium under both duopoly and oligopoly scenarios. By applying the contraction mapping approach, the uniqueness is assured when the following condition is satisfied \cite{Vives1999game}:
\begin{equation}\label{eq:unique-condition}
\begin{aligned}
\textstyle
  - \frac{  \partial^2{ \Urdbm(\Probam, \BProbam) } }{ \partial{ (-{\Probam}) }^2 } \geq \sum_{j\neq m} \frac{  \partial^2{ \Urdbm( \Probam, \BProbam)  } }{ \partial{ {(-\Probam) } }\partial{ \Proban  } }, \ \forall m\in\Mset.
\end{aligned}
\end{equation}

We can check that the MSCG game under both duopoly and oligopoly scenarios given $\gy_{\m} = \alpha_{\m} + (\beta_{\m} - \alpha_{\m}) \cdot {\Prob_{\m}}^{\gamma_{\m}}$ satisfies the above condition. Hence, we have:
%The following proposition further gives the uniqueness condition of the Nash equilibrium in MSCG.
\vspace{-2mm}
\begin{proposition}[Uniqueness under Both Duopoly and Oligopoly Scenarios]\label{thrm:NE-uniquness}
Given the positive network externally function (\ref{eq:network_externality_specially}), The MSCG with $\M \geq 2$ databases has a unique Nash equilibrium $\BProba^*$.
%if
%$$
% \textstyle - \frac{  \partial^2{ \Urdbm(\Probam, \BProbam) } }{ \partial{ (-{\Probam}) }^2 } \geq \sum_{j\neq m} \frac{  \partial^2{ \Urdbm( \Probam, \BProbam)  } }{ \partial{ {(-\Probam) } }\partial{ \Proban  } }, \quad \forall m\in\Mset.
%$$
\end{proposition}
\vspace{-1mm}
%We apply the contraction mapping approach to get the above uniqueness conditions. We can easily check that
%%Next we provide a specific example to illustrate these conditions more intuitively.....

Once we obtain the Nash equilibrium $\BProba^*$ of MSCG, we can immediately obtain the Nash equilibrium $\bp^*$ of the original PCG by (\ref{eq:price-market-share-oligopoly}).
Notice that we may not be able to derive the analytical Nash equilibrium of MSCG, as we use the generic function $\gy(\cdot)$.
Nevertheless, because the objective function of database $\m$, $\m \in \Mset$ is quasiconcave, we can numerically compute the Nash equilibrium of MSCG through several standard algorithms such as the  ellipsoid algorithm in \cite{jain2010quasi}.

\section{Numerical Results}\label{sec:simu}

In this section, we numerically illustrate the NE of the database competition game, and evaluate the system performance (e.g., the network profit and the databases' revenue) at the NE. We will focus on the impact of system parameters (i.e. the number of databases, the network effect, the positive network externality, and the database's operational cost) on system performance.
As a concrete example, we will use (\ref{eq:network_externality_specially}) to model the positive network externality.
%\com{explain reference support? }
Unless specified otherwise, we assume that $\RB = 2$, $\RS = 8$, $\alpha_{\m} = 4.8$, and $\beta_{\m} = 6$, $\m \in \Mset$. The databases' initial market shares satisfy $\Prob_M > \Prob_{M-1}>\ldots > \Prob_1$. \textbf{In all the simulation figures, we denote sensing service as S, basic service as B, and database $\m \in \Mset$ as $\m$.}

%%%%%%%%%%%%%%%%%%%%%%%%%%%%%%%%%%%%%%%%%%
\vspace{-3mm}
\subsection{System Performance vs Number of Databases}
%We first show the social welfare (i.e., the aggregated profit of all participating databases and {\eus}), each participating database's profit,

\begin{figure*}
  \centering
  \includegraphics[width=2.3in]{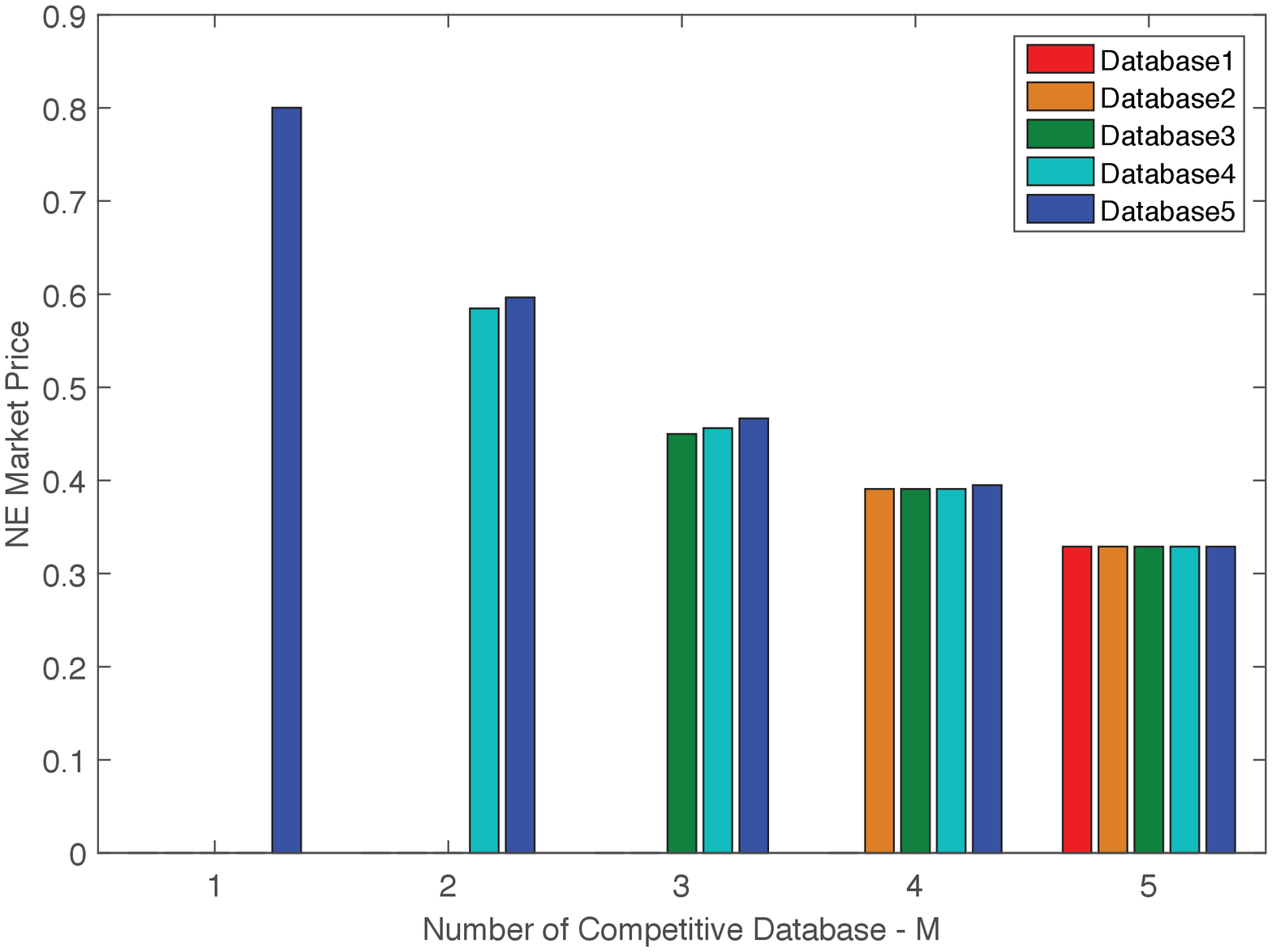}
  \includegraphics[width=2.25in]{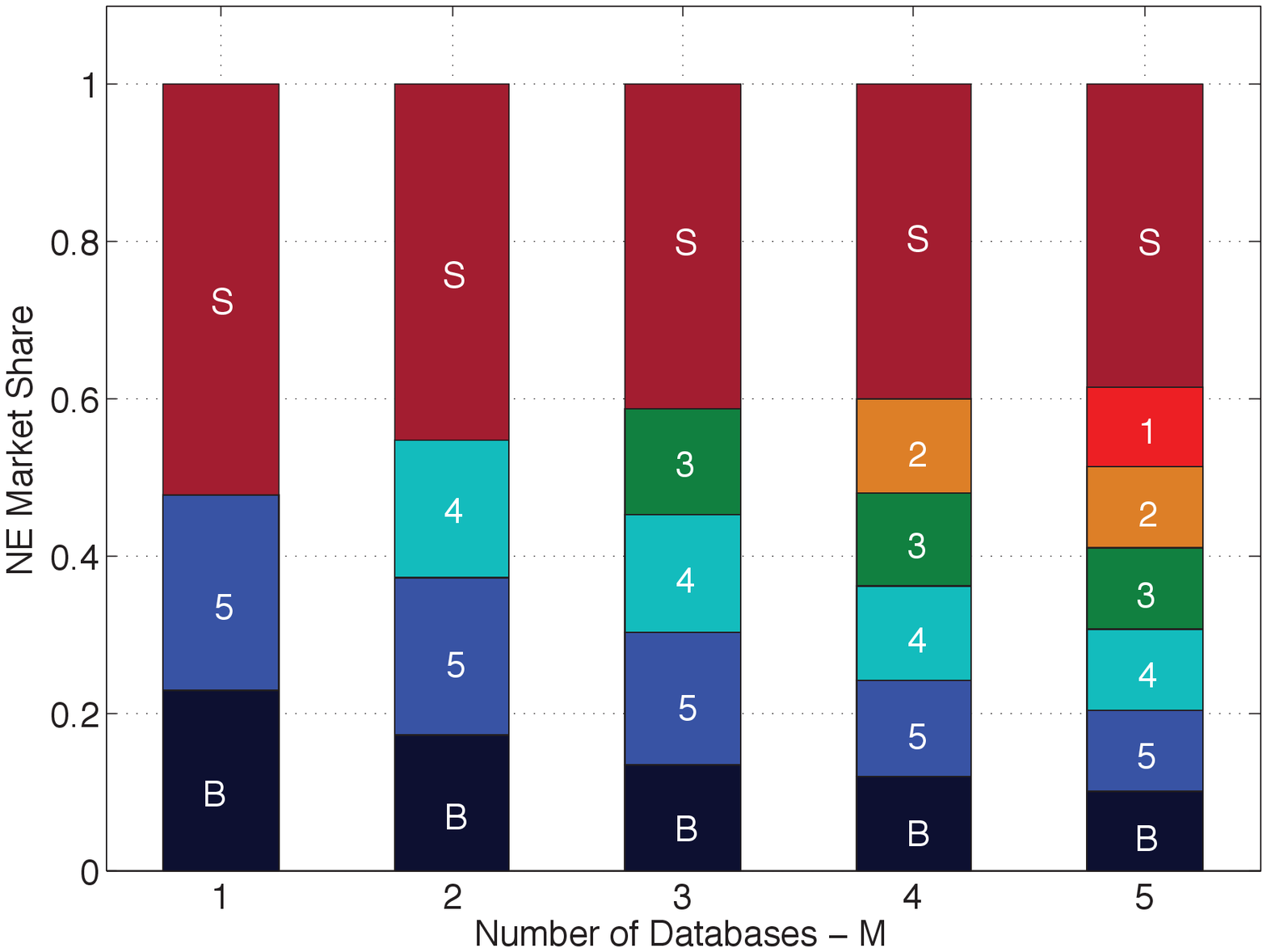}
    \includegraphics[width=2.3in]{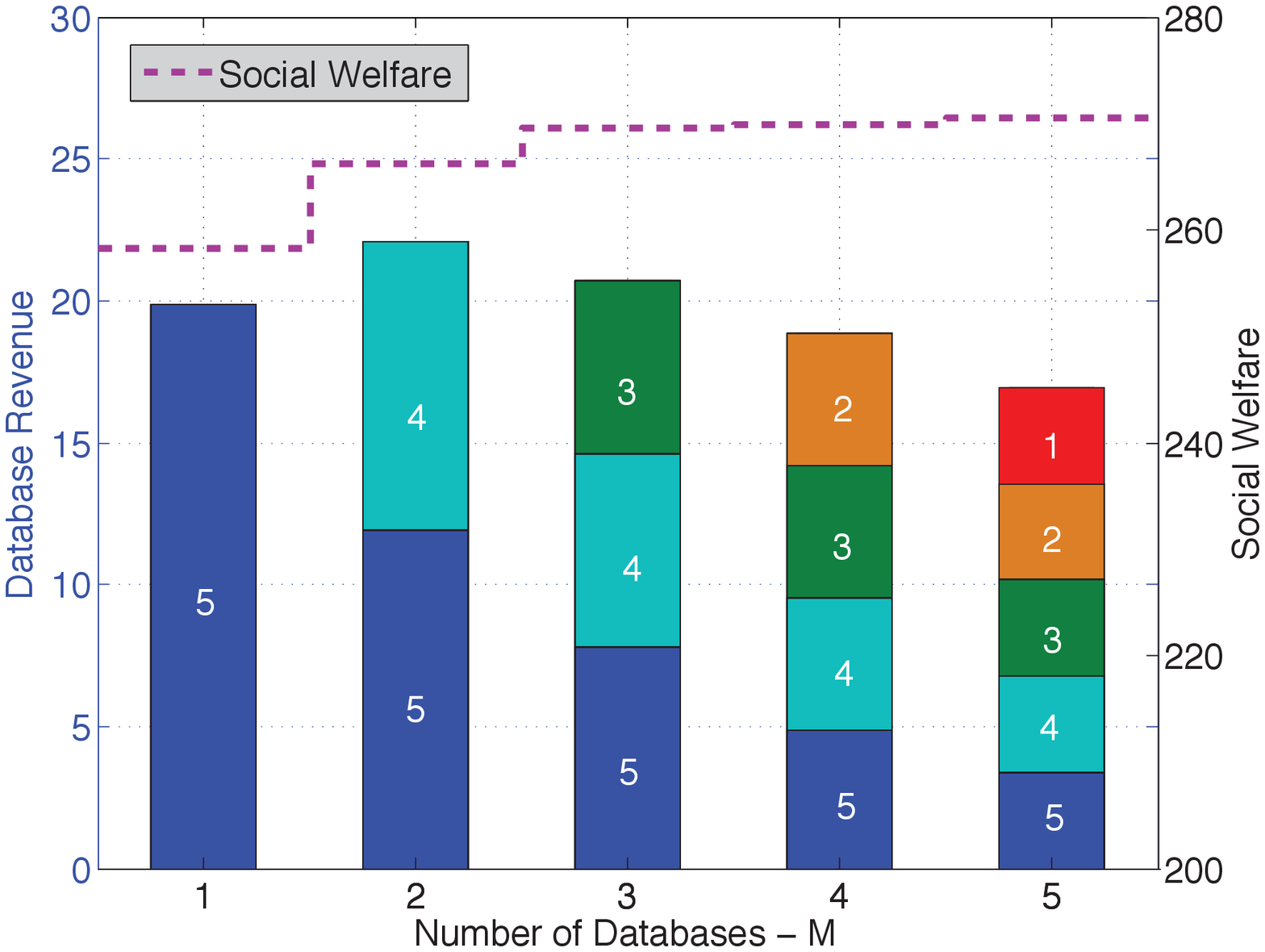}
\vspace{-3mm}
  \caption{(a) Price equilibrium, (b) Market share equilibrium, (c) The system performance vs the number of databases.}\label{fig:impact_db_number}
  %\com{This simulation result is different from the previous version?
%Have you double check that this is correct, and have revise ALL related discussions (even those that I did not edit previously)? }
  \vspace{-4mm}
\end{figure*}

Figure \ref{fig:impact_db_number} illustrates (a) market share equilibrium, (b) price equilibrium, and (c) the system performance achieved under different numbers of databases ($\M$ from $1$ to $5$). In this simulation, we fix the sensing cost as $\ps  = 2$, the network externality impact as $\gamma_{\m} = 0.4$, and the database's operation cost as $\cm = 0$, $\m \in \Mset$.
%\com{To be concise, it is enough to say "the number of databases" instead of "the number of competitive databases". Furthermore, competitive may imply something different (such as competitive market, where there are a large number of producers).
%I have changed some, but please global check:
%Replace "competitive" with "competing", if we really want to emphasize the competition among databases, especially during the initial discussions regarding the interactions among databases.
%or Remove "competitive", especially in later discussions and simulations. }
%\com{Put all figures on the page (or the immediate next page) where it is first  mentioned. }

Figure \ref{fig:impact_db_number}.a shows the price equilibrium achieved under different numbers of databases.
We can see that the equilibrium prices decrease with the number of databases, as the intensity of competition increases.
When the number of databases increases, the difference among the databases' initial market shares becomes smaller as $\sum_{\m \in \M} \Prob_{\m} = 1$. Hence, the difference among databases' price equilibrium also decreases as $\M$ becomes large.
%when $\M$ is large (e.g., $\M = 5$ in the figure) is small than that when $\M$ is small (e.g., $\M = 2$ in the figure).
%\com{Can we explain why initially the equilibrium prices are different among databases (e.g., $M=2$), and the difference almost disappears when M is large (e.g., $M=5$)? }

\begin{figure*}
	\centering
	\includegraphics[width=2.33in]{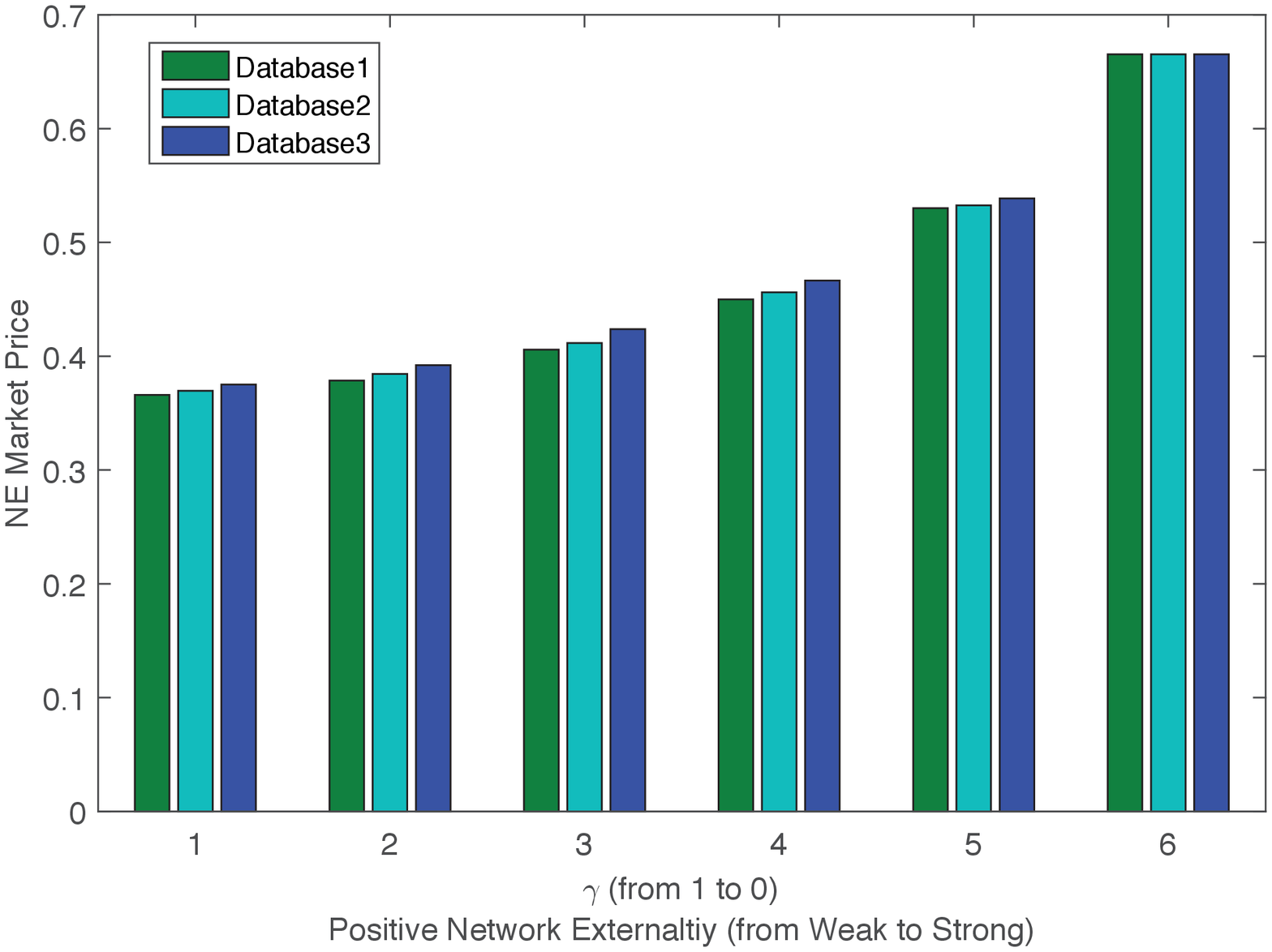}
	\includegraphics[width=2.3in]{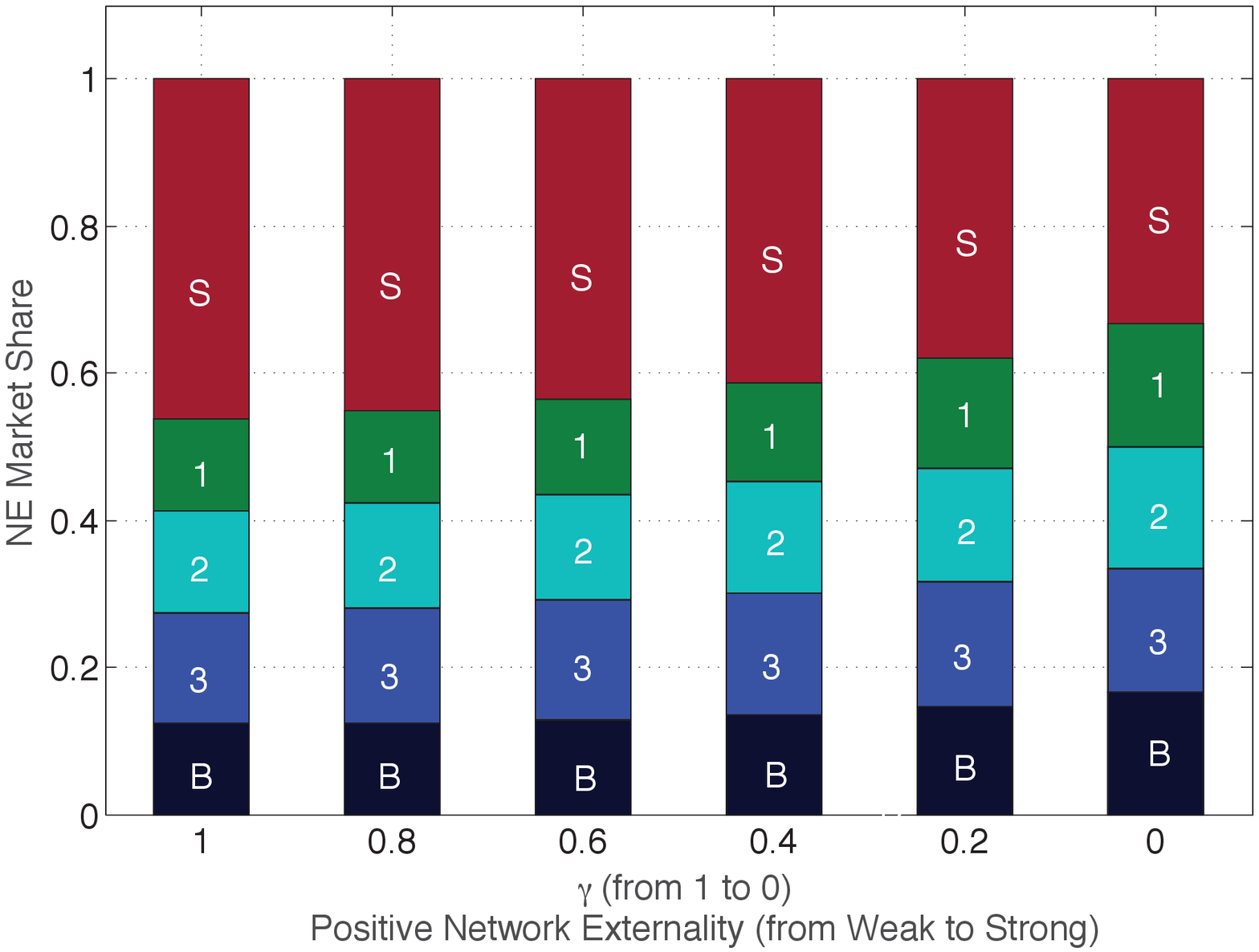}
	\includegraphics[width=2.3in]{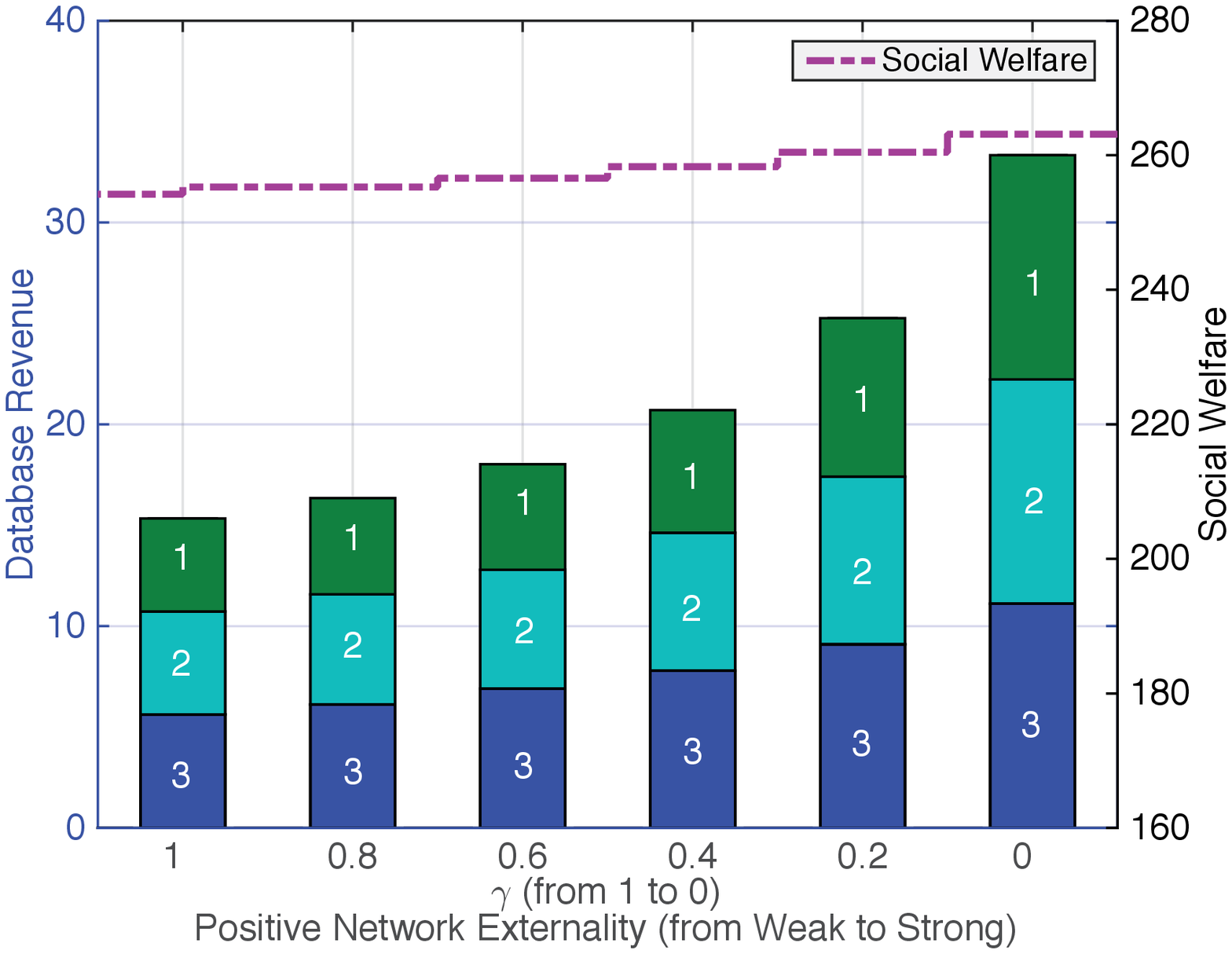}
	\vspace{-3mm}
	\caption{(a) Price equilibrium, (b) Market share equilibrium, (c) The system performance vs positive network externality. {A small value of $\gamma$ corresponds to a high level of positive network externality.}}\label{fig:impact_network_effect}
	%  \com{replace "network effect" with "positive network externality" in all three figures. }
	\vspace{-5mm}
\end{figure*}

Figure \ref{fig:impact_db_number}.b shows the equilibrium market share under different numbers of databases. Each bar denotes the market shares of the basic service (denoted as {``B"}), databases' advanced services (denoted as the database index $\m$), and sensing (denoted as {``S"}).
We can see that the databases' market shares increase with the number of databases.
%From figure \ref{fig:impact_db_number}.b, we can see that equilibrium market shares of basic service and sensing decrease with the number of databases.
%\rev{Note that the quality provided by subscribing to the advanced service is higher than that provided by the basic service. Also note that the cost of subscribing to the advanced service is lower than that of sensing.}
%\com{How are these related to the observations from Figure 6.b? }
As databases' equilibrium prices decrease with the number of databases, more WSDs will purchase the advanced services, hence increasing the market shares of the databases as $\M$ increases.

%Figure \ref{fig:impact_db_number}.c shows that the aggregated market share of competitive databases increases with the number of competitive databases.  Note that the quality provided by subscribing to the advanced service is higher than that provided by the basic service. Also note that the cost of subscribing to the advanced service is lower than that of sensing. Hence, the decreasing price attracts WSDs to choose advanced service.
%\com{Yuan: do we need to mention that the market share of particular database decreases with the number of competitive database? }

Figure \ref{fig:impact_db_number}.c shows each database's revenue and the total social welfare (i.e., the total revenue of all databases plus the total payoffs of {\eus}) achieved at the NE, given different numbers of databases.
Each bar denotes the aggregated revenue of $\M$ databases, while each sub-bar corresponds to the payoff of database $\m$. The dash red line denotes the value of social welfare. The left y-axis denotes the value of databases' revenues, and the right y-axis denotes the value of social welfare.

From Figure \ref{fig:impact_db_number}.c, we can see that the databases' aggregated revenue is a quasi-concave (i.e., first increasing and then decreasing) function of the number of databases $\M$.
This is because two things happen when $\M$ increases: (i) more intensive competition drives the equilibrium prices down for all databases, which reduces the revenue of each single database, (ii) low prices attract more WSDs to purchase the advanced services, which leads to the increase the overall databases' revenue. In this simulation, $M=2$ achieves the best trade-off and maximizes the databases' total revenue.

%Figure \ref{fig:impact_db_number}.c shows that the profit of particular database (e.g. database $1$) decreases with the number of competitive databases.  This is because the competition among the databases results in the decreasing of price. The decreasing price allows WSDs to enjoy the good quality of databases' advanced service with a low information price. Hence, the competition among databases benefits the social welfare. %\com{Yuan: It seems that this is comes from the impact of positive network externality.. However, how to elaborate the reason.....}

Figure \ref{fig:impact_db_number}.c shows that the social welfare increases with the number of databases. As the competition among databases reduce the equilibrium prices, more WSDs choose to use the advanced services, which offers a better quality of service than the basic service and a lower cost than the sensing. Overall this improves the social welfare.

\vspace{-3mm}
\subsection{System Performance vs Network Externality}

Figure \ref{fig:impact_network_effect} illustrates (a) market share equilibrium, (b) price equilibrium, and (c) the system performance achieved under different levels of network externality (e.g., $\gamma_1=\gamma_2=\gamma_3=\gamma$ changes from $1$ to $0$, hence the positive network externality changes from weak to strong).
%In this simulation, we assume that the sensing cost $\ps  = 2$ and the number of competitive database $\M = 3$.
According to (\ref{eq:network_externality_specially}), a small $\gamma_{\m}$ means that the value of $\gy(\Prob_{\m})$ can be reasonably large even with a small $\Prob_{\m}$.
Hence, a small value of $\gamma$ represents a high level of network externality.
In this simulation, we fix the sensing cost as $\ps  = 2$, the number of database $\M = 3$, and the database's operation cost as $\cm = 0$, $\m \in \Mset$.
%, while changing the network externality impact $\gamma$ from $\gamma=0$ to $\gamma=1$. We assume that $\gamma_1 = \gamma_2 = \gamma_3$ in this simulation.

Figure \ref{fig:impact_network_effect}.a shows the equilibrium prices of positive network externality.
%Each bar group denotes each competitive database's equilibrium price given the network externality value.
We can see that the databases' equilibrium prices increases with the level of network externality.
This is because a higher level of positive network externality will make the utility provided by the database's advanced service reasonably large even when the database has a small market share. This leads to a less intensive competition for the market share, hence drives the equilibrium prices up.
%will make the quality of database's advanced service
%
%This is because a higher level of positive network externality will make it more desirable for a database to have a larger market share, which can provide a significant higher quality of service. This leads to a more intensive competition for the market share, hence drives the equilibrium prices down.
%The reason is as follows. In the network with high level of network externality (i.e., the high value of $\gamma$),  attracting one more WSD will increase the quality of database's advanced service dramatically. The high quality service can help the database attract even more WSDs.
%%Meanwhile, losing one WSD will significant decrease the quality of the database, which in turn making more WSDs quit the database's advanced service.
%Hence, the increasing network externality makes the competition among the databases more severe. The severe competition makes each database decrease his advanced service price in order to earn profit.
Figure \ref{fig:impact_network_effect}.a also shows that the database $3$ always has the highest equilibrium price among all the databases. The reason is that
we assume the initial market shares of databases are $\Prob_3 > \Prob_2 > \Prob_1$. The advantage of having a larger initial market share leads to a higher equilibrium price for database $3$.

Figure \ref{fig:impact_network_effect}.b shows the equilibrium market share under different levels of positive network externality.
Each bar denotes the market share allocation among the basic service (denoted as {``B"}), database $m$'s advanced services (denoted as $m$), and sensing (denoted as {``S"}).
We can see that the databases' market shares increase with the level of positive network externality, as a high level of positive network externality makes the advanced services more attractive and attracts some high $\th$ value WSDs from sensing. Meanwhile, the market share of sensing decreases with the level of positive network externality.
Because the increasing equilibrium prices of the advanced services drive some low $\th$ value WSDs to choose the basic services, the market share of basic service increases with the level of positive network externality.

%Because the decreasing price of advanced service, some WSDs with low value of $\th$ are attracted by the advanced service, which leads to the decreasing market share of the basic service.
%On the other hand, the market share of sensing increases with the level of network externality.
%\rev{
%This is because, under higher level of network externality, one WSD's choice of service would result in the significantly change of quality of database's advanced service. Hence, some WSDs with high value of $\th$ changes from the advanced service to the sensing service due to the unstable quality of the advanced service.
%}
%\com{This explanation does not make sense. We are talking about equilibrium behavior, not the transit behavior. Why would the users worry about the "unstable" quality? }

\begin{figure*}
  \centering
  \includegraphics[width=2.2in]{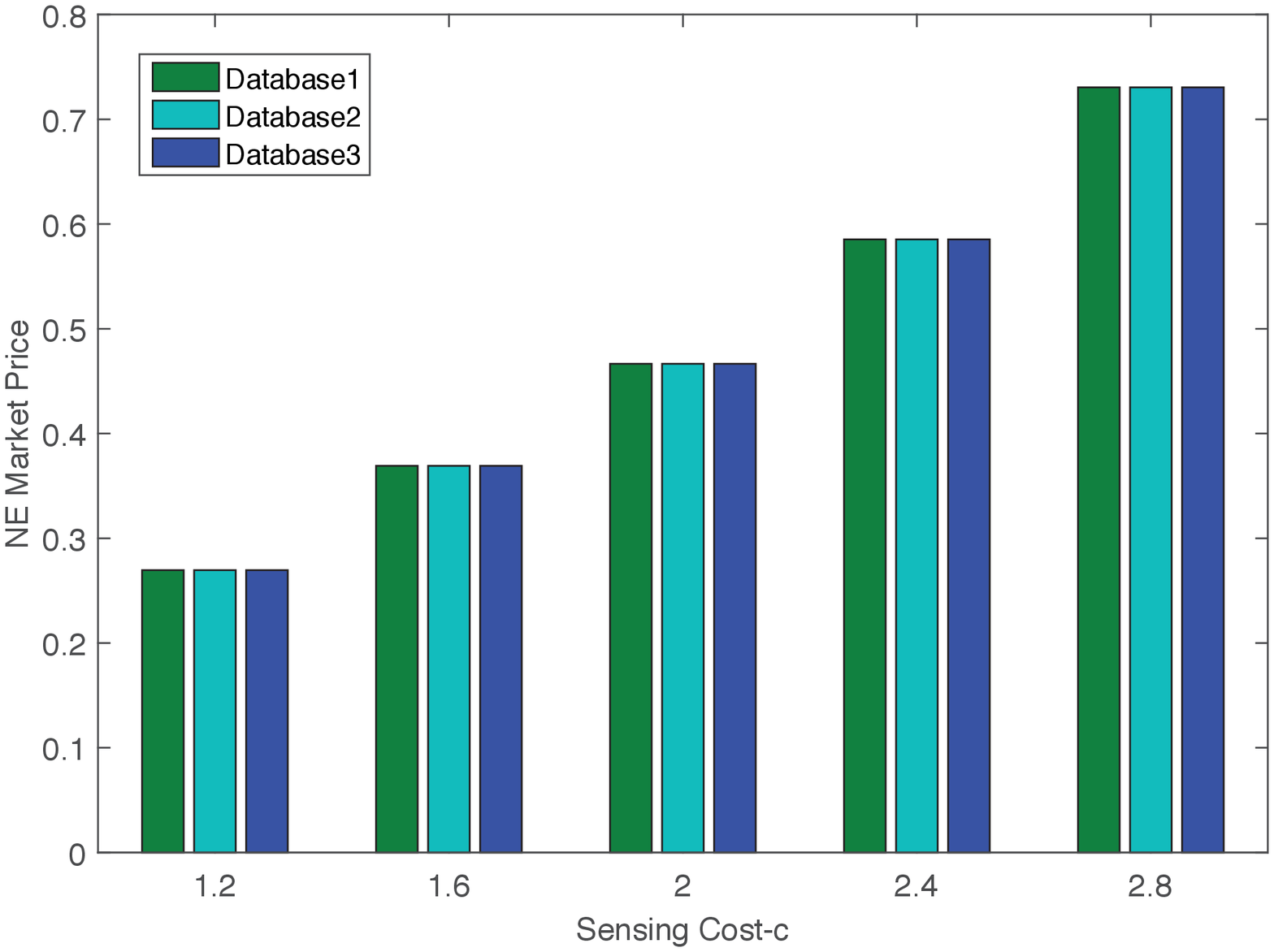}
  \includegraphics[width=2.2in]{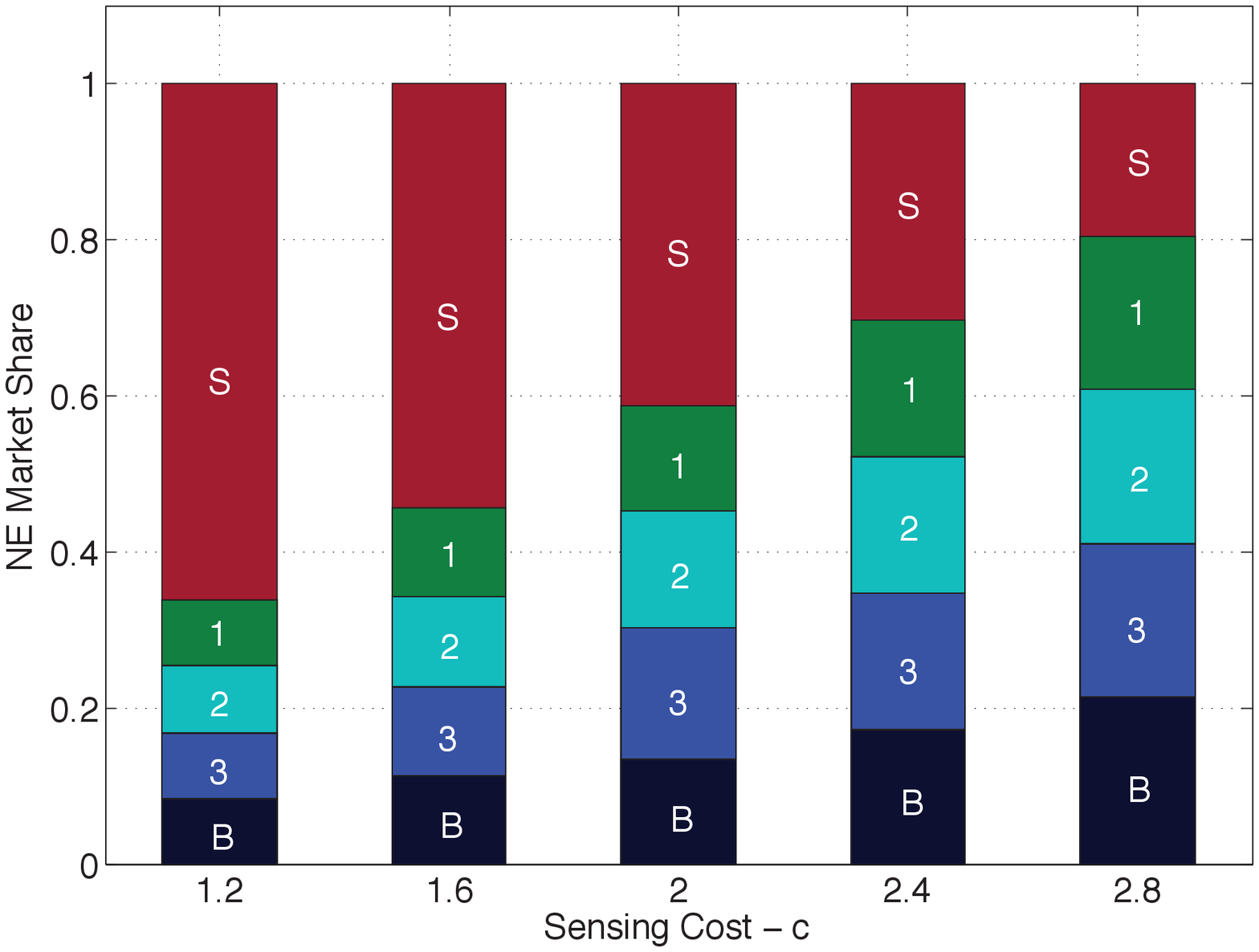}
    \includegraphics[width=2.33in]{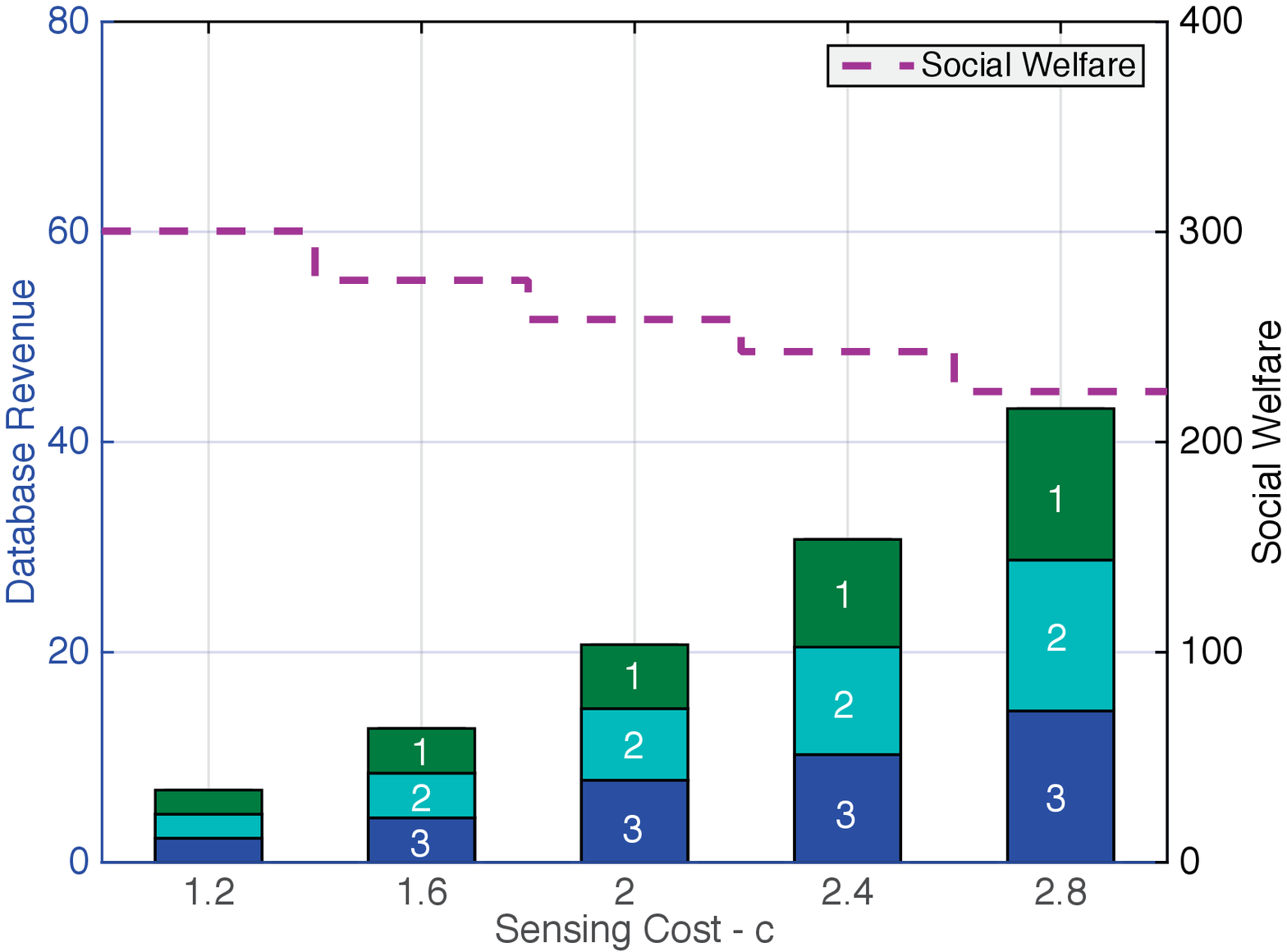}
\vspace{-3mm}
  \caption{(a) Price equilibrium, (b) Market share equilibrium, (c) The system performance vs WSDs' sensing cost}\label{fig:impact_sensing_cost}
  \vspace{-2mm}
\end{figure*}

Figure \ref{fig:impact_network_effect}.c shows each database's revenue and the total social welfare (i.e., the total revenue of all databases plus the total payoffs of {\eus}) achieved at the NE, under different values of network externality impact $\gamma$.
Each bar denotes the aggregated revenue of $3$ databases, while each sub-bar corresponds to the revenue of a particular database $\m$. The dash red line denotes the value of social welfare. The left y-axis denotes the value of database's revenue, and the right y-axis denotes the value of social welfare.
%From Figure \ref{fig:impact_network_effect}.c,
We can see that the databases' aggregated revenue increases with the network externality level. This is because when the level of positive network externality increases, high utility provided by the advanced service drives the equilibrium prices as well as the equilibrium market shares up for all databases.
%This is because high level of network externality \rev{results in the lower value of databases' service price.}\com{why? shown by the previous figures}.
%Moreover, the quality of database' advanced service is easily affected by one WSD's service choice under higher level of network externality. In such case, the quality of advanced service is unstable and thus result in the decreases of databases' market share.
%Hence, the aggregated profit of databases decreases with the network externality level.
The social welfare also increases with the network externality level. As the high level of positive network externality increases the quality of databases' service, more WSDs choose to use the advanced service,  which is cheaper than the sensing. Overall, this improves the social welfare.

\subsection{Performance vs Sensing Cost}
Figure \ref{fig:impact_sensing_cost} illustrates (a) price equilibrium, (b) market share equilibrium, and
(c) system performance achieved under different sensing cost $\ps$ (from $1.2$ to $2.8$).
%In this simulation, we assume that the network externality impact $\gamma_{\m} = 0.4$ and the number of competitive database $\M = 3$.
In this simulation, we fix the network externality impact $\gamma_1 =\gamma_2 = \gamma_3 = \gamma = 0.4$, the number of database $\M = 3$, and the database's operation cost as $\cm = 0$, $\m \in \Mset$.
%while changing the sensing cost $\ps$ from $\ps=1.2$ to $\ps=2.8$. We assume that $\gamma_1 = \gamma_2 = \gamma_3$ in this simulation.

Figure \ref{fig:impact_sensing_cost}.a shows the price equilibrium under different values of sensing cost.
We can see that the equilibrium market prices of databases increase with the sensing cost, as
a higher sensing cost allows databases to increases their prices without losing WSDs.

Figure \ref{fig:impact_sensing_cost}.b shows the equilibrium market shares achieved under different values of sensing cost.
Each bar denotes the market share allocation among the basic service (denoted as ``B"), database $m$'s advanced services (denoted as ``$m$''), and sensing (denoted as ``S").
As the sensing cost increases, the sensing services becomes less attractive, and hence
the market share of sensing decreases with the sensing cost.
Meanwhile,
the market shares of basic service and all databases' advanced services increase with the sensing cost.
%This is because, with the increasing value of sensing cost, less WSDs are willing to sense by themselves, and thus the market share of sensing service decreases with the sensing cost.
%Although increasing advanced service' price will loss some market share, the high service price can make sure the database obtaining high profit. Hence, the databases are not willing to decrease their advanced services' price, which in turn makes the market share of basic service increases.

Figure \ref{fig:impact_sensing_cost}.c shows each database's revenue and the total social welfare (i.e., the total revenue of all databases plus the total payoffs of  {\eus}) achieved at the NE, under different values of sensing cost $\ps$.
Each bar denotes the aggregated revenue of $3$ databases, while each sub-bar corresponds to the revenue of database $\m$. The dash red line denotes the value of social welfare. The left y-axis denotes the value of database's revenue, and the right y-axis denotes the value of social welfare.

From Figure \ref{fig:impact_sensing_cost}.c, we can see that the databases' aggregated revenue increases with the sensing cost, as the advanced services become more attractive to the {\eus}.
%This is because the increasing of sensing cost allows database to increase their advanced service price. Moreover, with the high value of sensing cost, WSDs are willing to choose advanced service instead of sense service. Hence, the market share of databases also increases.
However, as the sensing cost increasing, WSDs need to pay a higher price to enjoy a good quality of service, hence the social welfare decreases with sensing cost.
%\com{Yuan: how to explain the reason....}

%%%%%%%%%%%%%%%%%%%%%%%%%%%%%%%%%%%%%%%%%%
\subsection{Performance vs Operational Cost of Database}
\begin{figure*}
  \centering
  \includegraphics[width=2.2in]{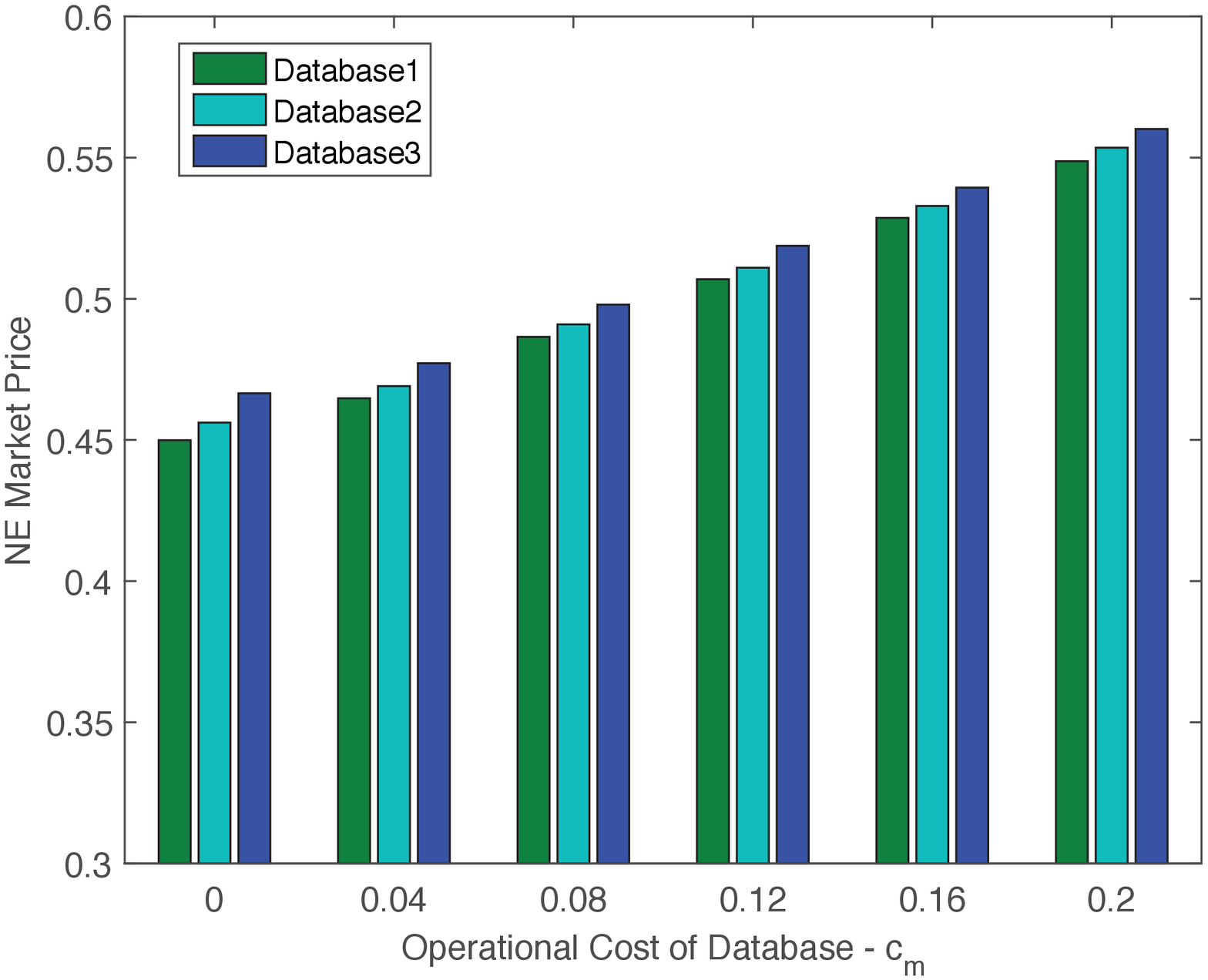}
  \includegraphics[width=2.2in]{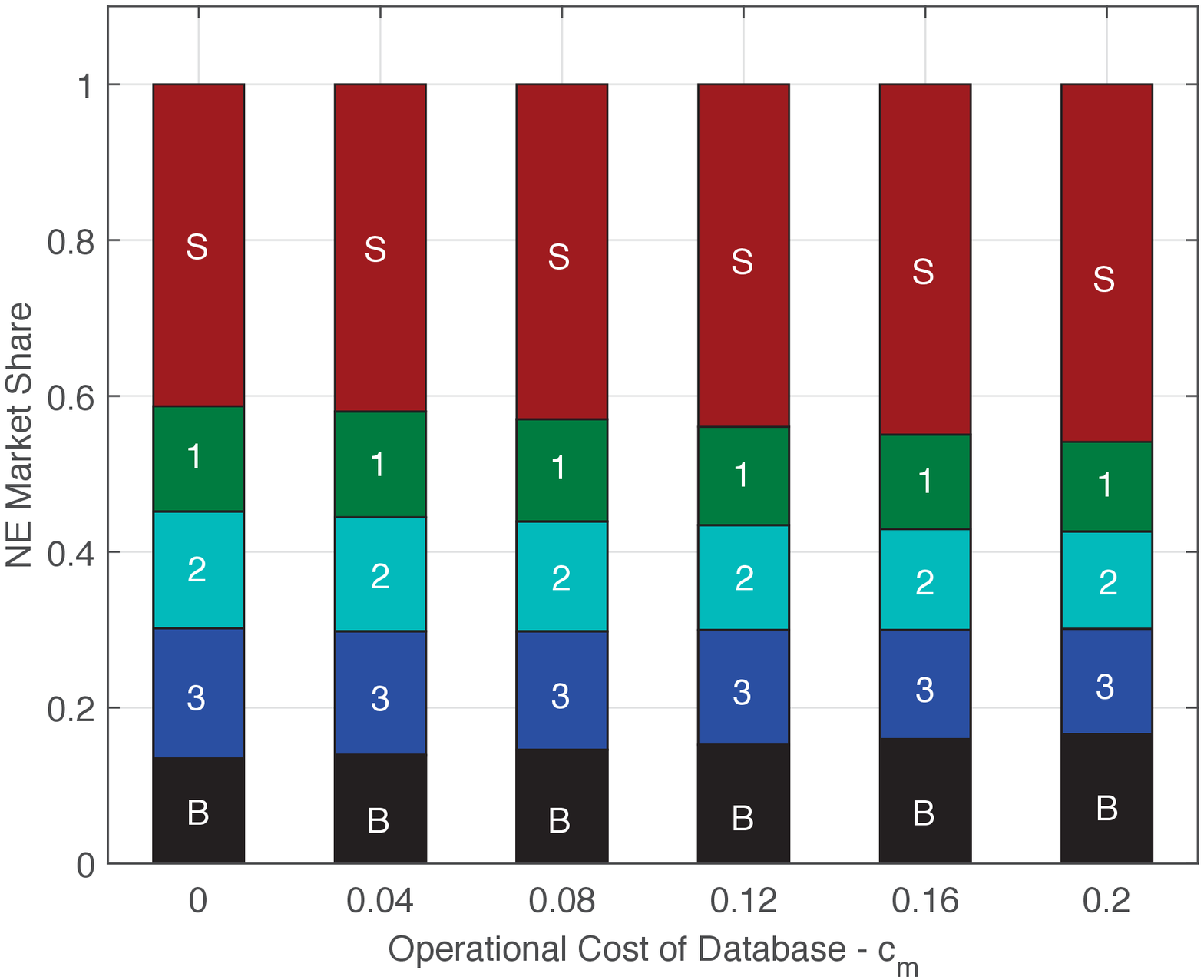}
    \includegraphics[width=2.33in]{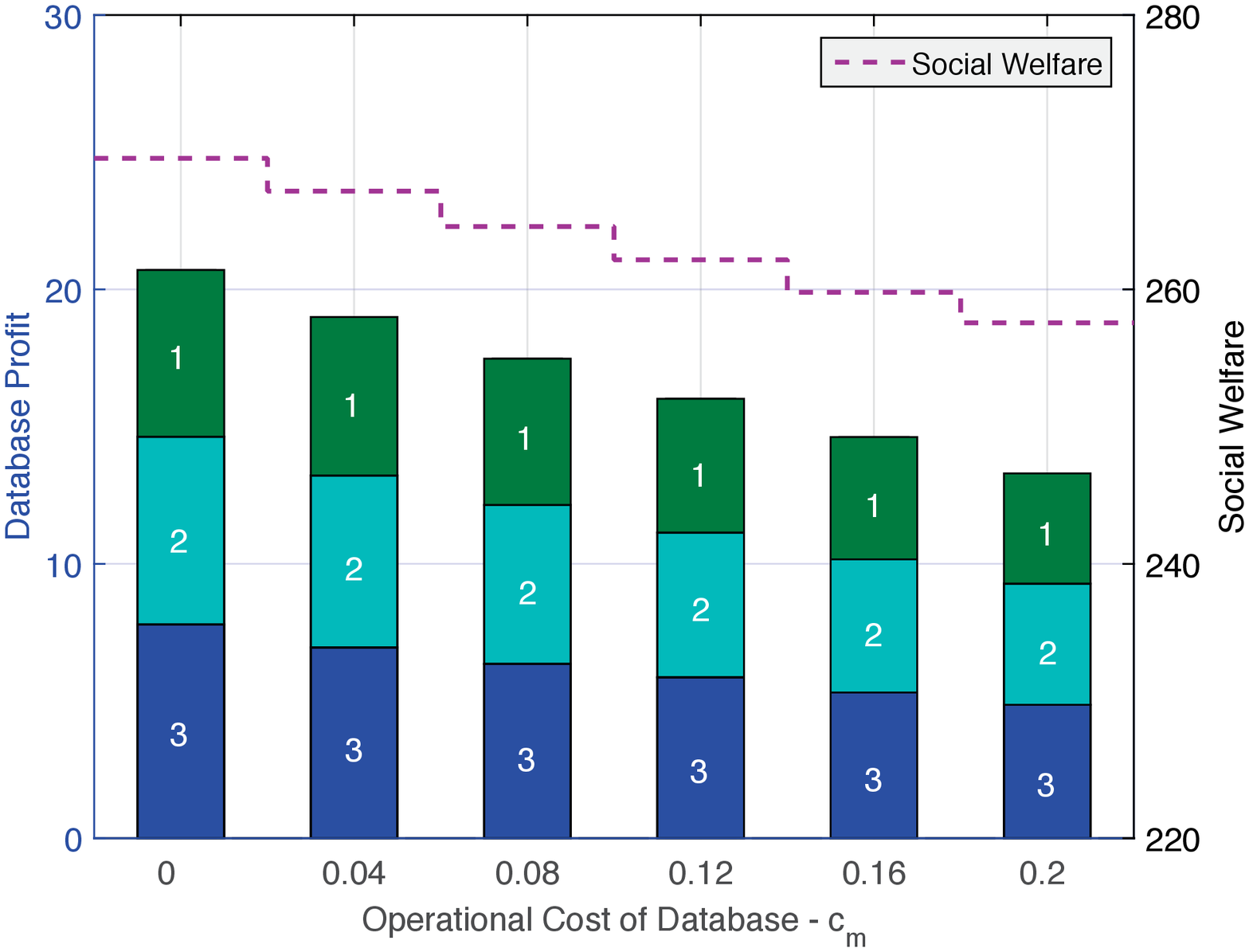}
\vspace{-3mm}
  \caption{(a) Price equilibrium, (b) Market share equilibrium, (c) the system performance vs operational cost of database.}\label{fig:impact_operation_cost}
  \vspace{-2mm}
\end{figure*}

Figure \ref{fig:impact_operation_cost} illustrates (a) the price equilibrium, (b) the market share equilibrium, and
(c) the system performance achieved under different operational costs of databases.
%Here we assume all databases has the same operational $\c_1 = \c_2 = \c_3 = \c_m$ (from $0$ to $0.2$).
%In this simulation, we assume that the network externality impact $\gamma_{\m} = 0.4$ and the number of competitive database $\M = 3$.
In this simulation, we consider a scenario of $\M=3$ databases with the same operational cost, i.e., $c_1 = c_2 = c_3$, and change such an operational cost from 0 to 0.2.
Moreover, we fix the network externality factor $\gamma_1 =\gamma_2 = \gamma_3 = 0.4$ and the sensing cost $\ps = 2$, and select slightly different initial market shares for three databases (i.e., $\Prob_3 > \Prob_2 > \Prob_1$).
In Figure \ref{fig:impact_operation_cost}.b, each bar with ``S'' or ``B'' denotes the percentage of WSDs choosing the sensing service or basic service, respectively, and each bar with number $m\in\{1,2,3\}$ denotes the percentage of WSDs choosing database $m$'s advanced service.
In Figure \ref{fig:impact_operation_cost}.c, each bar with number $m\in\{1,2,3\}$ denotes the database $m$'s profit.
%We denote sensing service as S, basic service as B, and database $\m \in \Mset$ as $\m$.
%while changing the sensing cost $\ps$ from $\ps=1.2$ to $\ps=2.8$. We assume that $\gamma_1 = \gamma_2 = \gamma_3$ in this simulation.

Figure \ref{fig:impact_operation_cost}.a shows the equilibrium market prices of databases under different operational costs.
We can see that the equilibrium market prices of all databases increase with the operational cost.
This is quite intuitive, as a higher operational cost drives the database to set a higher retail price in order to cover his operational cost.

Figure \ref{fig:impact_operation_cost}.b shows the equilibrium market shares achieved under different operational costs.
Each bar denotes the market share allocation among the basic service (denoted as ``B"), database $m$'s advanced services (denoted as ``$m$''), and sensing (denoted as ``S").
We have shown in Figure \ref{fig:impact_operation_cost}.a that with the increasing of database operational cost,
all databases will set higher equilibrium market prices to cover their operational costs.
This makes the advanced services become less attractive, and hence reduces the total market share of databases as shown in Figure \ref{fig:impact_operation_cost}.b.
Accordingly, the market shares of both basic service and sensing service increase.

Figure \ref{fig:impact_operation_cost}.c shows the databases' profits and the total social welfare (i.e., the total profit of all databases plus the total payoffs of {\eus}) achieved at the market equilibrium, under different operational costs.
Each bar denotes the aggregated profit of $3$ databases, while each sub-bar corresponds to the profit of database $\m$.
The dash red line denotes the value of social welfare.
The left y-axis denotes the value of database's profit, and the right y-axis denotes the value of social welfare.
From Figure \ref{fig:impact_operation_cost}.c, we can see that both the databases' aggregated profit and the social welfare decrease with the operational cost.
%, as the advanced services become less attractive to the {\eus} due to the increase of NE market price.
%Meanwhile, with the increasing of the operational cost, WSDs need to pay a higher market price to enjoy a good quality of service, hence the social welfare decreases accordingly.

\begin{figure*}
%\vspace{-2mm}
  \centering
  \includegraphics[width=2.2in]{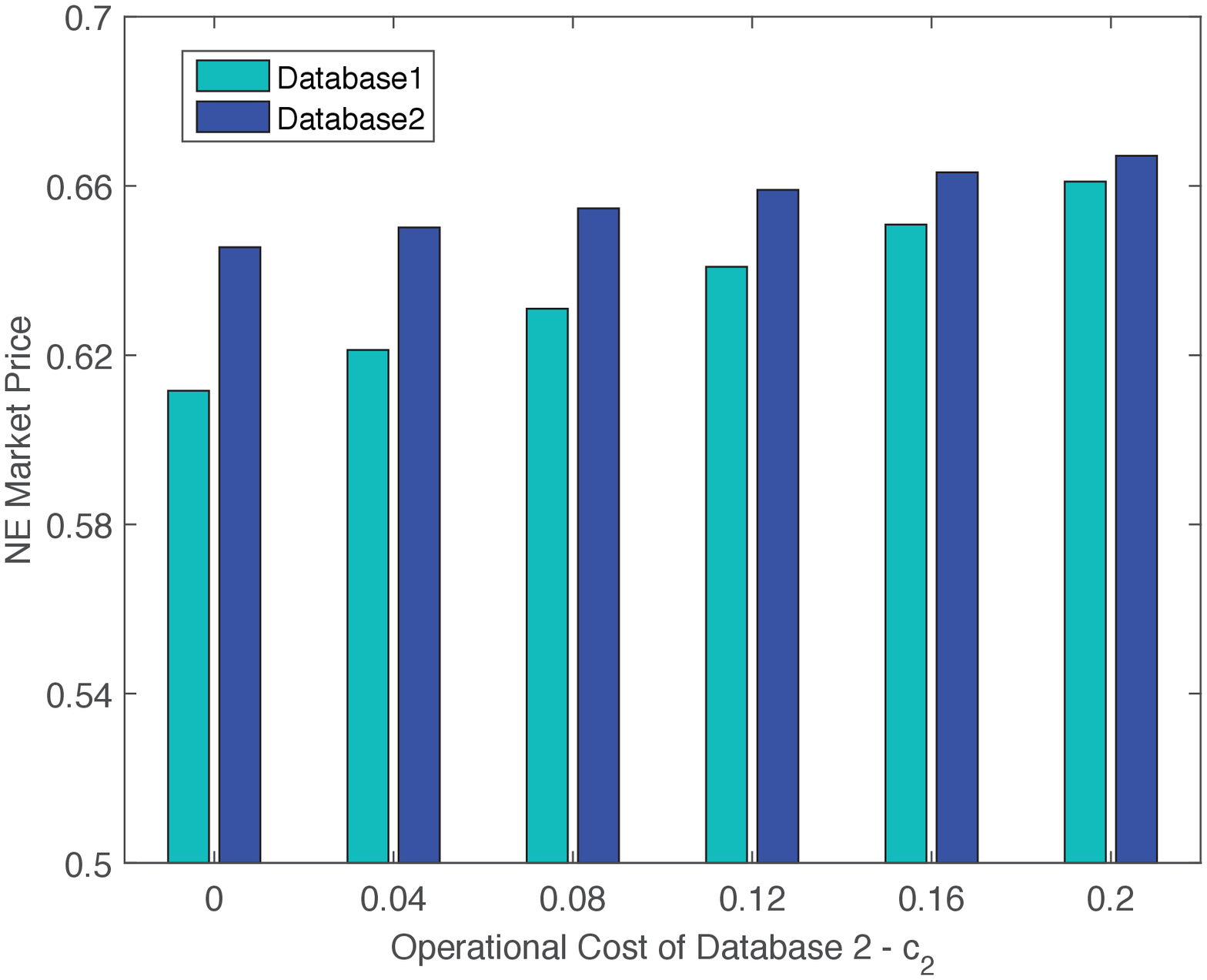}
  \includegraphics[width=2.2in]{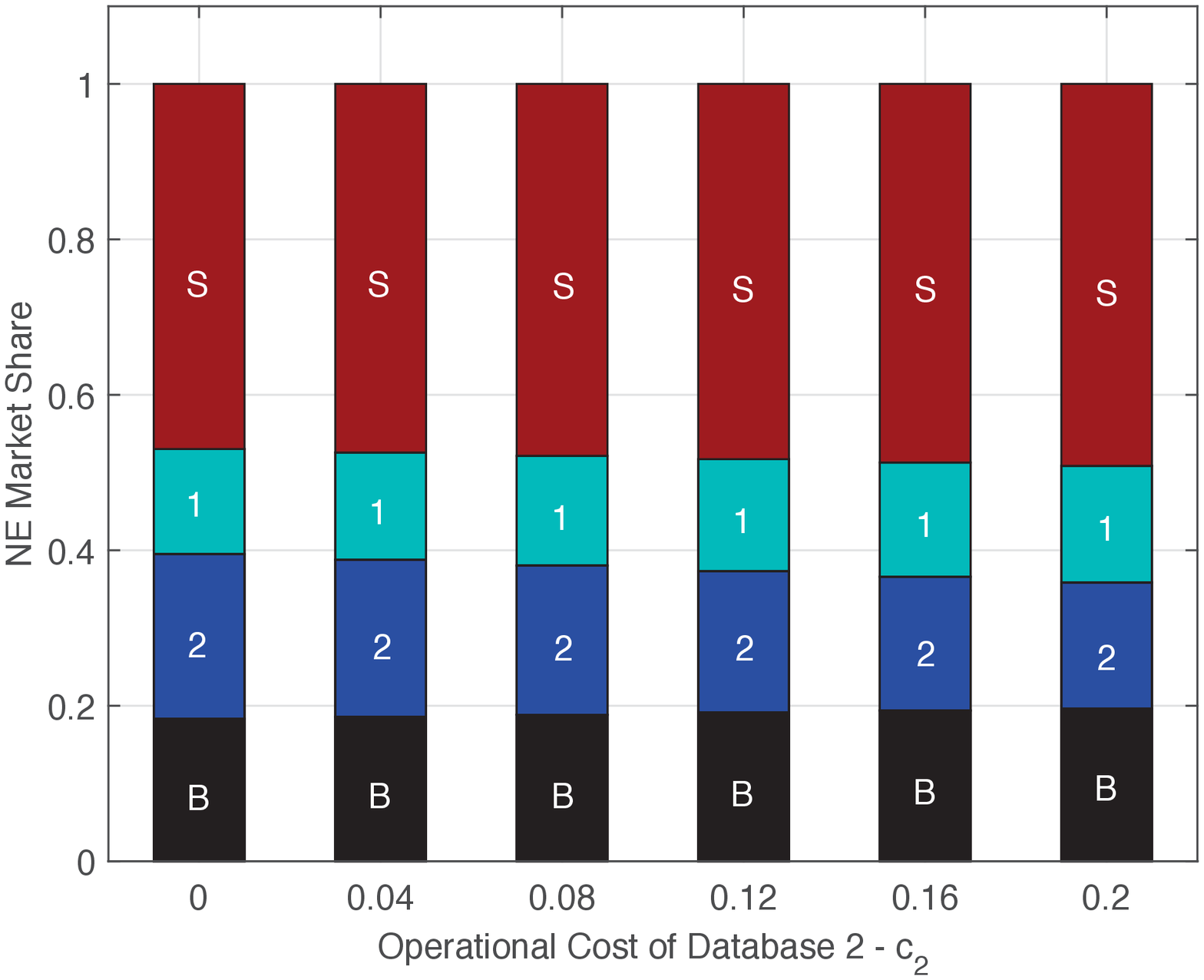}
    \includegraphics[width=2.33in]{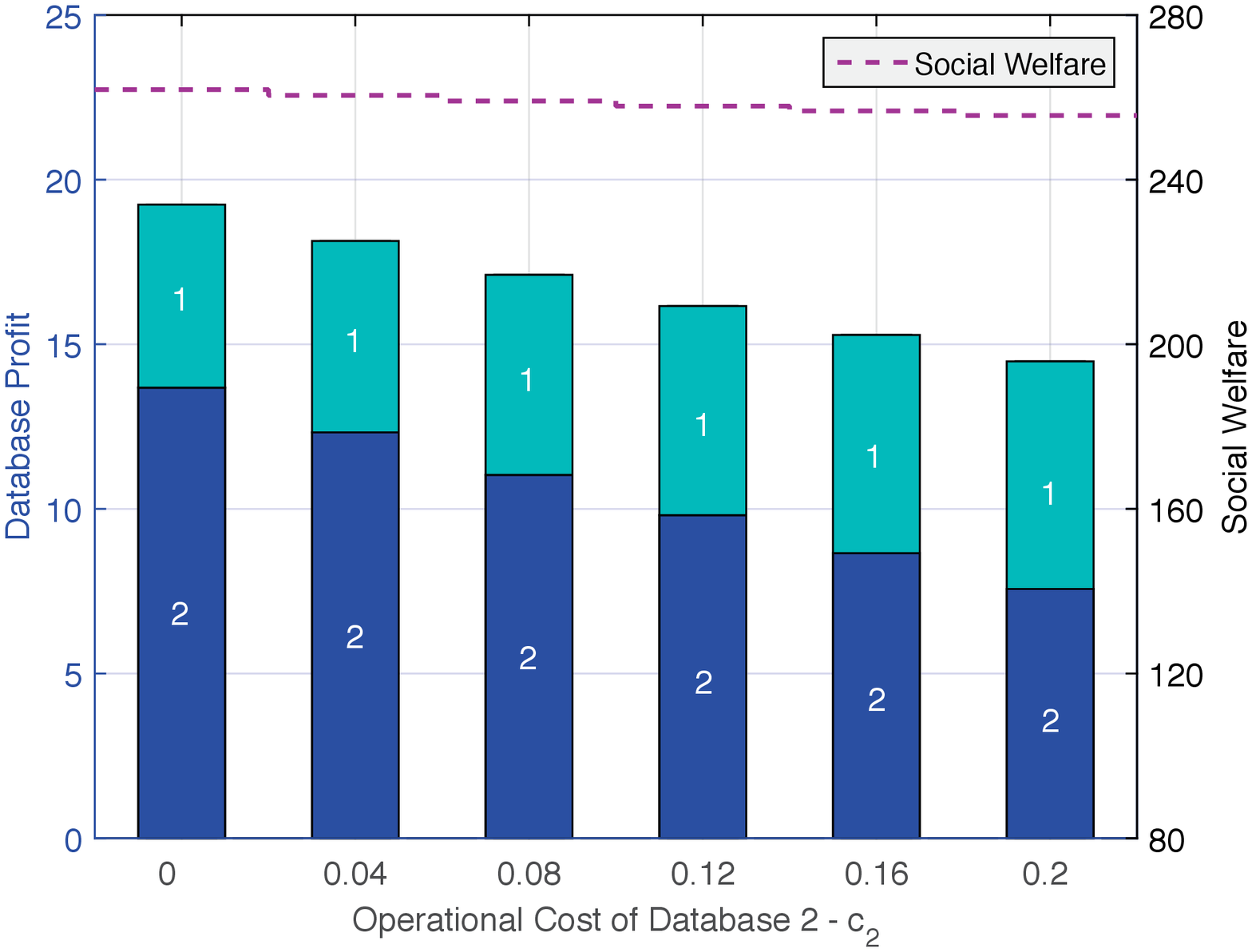}
\vspace{-3mm}
  \caption{(a) Price equilibrium, (b) Market share equilibrium, and (c) the system performance vs the operational cost of database $2$.}\label{fig:impact_operation_cost_v2}
  \vspace{-3mm}
\end{figure*}

We further consider the scenario of different operational costs for databases. To illustrate the results clearly, we consider a simple scenario of two databases with different operational costs. Specifically, we fix the operational cost of database 1 as $c_1 = 0.2$, while change the operational cost $c_2$ of database 2 from 0 to 0.2. The other parameters are same as those in the previous simulation.
Figure \ref{fig:impact_operation_cost_v2} illustrates (a) the price equilibrium, (b) the market share equilibrium, and
(c) the system performance, achieved under different operational costs of database 2.
%In this simulation, we fix the network externality impact $\gamma_1 =\gamma_2 = \gamma = 0.4$, the number of database $\M = 2$, the sensing cost $\ps = 2$, the operational cost of the database 1 $\c_1 = 0.2$.
%The databases' initial market shares satisfy $\Prob_2 > \Prob_1$. We denote sensing service as S, basic service as B, and database $\m \in \Mset$ as $\m$.
%while changing the sensing cost $\ps$ from $\ps=1.2$ to $\ps=2.8$. We assume that $\gamma_1 = \gamma_2 = \gamma_3$ in this simulation.

Figure \ref{fig:impact_operation_cost_v2}.a shows the price equilibrium under different operational costs of database 2.
We can see with the increasing of operational cost $c_2$, database 2 will select a higher equilibrium market price in order to cover its operational cost; accordingly,
%This is because a higher operational cost will make the database set a higher retail price to cover his operational cost.
database 1 can also set a higher  equilibrium market price to gain more profit, even thought its own operational cost remains unchanged.
%This is because the increase of database $2$ drive the database $1$ to increase his own price.
We can also see that the price difference between two databases decreases with $\c_2$, due to the decrease of their operation cost difference.
%Notice that database $1$ has a larger initial market share than database $2$.
%When the database $2$'s operational cost is small (e.g., $\c_2 = 0$), the database $1$ has a larger advantage over the database $2$ as the database $1$ has larger initial market share and small operational cost. This advantage allows the database $1$ to set a much higher retail price than than that of the database $2$.
%With the increase of $\c_2$, the advantage of database $1$ is not so obvious, and thus leads to the small NE market price difference between the two databases.
%At the extreme case, when both of the databases have the same operational cost (i.e., when $\c_2 = 0.2$), the database $1$ can set a slight higher retail price than that of the database $2$ due to database $1$ has larger initial market share.
%%as the operational cost between the database $1$ and $2$
%%with the increase of the database $2$ operational cost, the competition among the databases becomes less intensive, and thus allows the other databases increases their price.

{
Figure \ref{fig:impact_operation_cost_v2}.b shows the equilibrium market shares achieved under different operational costs of database 2.
Each bar denotes the market share allocation among the basic service (denoted as ``B"), database $m$'s advanced services (denoted as ``$m$''), and sensing (denoted as ``S").
We have shown in Figure \ref{fig:impact_operation_cost_v2}.a that with the increase of  the database $2$'s operational cost, both databases will set higher equilibrium market prices.
This makes the advanced services of both databases becomes less attractive, and hence reduces the market share of databases and increases the the market shares of sensing service and basic service.
Moreover, the market share of database 2 decreases, while that of database 1 slightly increases. This is because an increased operational cost of database 2 reduces its competitiveness, hence drives some of its market share to database 1.
%\rev{As the equilibrium market prices of two databases get close, the market share difference between two databases becomes small. Hence, we can see that the market share of the database $1$ increases with the database $2$'s operational cost, while the market share of the database $2$ decreases with the database $2$'s operational cost. As the database $2$ has larger initial market share, even when the two databases have the same operational cost (i.e., when $\c_2 = 0.2$), the database $2$ still has larger market share.
%}
%Hence, the database $2$ market share decreases with $\c_2$.
%Meanwhile, the market share of the database $1$ slightly increases with $\c_2$. This is because some median $\th$ value \eus~are driven to the database $1$'s advanced service as the difference between the two databases becomes small.
%However, due to the increase of the operational cost of database, most \eus~are attracted by either sensing service to enjoy a good quality of service (those \eus~with high $\th$ value) or basic service to enjoy a free service (those \eus~with low $\th$ value). Hence, we can see the increases
%\com{Huang: How do the market shares of database 1 and database 2 change?}

Figure \ref{fig:impact_operation_cost_v2}.c shows the databases' profits and the total social welfare achieved at the market equilibrium, under different operational costs of database $2$.
Each bar denotes the aggregated profit of $2$ databases, while each sub-bar corresponds to the profit of database $\m$.
The dash red line denotes the value of social welfare. The left y-axis denotes the value of database's profit, and the right y-axis denotes the value of social welfare.
From Figure \ref{fig:impact_operation_cost_v2}.c, we can see that both the databases' aggregated profit and the social welfare decrease with the operational cost.
Moreover, the profit of database 2 decreases due to the reduction of its market share, while that of database 1 slightly increases due to the slight increase of its market share.

%\vspace{-2mm}
\section{Conclusion}\label{sec:conclusion}

%Today geo-location database-assisted TV white space network has been considered as one of the promising commercial applications of green cognitive communication technology.
%% as the database can reduce the energy consumption due to spectrum sensing, and provide channel information to cognitive radio users so that they can make the best choices to balance the energy consumption and communication quality.
%As the success of such a network replies on a proper business model,
In this paper,
we propose an information market model called MINE GOLD, which enables the geo-location databases to sell information regarding the white space to WSDs.
%%In this paper, we
%%study a novel information market for the TV white space networks, where the geo-location databases sell information regarding the white space to WSDs
We characterize the positive network externality in
%the information market
the proposed  information market model, and study the user subscription dynamics and the associated market equilibrium.
Based on this, we further examine the databases' pricing decision from a game-theoretic perspective.
We discover several interesting insights of the databases' competition game in the information market.
For example, there exists an optimal number of databases to achieve the maximum total database revenue.
Moreover, a larger positive network externality will have a more positive impact on the system performance, both in terms of the databases' revenues and the social welfare.

The information market proposed in this paper mainly concerns the utilization of unlicensed TV channels, where WSDs share with others. In practice, some licensees are willing to lease their under-utilized licensed spectrum for extra profit, and WSDs can have exclusive usage right by leasing such spectrum. Hence, a joint market design involving both unlicensed and licensed TV channels will be an important future research direction.
%\vspace{-0.5mm}

%\vspace{-2mm}
%\input{Section_Ref}
%!TEX root = main_pure_information_journal.tex
%SourceDoc main_pure_information_journal.tex

%\input{section-bios}

%
%\bibliographystyle{IEEEtran}
%\bibliography{ref-db}

%\appendix
%\section{Appendix}\label{sec:appendix}
%\input{Section_Appendix_Proofs_Pure_Information}
%!TEX root = main_pure_information_journal.tex
%SourceDoc main_pure_information_journal.tex

\newpage

%\noindent
%\textbf{Technical Report} for ``\emph{MINE GOLD to Deliver Green Cognitive Communications}''
%\\
%\textbf{Authors}: Yuan luo, Lin Gao, and Jianwei Huang

\appendix
\section{Appendix}\label{sec:appendix}

\subsection{Property of Information Market}\label{sec:information_market_appendix}
In this section, we will discuss the properties of positive network externality in the information market.
For illustration purpose, we first define the advanced information as  \emph{the interference level on each channel}, then we characterize the information value to the WSDs, based on which we can further characterize the properties of the information market.

\subsubsection{Interference Information}
\label{sec:interference_info}
For each {\eu} $n\in \Nset$ operating on the TV \ch, each channel $\k$ is associated with an \emph{interference level}, denoted by  $\InfTot_{\n,\k}$, which reflects the aggregate interference from all other nearby devices (including TV stations
and other {\eus}) operating on this channel.
Due to the fast changing of wireless channels and the uncertainty of {\eus}' mobilities and activities, the interference $\InfTot_{\n,\k}$ is a random variable. We impose assumptions on the interference $\InfTot_{\n,\k}$ as follows.
\begin{assumption}\label{assum:iid}
For each {\eu} $n\in \Nset$, each channel $\k$'s interference level $\InfTot_{\n,\k}$ is \emph{temporal-independence} and \emph{frequency-independence}.
\end{assumption}

This assumption shows that
(i) the interference $\InfTot_{\n,\k}$ on channel $\k$ is independent identically distributed (iid) at different times, and (ii)
the interferences on different channels, $\InfTot_{\n,\k}, \k\in\Kset$, are also iid at the same time.\footnote{{Note that the iid assumption is a reasonable approximation of the practical scenario. This is because {\eus} with basic service will randomly choose one TV \ch, hence the number of such {\eus} per channel will follow the same distribution. For {\eus} with advanced service, they will go to the TV \ch~with the minimum realized interference. If the interference among each pair of users is iid over time, then statistically the number of such users in each channel will also follow the same distribution. Note that even though all channel quality distributions are the same, the realized instant qualities of different channels are different. Hence, the advanced information provided by the \db~is still valuable as such an advanced information is accurate interference information.} }
As we are talking about a general \eu~$n$, \textbf{we will omit the {\eu} index $n$ in the notations (e.g., write $\InfTot_{\n,\k}$ as $\InfTot_{\k}$), whenever there is no confusion caused.}
Let $H_{\InfTot}(\cdot)$ and $h_{\InfTot}(\cdot)$ denote the cumulative distribution function (CDF) and probability distribution function (PDF) of $\InfTot_{\k}$, $\forall \k\in\Kset$.\footnote{In this paper, we will conventionally  use $H_X(\cdot)$ and $h_X(\cdot)$ to denote the CDF and PDF of a random variable $X$, respectively.}~~~~

Usually, a particular {\eu}'s experienced interference $\InfTot_{\k}$ on a \ch~$\k$ consistss of the following three components:
\begin{enumerate}
\item
$\InfTV_{\k}$: the interference from licensed TV stations;
\item
$\InfEU_{\k,j}$: the interference from another {\eu} $j$ operating on the same channel $k$;
\item
$\InfOut_{\k}$: any other interference from outside systems.
\end{enumerate}
The total interference on channel $k$ is
$
\InfTot_{\k} = \InfTV_k + \InfEU_k + \InfOut_k
$, where  $\InfEU_k  \triangleq \sum_{j \in \Nkset} \InfEU_{\k,j}$ is the total interference from all other {\eus} operating on channel $k$ (denoted by $\Nkset$).
Similar to $\InfTot_{\k}$, we also assume that  $ \InfTV_k , \InfEU_k , \InfEU_{\k,j}$, and $ \InfOut_k$ are random variables with \emph{temporal-independence} (i.e., iid across time) and \emph{frequency-independence} (i.e., iid across frequency).
We further assume that $\InfEU_{\k,j}$ is \emph{user-independence}, i.e., $\InfEU_{\k,j}, j \in \Nkset$, are iid.
{It is important to note that \textbf{different {\eus} may experience different interferences $\InfTV_k$ (from TV stations), $ \InfEU_{\k,j} $ (from another \eu~operating on the same \ch), and $ \InfOut_k$ (from outside systems) on a channel $k$, as we have omitted the \eu~index $n$ for all these notations for clarity.}}

Next we discuss these interferences in more details.

\begin{itemize}
\item
Each \db~is able to compute the interference $\InfTV_k$ from TV stations to every {\eu} (on channel $k$), as it knows the locations and channel occupancies of all TV stations.~~~~

\item
Each \db~cannot compute the interference
$\InfOut_k$ from outside systems, due to the lack of outside interference source information.
Thus, the interference $\InfOut_{\k}$ will \emph{not} be included in a database's advanced information sold to {\eus}.~~~~~~~~

\item
Each \db~may or may not be able to compute the interference $\InfEU_{\k,j}$ from another {\eu} $j$, depending on whether {\eu} $j$ subscribes to the \db's advanced service.
Specifically, if {\eu} $j$ subscribes to the advanced service, the \db~can predict its channel selection (since the {\eu} is fully rational and will always choose the channel with the lowest interference level indicated by the \db~at the time of subscription), and thus can compute its interference to any other \eu.
However, if {\eu} $j$ only chooses the \db's basic service, the \db~cannot predict its channel selection, and thus cannot compute its interference to other \eus.~~~~
\end{itemize}

For convenience, we denote $\Nkset^{[m]},\m\in \Mset$, as the set of {\eus} operating on channel $k$ and subscribing to the database $\m$'s advance service (i.e., those choosing the strategy $\s = \m$), $\Nkset^{[\B]}$ as the set of {\eus} operating on channel $k$ and choosing the \db's basic service (i.e., those choosing the strategy $\s = \B$), and $\Nkset^{[\S]}$ as the set of {\eus} operating on channel $k$ and sensing all the available {\chs} (i.e., those choosing the strategy $\s = \S$).
That is, $\bigcup_{\m \in \Mset}\Nkset^{[\m]} \bigcup \Nkset^{[\B]} \bigcup \Nkset^{[\S]}  = \Nkset $.
Then, for a particular WSD, its experienced interference (on channel $k$) \textbf{known by database $\m$} is
\begin{equation}\label{eq:known_inf}
\begin{aligned}
\textstyle
\InfKnown^{\m}_{\k} \triangleq \InfTV_{\k} + \sum_{j \in \Nkset^{[\m]}} \InfEU_{\k,j},
\end{aligned}
\end{equation}
which contains the interference from TV licensees and all {\eus} (operating on channel $k$)  subscribing to the \db $\m$'s advanced service.
The \eu's experienced interference (on channel $k$) \textbf{\emph{not} known by the {\db} $\m$} is
% (and thus will not be included in the database $\dbb_i$'s advanced information) is
\begin{equation}\label{eq:unknown_inf}
\begin{aligned}
\textstyle
\InfUnknown^{\m}_{\k}  \triangleq & \InfOut_{\k} + \sum_{j \in \Nkset^{[\B]}} \InfEU_{\k,j} \\
& + \sum_{j \in \Nkset^{[\S]}} \InfEU_{\k,j} + \sum_{j \in \Nkset^{[i]}, i \neq m} \InfEU_{\k,j},
%\InfUnknown^{\m}_{\k}  \triangleq  \InfOut_{\k} + \sum_{j \in \Nkset^{[\B]}} \InfEU_{\k,j}
% + \sum_{j \in \Nkset^{[\S]}} \InfEU_{\k,j} + \sum_{j \in \Nkset^{[i]}, i \neq m} \InfEU_{\k,j},
\end{aligned}
\end{equation}
which contains the interference from outside systems and all {\eus} (operating on channel $k$) not subscribing to the database $\m$'s advanced service.
Obviously, both $\InfUnknown^{\m}_{\k}$ and $\InfKnown^{\m}_{\k}$ are also random variables with temporal- and frequency-independence.
Accordingly, the total interference on {\ch} $\k$ for a {\eu} can be written as
$\textstyle
\InfTot_{\k} = \InfKnown^{\m}_{\k} + \InfUnknown^{\m}_{\k}.$~~~~~~

\textbf{Since the \db~$\m$ knows only $\InfKnown^{\m}_{\k}$, it will provide this information (instead of the total interference $\InfTot_k$) as the {advanced service} to a subscribing {\eu}.}
It is easy to see that the more {\eus} subscribing to the {\db} $\m$'s advanced service, the more information the {\db} $\m$ knows, and the more accurate the {\db} $\m$'s information will be.

Next we can characterize the accuracy of a database's information explicitly. Note that $\Probam$ and $\Probs$ denote the fraction of {\eus} choosing the database $\m$'s advanced service and sensing service, respectively. Moreover, $(1 - \sum_{\m \in \Mset}\Probam - \Probs)$ denotes the fraction of {\eus} choosing the basic service.
%Hence, there are $(1 - \Probl) \cdot \N$ {\eus} in the network that we consider operating on the TV \chs.
Due to the Assumption \ref{assum:iid}, it is reasonable to assume that each channel $k\in\Kset$ will be occupied by an average of $\frac{\N}{\K}$ {\eus}.
%Let $\Prob_l$ denote the percentage of {\eus} subscribing to the advanced service of database $\dbb_l$ (called the \emph{market share} of database $\dbb_l$).
Then, among all $\frac{\N}{\K}$ {\eus} operating on channel $k$, there are, \emph{on average}, $\frac{\N}{\K}\cdot\Probam$ {\eus}  subscribing to the {\db} $\m$'s advanced service, $\frac{\N}{\K}\cdot\Probs$ {\eus}  choosing sensing service, and $\frac{\N}{\K}\cdot ( 1 - \sum_{\m\in\Mset}\Probam - \Probs )$ {\eus} choosing the  basic service.
That is, $| \Nkset | = \frac{\N}{\K} $, $| \Nkset^{[\m]} | = \frac{\N}{\K}\cdot\Probam$, $| \Nkset^{[\S]} | = \frac{\N}{\K}\cdot\Probs$, and $|\Nkset^{[\B]} | = \frac{\N}{\K}\cdot(1 - \sum_{\m \in \Mset}\Probam - \Probs)$.\footnote{{Note that the above discussion is from the aspect of expectation, and in a particular time period, the realized numbers of {\eus} in different channels may be different.}}
Finally, by the {user-independence} of $\InfEU_{\k,m}$, we can immediately calculate the distributions of $ \InfKnown^{\m}_{\k}$ and $ \InfUnknown^{\m}_{\k}$ under any given market share $\Probam$ $\m \in \Mset$ via (\ref{eq:known_inf}) and (\ref{eq:unknown_inf}).

\subsubsection{Information Value}
Now we evaluate the value of the \db~$\m$'s advanced information to {\eus}, which is
reflected by the {\eu}'s benefit (utility) that can be achieved from this information.

We first consider the expected  utility of a {\eu} when choosing the \db's basic service (i.e., $\s=\B$).
In this case, the {\eu} will randomly choosing a \tvch~based on the information provided in the free basic service,
and its expected data rate is
\begin{equation}\label{eq:rate-random-fixed}
\begin{aligned}
\textstyle
\R_{[\B]}  \textstyle = \Ex_{Z} [\rt(\InfTot)] = \int_{z} \rt(z) \mathrm{d} H_{\InfTot}(z), \\
%\RB   \textstyle \triangleq \fx = \ut(\R_0) = \ut\left(\int_{z} \rt(z) \mathrm{d} F_{\InfTot}(z)\right),
\end{aligned}
\end{equation}
%where $\R_0$ is the expected data rate and is given as follows.
%\begin{equation}\label{eq:rkgs}
%\begin{aligned}
%& \R_0  \textstyle = \Ex_{Z} [\rt(\InfTot)] = \int_{z} \rt(z) \mathrm{d} F_{\InfTot}(z),
%\end{aligned}
%\end{equation}
where $\rt(\cdot)$ is the transmission rate function (e.g.,
the Shannon capacity) under any given interference.
As shown in Section \ref{sec:interference_info}, each channel $k\in\Kset$ will be occupied by an average of $\frac{\N}{\K}$ {\eus} based on the Assumption \ref{assum:iid}. Hence,
$\R_{[\B]}  $ depends only on the distribution of the total interference $\InfTot_k$, while not on the specific distributions of $\InfKnown^{\m}_{\k}$ and $\InfUnknown^{\m}_{\k}$.
%and thus depends on the fraction of {\eus} operating on TV \chs~(i.e., $1 - \Probl$).
Then the expected utility provided by the basic service is:
\begin{equation}\label{eq:utility-random-fixed}
\begin{aligned}
\textstyle
\RB = \textstyle \ut\big(\R_{[\B]} \big),
% = \ut\left(\int_{z} \rt(z) \mathrm{d} F_{\InfTot}(z)\right),
\end{aligned}
\end{equation}
where $\ut(\cdot)$ is the utility function of the \eu.
We can easily check that the accuracy of the database $\m$'s information does not affect the utilities of theses WSDs not subscribing to the database $\m$'s advanced service.

Then we consider the expected  utility of a {\eu} when choosing the sensing service (i.e., $\s=\S$).
In this case, the {\eu} will sense all the available channels and select one with the lowest interference level. Hence, its expected data rate is
\begin{equation}\label{eq:rate-random-fixed}
\begin{aligned}
\textstyle
\R_{[\S]}  \textstyle = \Ex_{Z_{(1)}} [\rt(\InfTot)] = \int_{z} \rt(z) \mathrm{d} H_{\InfTot_{(1)}}(z), \\
%\RB   \textstyle \triangleq \fx = \ut(\R_0) = \ut\left(\int_{z} \rt(z) \mathrm{d} F_{\InfTot}(z)\right),
\end{aligned}
\end{equation}
%where $\R_0$ is the expected data rate and is given as follows.
%\begin{equation}\label{eq:rkgs}
%\begin{aligned}
%& \R_0  \textstyle = \Ex_{Z} [\rt(\InfTot)] = \int_{z} \rt(z) \mathrm{d} F_{\InfTot}(z),
%\end{aligned}
%\end{equation}
where $Z_{(1)} \triangleq \min\{Z_1,...,Z_K\}$ denotes the minimum interference on all channels, $H_{\InfTotMin}(z) = [1 - H_{\InfTot}(z)]^K$ is the CDF of $Z_{(1)}$, and $\rt(\cdot)$ is the transmission rate function (e.g., the Shannon capacity).
We can check that $\R_{[\S]} $ depends only on the distribution of the total interference $\InfTot_k$, while not on the specific distributions of $\InfKnown^{\m}_{\k}$ and $\InfUnknown^{\m}_{\k}$.
%and thus depends on the fraction of {\eus} operating on TV \chs~(i.e., $1 - \Probl$).
Then the expected utility provided by the sensing service is:
\begin{equation}\label{eq:utility-random-fixed}
\begin{aligned}
\textstyle
\RS= \textstyle \ut\big(\R_{[\S]}\big),
% = \ut\left(\int_{z} \rt(z) \mathrm{d} F_{\InfTot}(z)\right),
\end{aligned}
\end{equation}
where $\ut(\cdot)$ is the utility function of the \eu.
We can easily check that the accuracy of the database $\m$'s information does not affect the utilities of theses WSDs not subscribing to the database $\m$'s advanced service.
We, therefore, have the Assumption \ref{assume:random}.

%We can easily check that
%the more {\eus} operating on the TV \chs, the higher value of $\InfTot_k$ is, and thus the lower expected utility provided by the basic service. Hence, the basic service's expected utility reflects the congestion level of the TV \chs.
%%We use the function $\fx(\cdot)$ to characterize the congestion effect and have $\fx(1 - \Probl) = \RB( 1 - \Probl)$.
%%while not on the specific distributions of $\InfKnown_k$ and $\InfUnknown_k$. This implies that the accuracy of the \db's information does not affect the utilities of those {\eus} choosing the basic service. However, as t

Then we consider the expected utility of a {\eu} when subscribing to the \db~$\m$'s advance service (i.e., $\s = \m, \m \in \Mset$).
In this case, the {\db} $\m$ returns the interference $\{\InfKnown^{\m}_{k}\}_{k\in\Kset}$ to the {\eu} subscribing to the advanced service, together with the basic information such as the available channel list.
For a rational {\eu}, it will always choose the channel with the minimum $\InfKnown^{\m}_{k}$ (since $\{\InfUnknown^{\m}_{k}\}_{k\in\Kset}$ are iid).
Let $\InfKnownMin^{[\m]} =  \min\{ \InfKnown^{\m}_{1},  \ldots, \InfKnown^{\m}_{K} \} $ denote the minimum interference indicated by the \db~$\m$'s advanced information.
Then, the actual interference experienced by a {\eu} (subscribing to the \db~$\m$'s advanced service) can be formulated as the sum of two random variables, denoted by $\InfTotA^{\m} = \InfKnownMin^{\m} + \InfUnknown^{\m}$. Accordingly, the {\eu}'s expected data rate under the strategy $\s= \m$ can be computed by
\begin{equation}\label{eq:rate-pay-fixed}
\begin{aligned}
\Ra^{\m}(\Probam) & \textstyle = \Ex_{\InfTotA^{\m}} \big[ \rt \left( \InfTotA^{\m} \right) \big] = \int_z \rt(z)  \mathrm{d} H_{\InfTotA^{\m}}(z),
%\\
%\RA & \textstyle  \triangleq \ut\left(\Ra\right) = \ut\left(\int_z \rt(z)  \mathrm{d} F_{\InfTotA}(z)	\right) ,
\end{aligned}
\end{equation}
where
$H_{\InfTotA^{\m}}(z)$ is the CDF of $\InfTotA^{\m}$. It is easy to see that $\Ra^{\m}$ depends on the distributions of $\InfKnown^{\m}_k$ and $\InfUnknown^{\m}_k$, and thus depend on the market share $\Probam$.
Thus, we will write $\Ra^{\m}$ as $\Ra^{\m}(\Probam)$. Accordingly, the advanced service's utility is:
 \begin{equation}\label{eq:utility-pay-fixed}
\begin{aligned}
%\Ra & \textstyle = \Ex_{\InfTotA} \big[ \rt \left( \InfTotA \right) \big] = \int_z \rt(z)  \mathrm{d} F_{\InfTotA}(z),
%\\
\RA^{\m}(\Probam) & \textstyle  \triangleq \ut\bigg( \Ra^{\m}( \Probam ) \bigg)
%= \ut\left(\int_z \rt(z)  \mathrm{d} F_{\InfTotA}(z)	\right) ,
\end{aligned}
\end{equation}

%Note that the congestion effect also affects the value of $\RA$. However, compared with the utility of {\eu} choosing basic service, the benefit of a {\eu} subscribing to the \db's advanced information is coming from the $\InfKnownMin^{[l]}$, i.e., the minimum interference indicated by the \db's advanced information. As the value of $\InfKnownMin^{[l]}$ depends on the $\Proba$ only, we can get the approximation $\RA = \fx(1 - \Probl) + \gy(\Proba)$, where function $\gy(\cdot)$ characterize the benefit brought by $\InfKnownMin^{[l]}$ and denotes the positive network effect.

By further checking the properties of $\RB$, $\RS$, and $\RA^{\m}(\Probam)$, we have the Assumption \ref{assume:advance}-\ref{assum:positive}.

\subsection{Proof for Lemma \ref{lemma:market-share}}

\begin{proof}

By solving (\ref{eq:utility_function}), we can get  three thresholds, denoted by
$$
\textstyle
\thsb^{\t} \eq \frac{ \ps}{ \RS- \RB },
~~~~
\thab^{\t} \eq \frac{ \po}{ \RAo(\Probao^{\t})- \RB },
~~~~
\thsa^{\t} \eq \frac{\ps-\po}{\RS - \RAo(\Probao^{\t})}.
$$
Consider two cases: (i) $\thsb^{\t} > \thab^{\t}$, and (ii) $\thsb^{\t} \leq \thab^{\t}$.

(i) When  $\thsb^{\t} > \thab^{\t}$, it is easy to check that
$$
\thsa^{\t} - \thsb^{\t} = \frac{\ps \cdot \big( \RAo(\Probao^{\t})-\RB \big) - \po\cdot (\RS-\RB)}{\big(\RS-\RAo(\Probao^{\t})\big) \cdot (\RS-\RB)} > 0 ,
$$
since $\RS > \RAo\big(\Probao^{\t}\big) $, $\RS > \RB $, and
$\ps\cdot(\RAo(\Probao^{\t})-\RB) > \po\cdot(\RS-\RB) $ as $\thsb^{\t} > \thab^{\t}$.
Hence, we have:
$\thsa^{\t} > \thsb^{\t} > \thab^{\t} $.
%Moreover, when  $\thsb^{\t} > \thab^{\t}$,
%(i)
%the best strategy of users with $\th \in [0, \thab)$ is to choose the basic service,
%(ii) the best strategy of users with  $\th \in (\thab, \thla)$ is
%to choose the advanced service,
Accordingly, the newly derived market share is:
$$\Probao^{\t} = \thsa^{\t} - \thab^{\t},$$
%and (iii) the best strategy of users with  $\th \in (\thla, 1]$ is
%to choose  the leasing service.

(ii) When $\thsb^{\t} \leq \thab^{\t}$, we can similarly check that
$\thsa^{\t} - \thsb^{\t} < 0$, and hence $\thsa^{\t} < \thsb^{\t} < \thab^{\t}.$
 %when  $\thlb < \thab$, (i)
% the best strategy of users with $\th \in [0, \thlb)$
% is to choose the basic service,
%(ii)  the best strategy of users with $\th \in (\thlb, 1]$
% is to choose the leasing service,
% and (iii) no user is willing to choose the advanced service.
Accordingly, the newly derived market share $\Probao^{\t}$ is
$$\Probao^{\t} = 0.$$

Formally, based on the above (i) and (ii),  we can get the result in (\ref{eq:user-prob-1}).
\end{proof}

\subsection{Proof for Proposition \ref{lemma:existence-eq_pt_mon}}\label{lemma:existence-eq_pt_mon-proof}
\begin{proof}
By Definition \ref{def:stable-pt}, $\Probao^{*}$ is an equilibrium point, if and only if it is a solution of (\ref{eq:equilibriu_pt_mono}).
We consider two cases: (i) $\thsb > \thab$ and (ii)
$\thsb \leq \thab$.

(i) If $\thsb > \thab$, the solution $\Probao$ should satisfy
\begin{equation}\label{eq:user-profit-new1}
\begin{aligned}
&\triangle \Probao (\Probao)  =  \frac{\ps-\po}{\RS - \RAo(\Probao)} - \frac{ \po}{ \RAo(\Probao^{\t})- \RB } - \Probao = 0.
\end{aligned}
\end{equation}
It is easy to check that $\left.\triangle \Probao (\Probao) \right|_{\Probao= 0}  > 0$ and $\left.\triangle\Probao (\Probao) \right|_{\Probao = 1}  < 0$.
As $\triangle \Probao (\Probao)$ is continuous on $[0,1]$, we have: $\triangle \Probao (\Probao) = 0$ has at least one root on  $[0,1]$.

(ii) If $\thsb \leq \thab$, the solution $\Probao^{*} = 0$  directly.
%should satisfy that:
%\begin{equation}\label{eq:user-profit-new21}
%\begin{aligned}
%%\textstyle
%&\triangle \Probao (\Probao)  =  1 - \frac{ \po}{ \RAo(\Probao^{\t})- \RB } - \Probao = 0;
%\end{aligned}
%\end{equation}
%Using the similar method, we can check that $\left.\triangle \Probao (\Probao) \right|_{\Probao= 0}  > 0$ and $\left.\triangle\Probao (\Probao) \right|_{\Probao = 1}  < 0$. Hence, we can get the conclusion that $\triangle \Probao (\Probao) = 0$ has a root on the domain of $\Probao$, given $\triangle \Probao (\Probao)$ is continuous on $[0,1]$.
\end{proof}

\subsection{Proof for Proposition \ref{lemma:uniqueness-eq_pt_mon}}\label{lemma:uniqueness-eq_pt_mon-proof}
\begin{proof}
To prove the uniqueness, we   need to show that the function $\triangle \Probao (\Probao)$ in (\ref{eq:user-profit-new1}) is strictly decreasing on $[0,1]$. First, we can easily check that if
\begin{equation}\label{eq:user-uniquenee-cod}
\begin{aligned}
\gy^{\prime}(\Probao) \cdot \left[ \frac{ \ps - \po }{ ( \RS -  \RAo(\Probao) )^2} + \frac{ \po }{ ( \RAo(\Probao) - \RB )^2 } \right]   <  1,
\end{aligned}
\end{equation}
then the first derivative of   $\triangle \Probao (\Probao)$ is negative, hence $\triangle \Probao (\Probao)$ in (\ref{eq:user-profit-new1}) is strictly decreasing on $[0,1]$.

We further notice that   $\frac{ \ps - \po }{ ( \RS -  \RAo(\Probao)  )} \leq 1$ and $\frac{ \po }{ \RAo (\Probao)  - \RB} \leq 1$, and
\begin{equation}\label{eq:user-uniquenee-cod-simple}
\begin{aligned}
\textstyle
& \gy^{\prime}(\Probao) \cdot \left[ \frac{ \ps - \po }{ ( \RS - \RAo(\Probao) )^2} + \frac{ \po }{ ( \RAo(\Probao) - \RB )^2  } \right] \\
%& \leq \max{ ( \frac{ \pl - \pa }{ ( L - \fx(\Probl) - \gy(\Proba) )} , \frac{ \pa }{ \gy^2(\Proba) } ) \cdot \gy^{'}(\Proba) \cdot [ \frac{1}{ L - \fx(\Probl) - \gy(\Proba)} + \frac{1}{ \gy(\Proba) } ] }
& \leq \max \bigg\{ \frac{ \ps- \po }{ ( \RS -  \RAo(\Probao) )} , \frac{ \po }{( \RAo(\Probao) - \RB ) } \bigg\} \cdot \frac{ \RAo^{\prime}(\Probao) }{ \RAo(\Probao) - \RB   } \cdot \frac{ \RS - \RB }{ \RS - \RAo(\Probao) }  .
\end{aligned}
\end{equation}
%\begin{equation}\label{eq:user-uniquenee-cod-simple}
%	\begin{aligned}
%		\textstyle
%		& \gy^{\prime}(\Probao) \cdot \left[ \frac{ \ps - \po }{ ( \RS - \RAo(\Probao) )^2} + \frac{ \po }{ ( \RAo(\Probao) - \RB )^2  } \right] \\
%		%& \leq \max{ ( \frac{ \pl - \pa }{ ( L - \fx(\Probl) - \gy(\Proba) )} , \frac{ \pa }{ \gy^2(\Proba) } ) \cdot \gy^{'}(\Proba) \cdot [ \frac{1}{ L - \fx(\Probl) - \gy(\Proba)} + \frac{1}{ \gy(\Proba) } ] }
%		& \leq \max \bigg\{ \frac{ \ps- \po }{ ( \RS -  \RAo(\Probao) )} , \frac{ \po }{( \RAo(\Probao) - \RB ) } \bigg\} \\
%		& \quad{} \cdot \frac{ \RAo^{\prime}(\Probao) }{ \RAo(\Probao) - \RB   } \cdot \frac{ \RS - \RB }{ \RS - \RAo(\Probao) }  .
%	\end{aligned}
%\end{equation}
Hence,
%$\triangle \Probao (\Probao)$ in (\ref{eq:user-profit-new1})  is strictly decreasing on $[0,1]$ when
the condition in (\ref{eq:user-uniquenee-cod}) is satisfied when
\begin{equation}\label{eq:user-uniquenee-cod-simple2}
\max_{\Probao \in [0,1]} \frac{ \RAo^{\prime}(\Probao) }{ \RAo(\Probao) - \RB } \cdot \frac{  \RS - \RB }{ \RS - \RAo(\Probao) }  \leq 1,
\end{equation}
%where $\kappa_2 = \frac{1}{\max_{\Probao \in [0,1]} \thsa(\Probao)}$.
%\com{???}
\end{proof}

\subsection{Proof for Theorem \ref{thrm:stable-eq_pt-mon}}\label{thrm:stable-eq_pt-mon-proof}
\begin{proof}
 By Propositions  \ref{lemma:existence-eq_pt_mon} and   \ref{lemma:uniqueness-eq_pt_mon}, together with the non-decreasing of $\gy(\cdot)$, we can get the conclusion.
\end{proof}

\subsection{Proof for Proposition \ref{lemma:game_tranform_mono}}\label{lemma:game_tranform_mono-proof}
\begin{proof}
By Proposition \ref{lemma:uniqueness-eq_pt_mon}, there exist an one-to-one mapping between the information price and market share.
Hence,
we can get the conclusion immediately.
\end{proof}

\subsection{Proof for Proposition \ref{lemma:db-profit-mono}}\label{lemma:db-profit-mono-proof}
\begin{proof}
The proof for the existence is straightforward. Next we show there exists an unique optimal price under both the low sensing cost region and high sensing cost region.

We first look at the \db's revenue under high sensing cost region:
$\Udbo(\Probao)= ( \po - \c_1 ) \cdot \Probao(\po) $, where $ \Probao(\po)$ is given by
$$
  1 - \frac{ \po }{ \RA(\Probao) - \RB } - \Probao = 0.
$$
The second order derivative of $\Udbo(\p)$ with respect to $\po$ is
$$
\frac{ \partial^2{\Udbo} }{ \partial{\po^2}   } = 2 \cdot \frac{ \mathrm{d} \Probao }{\mathrm{d} \po  } + ( \po - \c_1) \cdot \frac{ \mathrm{d^2} \Probao }{\mathrm{d} \po^2  }.
$$
It is easily to verify that
$$
\frac{ \mathrm{d^2} \Probao }{\mathrm{d} \po^2  }  =  - \po \cdot \frac{ [\RAo(\Probao) - \RB] \RAo'(\Probao) }{ [ \po \RAo'( \Probao ) - ( \RAo(\Probao) - \RB)^2 ]^4  } \leq 0 .
$$
where $\RAo'( \Probao ) = { \mathrm{d} \RAo(\Probao) }/{\mathrm{d} \po  }>0$.
%Note that $\p A^{'}( \Prob ) - (A(\Prob) - B)^2 < 0$ for $\Prob > \Probb$ and $A(\Prob) > B$. Thus, we have $\frac{ \partial^2{\Udb} }{ \partial{\p^2}   } \leq 0$.
Moreover,  $\Probao(\po)$ decreases with $\po$.
Therefore, we have $\frac{ \partial^2{\Udbo} }{ \partial{\po^2}   } < 0$,
and thus there exist a unique solution that maximizes $\Udbo(\po)$.

Similarly, we can prove that under the high price region, there exists an unique solution that maximizes $\Udbo(\po)$.
%, as long as $ 2 \RAo(\Probao) \leq \RS + \RB $.
%We further note that $\Probd < \Probb$, where $\Probb$ is the critical point given by $ \frac{ \c - \p }{ S - A(\Prob) } = 1 $. Hence, $ 2 A(\Prob) \leq S + B $ is always held.
\end{proof}

\subsection{Proof for Proposition \ref{lemma:market-share-oligopoly}}\label{lemma:market-share-oligopoly-proof}
\begin{proof}
Given market shares
$\{ \Probao^{\t},\Probaoo^{\t},\ldots, \ProbaM^{\t} \}$ in slot $\t$ with
 $\ProbaM^{\t} > \ProbaMM^{\t} > \ldots > \Probao^{\t}$, we have: $ \RS > \RA_{\M}(\ProbaM^{\t}) > \RA_{\M-1}(\ProbaMM^{\t}) > \ldots > \RA_{1}(\Probao^{\t})$.
Notice that no {\eu} is willing to choose a service with a lower QoS and a higher price.
Hence, we can focus on the non-trivial scenario with $\ps > \p_{\M} > \p_{\M-1} > \ldots > \p_{1} $, under which we can get the conclusion immediately by
solving (\ref{eq:eu-choice-random-oligopoly}) - (\ref{eq:eu-choice-db-oligopoly}).
\end{proof}

\subsection{Proof for Theorem \ref{thrm:stable-eq_pt-oligopoly}}\label{thrm:stable-eq_pt-oligopoly-proof}
\begin{proof}
By Definition \ref{def:stable-pt-oligopoly} , $\BProb^{*}$ is an equilibrium point, if and only if it it a solution of (\ref{eq:equilibrium_pt_set}).
Given market shares
$\{ \Probao^{\t},\Probaoo^{\t},\ldots, \ProbaM^{\t} \}$ in slot $\t$ with
 $\ProbaM^{\t} > \ProbaMM^{\t} > \ldots > \Probao^{\t}$, the solution $\BProb$ should satisfy
% \begin{equation}\label{eq:NE-pt-oligopoly-proof}
%\left\{
%\begin{aligned}
%&\frac{ \ps - \pM}{ \RS- \RAM(\ProbaM) }- \frac{\p_{\M} -\p_{\M-1}}{\RA_{\M}(\Proba_{\M}) - \RA_{\M-1}(\Proba_{\M - 1})} - \ProbaM = 0,
%\\
%&\frac{\p_{\m+1} -\p_{\m}}{\RA_{\m+1}(\Proba_{\m+1}) - \RA_{\m}(\Proba_{\m })} - \Probam \\
%& \quad{} - \frac{\p_{\m} -\p_{\m-1}}{\RA_{\m}(\Proba_{\m}) - \RA_{\m-1}(\Proba_{\m - 1})}  = 0,\ \forall \m = 2,\ldots, \M-1, \\
%%& \qquad{} \qquad{} \qquad{} \qquad{} - \Probam = 0,\quad \forall \m = 2,\ldots, \M-1,
%%\\
%& \frac{\p_{2} -\p_{1}}{\RA_{2}(\Proba_{2}) - \RA_{1}(\Proba_{1})}  - \frac{ \po}{ \RAo(\Probao)- \RB } - \Probao = 0.
%\end{aligned}
%\right.
%\end{equation}
\begin{equation}\label{eq:NE-pt-oligopoly-proof}
	\left\{
	\begin{aligned}
		&\frac{ \ps - \pM}{ \RS- \RAM(\ProbaM) }- \frac{\p_{\M} -\p_{\M-1}}{\RA_{\M}(\Proba_{\M}) - \RA_{\M-1}(\Proba_{\M - 1})} - \ProbaM = 0,
		\\
		&\frac{\p_{\m+1} -\p_{\m}}{\RA_{\m+1}(\Proba_{\m+1}) - \RA_{\m}(\Proba_{\m })}  - \frac{\p_{\m} -\p_{\m-1}}{\RA_{\m}(\Proba_{\m}) - \RA_{\m-1}(\Proba_{\m - 1})} - \Probam  = 0,\ \forall \m = 2,\ldots, \M-1, \\
		& \frac{\p_{2} -\p_{1}}{\RA_{2}(\Proba_{2}) - \RA_{1}(\Proba_{1})}  - \frac{ \po}{ \RAo(\Probao)- \RB } - \Probao = 0.
	\end{aligned}
	\right.
\end{equation}
By solving the above equation, we can get the conclusion immediately.
 \end{proof}

\subsection{Proof for Proposition \ref{prop:unique_NE_oligopoly}}\label{prop:unique_NE_oligopoly-proof}
%\com{How to prove it....using figures?}
\begin{proof}
We first notice that the best response dynamics must converge to a market share equilibrium, as the changing of each database's market share is monotonic. This is due  to the positive externality of information market, under which a database with an increased market share in a slot tends to get more market share in the future.

Next, we can easily find that given a particular market share set in a   slot,  the best response dynamics will evolve to a fixed newly derived market share set in the next  slot, and eventually converge  to a unique market share equilibrium.
\end{proof}

\subsection{Proof for Proposition \ref{lemma:game_tranform}}\label{lemma:game_tranform-proof}
\begin{proof}
If $\BProba^*$ is a Nash equilibrium of MSCG, then we have:
\begin{equation}\label{eq:db-share-dynamic-proof}
\Probam^* = \arg \max_{\Probam} \Urdbm(\Probam, \BProbam^{*})
\end{equation}
By (\ref{eq:price-market-share-oligopoly}), we further have:
\begin{equation}\label{eq:price-market-share-rs-proof}
\textstyle
 \pm   =  \sum_{\m = 1}^{\M+1} \left[ \left( 1 - \sum_{\n = \m}^{\M+1}\Proba_{\n} \right) \cdot (  \gy(\Proba_{\m}) - \gy(\Proba_{\m-1}) ) \right]
 \end{equation}
Hence, we can easily check that $\bp^*$ is a Nash equilibrium of PCG, where
\begin{equation}\label{eq:db-price-dynamic-proof}
  \pm^{*} = \arg \max_{\pm \geq 0}\ \Udbm(\pm , \bpm^{*}),~~\forall \m \in \Mset.
  \end{equation}
\end{proof}

\subsection{Proof for Lemma \ref{thrm:NE-existence-two}}\label{thrm:NE-existence-two-proof}
\begin{proof}

For convenience, we first give the formal definition of supermodular game  and some related important concepts \cite{topkis1998supermodular}. A real $n-$diensional set $\mathcal{V}$ is a \emph{sublattice} of $\mathbb{R}^{n}$ if for any two elements $a, b \in \mathcal{V}$, the component-wise minimum, $a \vee b$, and the component-wise maximum, $a \wedge b$, are also in $\mathcal{V}$. Particularly, a compact sublattice has a smallest and largest element. A function $f(x_1, \ldots, f_N)$ has increasing differences in $(x_i, x_j)$ for all $i \neq j$ if $f(x_1, \ldots, x^{1}_i, \ldots, x_N) - f(x_1, \ldots, x^{2}_i, \ldots, x_N)$ is increasing in $x_j$ for all $x^{2}_i - x^{1}_i > 0$.\footnote{If the function $f$ is twice differentiable, the property is equivalent to $\partial^2{f}/\partial{x_i}\partial{x_j}\geq 0$} The formal definition of a supermodular game is given below:
\begin{definition}[Supermodular Game \cite{topkis1998supermodular}]\label{def:supermodular}
A noncooperative game $( \mathcal{M}, \{S \}_{ m \in \mathcal{M} }, \{ U_m \}_{ m \in \mathcal{M} } )$ is called a supermodular game if the following conditions are all satisfied:
\begin{itemize}
  \item  The strategy set $ {S_m} $ is a nonempty and compact sublattice of real number.
  \item  The payoff function $U_m$ is supermodular in player $m$'s own strategy.\footnote{A function is always supermodular in a single variable.}
  \item  The payoff function $U_m$ has increasing differences in all sets of strategies.
\end{itemize}
\end{definition}

To prove the existence of equilibrium under duopoly scenario, we only need to prove that the MSCG is a supermodular game under duopoly scenario with respect to $\Proba_{1}$ and $-\Proba_{2}$.
Since the two databases with a single instrument $\BProb = ( \Proba_1, \Proba_2 )$ chosen from a compact set $[0,1]^2$, it suffices to show that both $\Urdb_1(\Proba_1 , -\Proba_2)$ and $\Urdb_2(\Proba_1 , -\Proba_2)$ have increasing difference in $( \Proba_1, \Proba_2 )$. By  (\ref{eq:price-market-share-oligopoly}) and (\ref{eq:db-profit-oligpoly-market-share}), we have:
\begin{equation}
\begin{aligned}
\frac{ \partial^2{ \Urdb_1(\Proba_1 , -\Proba_2) }}{ \partial{({\Proba_1})}\partial{-\Proba_2} } = \frac{\mathrm{d}{\RA_1}}{\mathrm{d}{\Proba_1}} \cdot \Proba_1 + (\RA_2 - \RB) \geq 0
\end{aligned}
\end{equation}
\begin{equation}
\begin{aligned}
\frac{ \partial^2{  \Urdb_2(\Proba_1 , -\Proba_2) }}{ \partial{({\Proba_1})}\partial{-\Proba_2} }  = \RA_1 - \RB \geq 0~~~~~~~~~~~~~~~~
\end{aligned}
\end{equation}
%where $\gy^{\prime}(\Proba)$ is the first-order derivative of $\gy(\cdot)$ with respect to $\Proba$.
%Note that the second term in the right-hand side of the above formula is independent of  $\p_j$ for all $j \neq m$.
%We further notice that  $\frac{\partial{\log{ \th_m ( \bp ) }}}{ \partial{\p_m}\partial{\p_j} }  \geq 0$.
%Thus, we have: $ \frac{ \partial^2{\log{ \Ur_m }(\bp) }}{ \partial{\p_m}\partial{\p_j} } \geq 0 $, i.e., the payoff function $\log{ \Ur_m }(\bp)$ has increasing differences in all sets of strategies.
Hence, we can conclude that  MSCG is a supermodular game with respect to $\Proba_1$ and $-\Proba_2$.
\end{proof}

\subsection{Proof for Proposition \ref{thrm:NE-existence}}\label{thrm:NE-existence-proof}
\begin{proof}

To prove the existence of equilibrium under oligopoly scenario,  we only need to prove that $\Urdbm(\Probam , \BProbam)$, $\forall \m \in \Mset$, is quasi-concave in $\Probam$ \cite{Fudenberg1991game}.

To prove  $\Urdbm(\Probam , \BProbam)$ is quasi-concave in $\Probam$, it is sufficient to show that $\frac{\partial {\Urdbm(\Probam , \BProbam)}}{\partial{\Probam} }$ changes the sign once.
We first notice that $\Probam$ is chosen from $[0, 1]$, and
%\begin{align}
%& \lim_{\Probam \rightarrow 0} \frac{\partial {\Urdbm(\Probam , \BProbam)}}{\partial{\Probam} }  >0 , \nonumber\\
%& \quad{} \mbox{~and~}
%\lim_{\Probam \rightarrow +1}\frac{\partial {\Urdbm(\Probam , \BProbam)}}{\partial{\Probam} } <0.
%\end{align}
\begin{align}
	& \lim_{\Probam \rightarrow 0} \frac{\partial {\Urdbm(\Probam , \BProbam)}}{\partial{\Probam} }  >0 , \mbox{~and~}
	\lim_{\Probam \rightarrow +1}\frac{\partial {\Urdbm(\Probam , \BProbam)}}{\partial{\Probam} } <0.
\end{align}
Then we consider the second order derivative of $\Urdbm(\Probam , \BProbam) $ with respect to $\Probam$. We have:
%\begin{equation}
%\begin{aligned}
%\frac{\partial^2 {\Urdbm (\Probam )}}{\partial{\Probam^2} } &= - (\gamma_{\m}^2 + 3 \gamma_{\m} + 1) \cdot \Probam \\
%& \qquad{} + \gamma_{\m} \cdot (\gamma_{\m}+1)\cdot \bigg(  1 - \sum_{h = \m+1}^{M} \Proba_{h} \bigg),
%\end{aligned}
%\end{equation}
\begin{equation}
	\begin{aligned}
		\frac{\partial^2 {\Urdbm (\Probam )}}{\partial{\Probam^2} } &= - (\gamma_{\m}^2 + 3 \gamma_{\m} + 1) \cdot \Probam  + \gamma_{\m} \cdot (\gamma_{\m}+1)\cdot \bigg(  1 - \sum_{h = \m+1}^{M} \Proba_{h} \bigg),
	\end{aligned}
\end{equation}
which is a linear function of $\Probam$.
We can further check that
%$\frac{\partial^2 {\Urdbm (\Probam )}}{\partial{\Probam^2} }$ is linear in $\Probam$, and
%\begin{align}
%& \lim_{\Probam \rightarrow 0}\frac{\partial^2 {\Urdbm(\Probam , \BProbam)}}{\partial{\Probam^2} }  >0 , \nonumber \\
%& \quad{} \mbox{~and~}
%\lim_{\Probam \rightarrow +1}\frac{\partial^2 {\Urdbm(\Probam , \BProbam)}}{\partial{\Probam^2} } <0 .
%\end{align}
\begin{align}
	\lim_{\Probam \rightarrow 0}\frac{\partial^2 {\Urdbm(\Probam , \BProbam)}}{\partial{\Probam^2} }  >0 , \nonumber \mbox{~and~}
	\lim_{\Probam \rightarrow +1}\frac{\partial^2 {\Urdbm(\Probam , \BProbam)}}{\partial{\Probam^2} } <0 .
\end{align}
%Hence, the value of ${\partial^2 {\Urdbm(\Probam , \BProbam)}}/{\partial{\Probam^2} }$ is first positive, then becomes negative.
%However, as ${\partial^2 {\Urdbm(\Probam , \BProbam)}}/{\partial{\Probam^2} }$ is the linear function of $\Probam$,
Hence, the second order derivative of $\frac{\partial^2 {\Urdbm (\Probam )}}{\partial{\Probam^2} }$ is first positive when $\Probam$ is less than a threshold, and then changes to negative when $\Probam$ is larger than a threshold.
This implies that the first order derivative
$
{\partial {\Urdbm(\Probam , \BProbam)}}/{\partial{\Probam} }
$
first increases (from a positive value), and then decreases to a negative value, hence it changes  the sign only once.
%\rev{
%where xxxx, and
%case 1: then $\frac{\partial^2 {\Urdbm (\Probam )}}{\partial{\Probam^2} }< 0$, and ${\partial {\Urdbm(\Probam , \BProbam)}}/{\partial{\Probam} }$ changes the sign once;
%case 2: xxxxx, then the value of ${\partial {\Urdbm(\Probam , \BProbam)}}/{\partial{\Probam} }$ is first negative, then becomes positive. However, as ${\partial {\Urdbm(\Probam , \BProbam)}}/{\partial{\Probam} }$ is the linear function of $\Probam$, then ${\partial {\Urdbm(\Probam , \BProbam)}}/{\partial{\Probam} }$ only changes the sign once.
%}
\end{proof}

\subsection{Proof for Proposition \ref{thrm:NE-uniquness}}\label{thrm:NE-uniquness-proof}
\begin{proof}
To prove the uniqueness of NE, we only need to verify that:
\begin{equation}
\begin{aligned}
  - \frac{  \partial^2{ \Urdbm(\Probam, \BProbam) } }{ \partial{ (-{\Probam}) }^2 } \geq \sum_{j\neq m} \frac{  \partial^2{ \Urdbm( \Probam, \BProbam)  } }{ \partial{ {(-\Probam) } }\partial{ \Proban  } }, \quad \forall m\in\Mset.
\end{aligned}
\end{equation}

The above condition is usually called the dominant diagonal condition.
Hence, as long as $\Urdbm(\Probam, \BProbam)$  satisfy the above condition,  the MSCG has a unique NE, so does the original PCG \cite{topkis1998supermodular}.
\end{proof}

\end{document}